\documentclass[twocolumn]{aastex63}

\usepackage[utf8]{inputenc}
\usepackage{natbib}
\usepackage{graphicx}
\usepackage{amsmath}
\usepackage{booktabs}
\usepackage{longtable}
\usepackage{xspace}
\usepackage{hyperref}
\usepackage[T1]{fontenc}
\usepackage{lipsum}
\usepackage{comment}
\usepackage{xcolor}
\usepackage{multirow}
\usepackage{IEEEtrantools}

\newcommand{\mearth}{\mbox{M$_{\Earth}$}}
\newcommand{\rearth}{\mbox{R$_{\Earth}$}}
\newcommand{\gcc}{\mbox{$\text{g cm}^{-3}$}}

\shorttitle{TKS XX: Mass Catalog}
\shortauthors{Polanski et al.}

\begin{document}

\title{The TESS-Keck Survey XX: 15 New TESS Planets and a Uniform RV Analysis of all Survey Targets}


\author[0000-0001-7047-8681]{Alex S. Polanski}
\altaffiliation{IPAC Visiting Graduate Research Fellow}
\affil{Department of Physics and Astronomy, University of Kansas, Lawrence, KS 66045, USA}

\author[0000-0001-8342-7736]{Jack Lubin}
\affiliation{Department of Physics and Astronomy, University of California, Irvine, CA 92697, USA}
\affiliation{Department of Physics \& Astronomy, University of California Los Angeles, Los Angeles, CA 90095, USA}

\author[0000-0001-7708-2364]{Corey Beard}
\altaffiliation{NASA FINESST Fellow}
\affiliation{Department of Physics and Astronomy, University of California, Irvine, CA 92697, USA}

\author[0000-0001-8898-8284]{Joseph M. Akana Murphy}
\altaffiliation{NSF Graduate Research Fellow}
\affiliation{Department of Astronomy and Astrophysics, University of California, Santa Cruz, CA 95060, USA}

\author[0000-0003-3856-3143]{Ryan Rubenzahl}
\altaffiliation{NSF Graduate Research Fellow}
\affiliation{Department of Astronomy, California Institute of Technology, Pasadena, CA 91125, USA}

\author[0000-0002-0139-4756]{Michelle L. Hill}
\affiliation{Department of Earth and Planetary Sciences, University of California, Riverside, CA 92521, USA}

\author{Ian J. M. Crossfield}
\affil{Department of Physics and Astronomy, University of Kansas, Lawrence, KS 66045, USA}


\author[0000-0003-1125-2564]{Ashley Chontos}
\altaffiliation{Henry Norris Russell Fellow}
\affiliation{Department of Astrophysical Sciences, Princeton University, 4 Ivy Lane, Princeton, NJ 08540, USA}
\affiliation{Institute for Astronomy, University of Hawai`i, 2680 Woodlawn Drive, Honolulu, HI 96822, USA}

\author[0000-0003-0149-9678]{Paul Robertson}
\affiliation{Department of Physics and Astronomy, University of California, Irvine, CA 92697, USA}

\author[0000-0002-0531-1073]{Howard Isaacson}
\affiliation{{Department of Astronomy, University of California Berkeley, Berkeley CA 94720, USA}}
\affiliation{Centre for Astrophysics, University of Southern Queensland, Toowoomba, QLD, Australia}

\author[0000-0002-7084-0529]{Stephen R. Kane}
\affiliation{Department of Earth and Planetary Sciences, University of California, Riverside, CA 92521, USA}

\author[0000-0002-5741-3047]{David R. Ciardi}
\affiliation{NASA Exoplanet Science Institute/Caltech-IPAC, 1200 E. California Blvd., Pasadena, CA 91125, USA}

\author[0000-0002-7030-9519]{Natalie M. Batalha}
\affiliation{Department of Astronomy and Astrophysics, University of California, Santa Cruz, CA 95060, USA}

\author[0000-0001-8189-0233]{Courtney Dressing}
\affiliation{{Department of Astronomy, University of California Berkeley, Berkeley CA 94720, USA}}

\author[0000-0003-3504-5316]{Benjamin Fulton}
\affiliation{NASA Exoplanet Science Institute/Caltech-IPAC, 1200 E. California Blvd., Pasadena, CA 91125, USA}

\author[0000-0001-8638-0320]{Andrew W. Howard}
\affiliation{Department of Astronomy, California Institute of Technology, Pasadena, CA 91125, USA}

\author[0000-0001-8832-4488]{Daniel Huber}
\affiliation{Institute for Astronomy, University of Hawai`i, 2680 Woodlawn Drive, Honolulu, HI 96822, USA}
\affiliation{Sydney Institute for Astronomy (SIfA), School of Physics, University of Sydney, NSW 2006, Australia}

\author[0000-0003-0967-2893]{Erik A. Petigura}
\affiliation{Department of Physics \& Astronomy, University of California Los Angeles, Los Angeles, CA 90095, USA}

\author[0000-0002-3725-3058]{Lauren M. Weiss}
\affiliation{Department of Physics and Astronomy, University of Notre Dame, Notre Dame, IN, 46556, USA}


\author[0000-0002-9751-2664]{Isabel Angelo}
\affiliation{Department of Physics and Astronomy, University of California, Los Angeles, CA 90095, USA}
\affiliation{Mani L. Bhaumik Institute for Theoretical Physics, University of California, Los Angeles, CA 90095, USA}

\author[0000-0003-0012-9093]{Aida Behmard}
\affiliation{Department of Astrophysics, American Museum of Natural History, 200 Central Park West, Manhattan, NY 10024, USA}

\author[0000-0002-3199-2888]{Sarah Blunt}
\affiliation{Department of Astronomy, California Institute of Technology, Pasadena, CA 91125, USA}
\affiliation{CIERA, Northwestern University, Evanston IL 60201}

\author[0000-0002-4480-310X]{Casey L. Brinkman}
\affiliation{Institute for Astronomy, University of Hawai'i, Honolulu, HI 96822, USA}

\author[0000-0002-8958-0683]{Fei Dai} 
\affiliation{Division of Geological and Planetary Science, California Institute of Technology, Pasadena, CA 91125, USA}
\affiliation{Department of Astronomy, California Institute of Technology, Pasadena, CA 91125, USA}
\affiliation{NASA Sagan Fellow}

\author[0000-0002-4297-5506]{Paul A.\ Dalba}
\affiliation{Department of Astronomy and Astrophysics, University of California, Santa Cruz, CA 95060, USA}

\author[0000-0002-3551-279X]{Tara Fetherolf}
\affiliation{Department of Physics, California State University, San Marcos, CA 92096, USA}
\affiliation{Department of Earth and Planetary Sciences, University of California, Riverside, CA 92521, USA}

\author[0000-0002-8965-3969]{Steven Giacalone}
\altaffiliation{NSF Astronomy and Astrophysics Postdoctoral Fellow}
\affiliation{Department of Astronomy, California Institute of Technology, Pasadena, CA 91125, USA}

\author[0000-0001-8058-7443]{Lea A.\ Hirsch}
\affiliation{University of Toronto  Mississauga, 3359 Mississauga Road
Mississauga ON L5L 1C6
Canada}

\author[0000-0002-5034-9476]{Rae Holcomb}
\affiliation{Department of Physics and Astronomy, University of California, Irvine, CA 92697, USA}

\author[0000-0002-6115-4359]{Molly R. Kosiarek}
\altaffiliation{NSF Graduate Research Fellow}
\affiliation{Department of Astronomy and Astrophysics, University of California, Santa Cruz, CA 95064, USA}

\author[0000-0002-7216-2135]{Andrew W. Mayo}
\affiliation{{Department of Astronomy, University of California Berkeley, Berkeley CA 94720, USA}}

\author[0000-0003-2562-9043]{Mason G.\ MacDougall}
\affiliation{Department of Physics \& Astronomy, University of California Los Angeles, Los Angeles, CA 90095, USA}

\author[0000-0003-4603-556X]{Teo Mo\v{c}nik}
\affiliation{Gemini Observatory/NSF's NOIRLab, 670 N. A'ohoku Place, Hilo, HI 96720, USA}

\author[0000-0001-9771-7953]{Daria Pidhorodetska} 
\affiliation{Department of Earth and Planetary Sciences, University of California, Riverside, CA 92521, USA}

\author[0000-0002-7670-670X]{Malena Rice}
\affiliation{Department of Astronomy, Yale University, New Haven, CT 06511, USA}

\author{Lee J.\ Rosenthal}
\affiliation{Department of Astronomy, California Institute of Technology, Pasadena, CA 91125, USA}

\author[0000-0003-3623-7280]{Nicholas Scarsdale}
\affiliation{Department of Astronomy and Astrophysics, University of California, Santa Cruz, CA 95060, USA}

\author[0000-0002-1845-2617]{Emma V. Turtelboom}
\affiliation{{Department of Astronomy, University of California Berkeley, Berkeley CA 94720, USA}}

\author[0000-0003-0298-4667]{Dakotah Tyler}
\affiliation{Department of Physics and Astronomy, University of California, Los Angeles, CA 90095, USA}

\author[0000-0002-4290-6826]{Judah Van Zandt}
\affiliation{Department of Physics \& Astronomy, University of California Los Angeles, Los Angeles, CA 90095, USA}

\author[0000-0001-7961-3907]{Samuel W.\ Yee}
\affiliation{Department of Astrophysical Sciences, Princeton University, 4 Ivy Lane, Princeton, NJ 08540, USA}
\affiliation{Center for Astrophysics \textbar \ Harvard \& Smithsonian, 60 Garden St, Cambridge, MA 02138, USA}
\altaffiliation{51 Pegasi b Fellow}


\author[0000-0002-1221-5346]{David R. Coria}
\affil{Department of Physics and Astronomy, University of Kansas, Lawrence, KS 66045, USA}

\author[0000-0001-6441-0242]{Shannon D. Dulz}
\affiliation{Department of Physics and Astronomy, University of Notre Dame, Notre Dame, IN, 46556, USA}

\author[0000-0001-8732-6166]{Joel D. Hartman}
\affiliation{Department of Astrophysical Sciences, Princeton University, 4 Ivy Lane, Princeton, NJ 08540, USA}

\author[0000-0002-5812-3236]{Aaron Householder}
\affiliation{Department of Earth, Atmospheric and Planetary Sciences, Massachusetts Institute of Technology, Cambridge, MA 02139, USA}
\affil{Kavli Institute for Astrophysics and Space Research, Massachusetts Institute of Technology, Cambridge, MA 02139, USA}

\author[0000-0002-6052-3562]{Sarah Lange}
\affiliation{Department of Astronomy and Astrophysics, University of California, Santa Cruz, CA 95064, USA}

\author[0000-0001-5312-649X]{Andrew Langford}
\altaffiliation{NSF Graduate Research Fellow}
\affil{Institute of Astronomy, University of Cambridge, Madingley Road, Cambridge, CB3 0HA, UK}
\affil{School of Aeronautical and Astronautical Engineering, Purdue University, West Lafayette, IN, USA}

\author[0000-0003-3179-5320]{Emma M. Louden}
\affiliation{Department of Astronomy, Yale University, New Haven, CT 06511, USA}


\author[0000-0002-9337-0902]{Jared C. Siegel}
\affiliation{Department of Astrophysical Sciences, Princeton University, 4 Ivy Lane, Princeton, NJ 08540, USA}



\author[0000-0002-0388-8004]{Emily A. Gilbert}
\affiliation{Jet Propulsion Laboratory, California Institute of Technology, 4800 Oak Grove Drive, Pasadena, CA 91109, USA}

\author[0000-0002-9329-2190]{Erica J. Gonzales}
\altaffiliation{NSF Graduate Research Fellow}
\affiliation{Department of Astronomy and Astrophysics, University of California, Santa Cruz, CA 95064, USA}

\author{Joshua E. Schlieder}
\affiliation{NASA Goddard Space Flight Center, 8800 Greenbelt Road, Greenbelt, MD 22071, USA}

\author[0000-0001-6037-2971]{Andrew W. Boyle}
\affiliation{Department of Astronomy, California Institute of Technology, Pasadena, CA 91125, USA}

\author[0000-0002-8035-4778]{Jessie L. Christiansen}
\affiliation{NASA Exoplanet Science Institute/Caltech-IPAC, 1200 E. California Blvd., Pasadena, CA 91125, USA}

\author[0000-0002-2361-5812]{Catherine A. Clark}
\affiliation{Jet Propulsion Laboratory, California Institute of Technology, 4800 Oak Grove Drive, Pasadena, CA 91109, USA}
\affiliation{NASA Exoplanet Science Institute/Caltech-IPAC, 1200 E. California Blvd., Pasadena, CA 91125, USA}

\author[0000-0002-3853-7327]{Rachel B. Fernandes}
\altaffiliation{President's Postdoctoral Fellow}
\affil{Department of Astronomy \& Astrophysics, 525 Davey Laboratory, The Pennsylvania State University, University Park, PA 16802, USA}
\affil{Center for Exoplanets and Habitable Worlds, 525 Davey Laboratory, The Pennsylvania State University, University Park, PA 16802, USA}

\author[0000-0003-2527-1598]{Michael B. Lund}
\affil{NASA Exoplanet Science Institute/Caltech-IPAC, 1200 E. California Blvd., Pasadena, CA 91125, USA}


\author[0000-0002-2454-768X]{Arjun B. Savel}
\affil{Department of Astronomy, University of Maryland, College Park, MD 20742, USA}

\author[0000-0001-6171-7951]{Holden Gill}
\affiliation{{Department of Astronomy, University of California Berkeley, Berkeley CA 94720, USA}}

\author{Charles Beichman}
\affiliation{NASA Exoplanet Science Institute/Caltech-IPAC, 1200 E. California Blvd., Pasadena, CA 91125, USA}

\author{Rachel Matson}
\affiliation{U.S. Naval Observatory, 3450 Massachusetts Avenue NW, Washington, D.C. 20392, USA}

\author[0000-0003-0593-1560]{Elisabeth C. Matthews}
\affiliation{Max-Planck-Institut f\"ur Astronomie, K\"onigstuhl 17, 69117 Heidelberg, Germany}


\author[0000-0001-9800-6248]{E. Furlan}
\affiliation{NASA Exoplanet Science Institute/Caltech-IPAC, 1200 E. California Blvd., Pasadena, CA 91125, USA}

\author[0000-0002-2532-2853]{Steve~B.~Howell}
\affil{NASA Ames Research Center, Moffett Field, CA 94035, USA}

\author[0000-0003-1038-9702]{Nicholas~J.~Scott}
\affil{NASA Ames Research Center, Moffett Field, CA 94035, USA}


\author[0000-0002-0885-7215]{Mark~E.~Everett}
\affiliation{NSF’s National Optical-Infrared Astronomy Research Laboratory, 950 N. Cherry Ave., Tucson, AZ 85719, USA}

\author[0000-0002-4881-3620]{John~H.~Livingston}
\affiliation{Astrobiology Center, NINS, 2-21-1 Osawa, Mitaka, Tokyo 181-8588, Japan}
\affiliation{National Astronomical Observatory of Japan, NINS, 2-21-1 Osawa, Mitaka, Tokyo 181-8588, Japan}
\affiliation{Astronomical Science Program, Graduate University for Advanced Studies, SOKENDAI, 2-21-1, Osawa, Mitaka, Tokyo, 181-8588, Japan}

\author[0009-0003-9919-0616]{Irina O. Ershova}
\affiliation{Faculty of Space Research, Moscow State University, 1 Leninskie Gory, bld. 52, Moscow, Russia}

\author[0009-0003-4203-9667]{Dmitry V. Cheryasov}
\affiliation{Sternberg Astronomical Institute, Lomonosov Moscow State University, 119992, Universitetskij prospekt 13, Moscow, Russia}

\author{Boris Safonov}
\affiliation{Sternberg Astronomical Institute, Lomonosov Moscow State University, 119992, Universitetskij prospekt 13, Moscow, Russia}


\author[0000-0003-3742-1987]{Jorge Lillo-Box}
\affiliation{Centro de Astrobiolo\'ia, CSIC-INTA, ESAC Campus, Camino Bajo del Castillo s/n, 28692 Villanueva de la Ca\~nada, Madrid, Spain}

\author[0000-0002-5971-9242]{David Barrado}
\affiliation{Centro de Astrobiolo\'ia, CSIC-INTA, ESAC Campus, Camino Bajo del Castillo s/n, 28692 Villanueva de la Ca\~nada, Madrid, Spain}

\author[0000-0001-9526-9499]{Mar\'ia Morales-Calder\'on}
\affiliation{Centro de Astrobiolo\'ia, CSIC-INTA, ESAC Campus, Camino Bajo del Castillo s/n, 28692 Villanueva de la Ca\~nada, Madrid, Spain}

\submitjournal{Astrophysical Journal Supplements}
\published{May 23, 2024}

\begin{abstract}
\noindent The Transiting Exoplanet Survey Satellite (TESS) has discovered hundreds of new worlds, with TESS planet candidates now outnumbering the total number of confirmed planets from \textit{Kepler}. Owing to differences in survey design, TESS continues to provide planets that are better suited for subsequent follow-up studies, including mass measurement through radial velocity (RV) observations, compared to Kepler targets.
In this work, we present the TESS-Keck Survey's (TKS) Mass Catalog: a uniform analysis of all TKS RV survey data which has resulted in mass constraints for 126 planets and candidate signals. This includes 58 mass measurements that have reached $\geq5\sigma$ precision. We confirm or validate 32 new planets from the TESS mission either by significant mass measurement (15) or statistical validation (17), and we find no evidence of likely false positives among our entire sample. This work also serves as a data release for all previously unpublished TKS survey data, including 9,204 RV measurements and associated activity indicators over our three year survey. We took the opportunity to assess the performance of our survey, and found that we achieved many of our goals including measuring the mass of 38 small ($<4\rearth$) planets, nearly achieving the TESS mission's basic science requirement. In addition, we evaluated the performance of the Automated Planet Finder (APF) as survey support and observed meaningful constraints on system parameters due to its more uniform phase coverage. Finally, we compared our measured masses to those predicted by commonly used mass-radius relations and investigated evidence of systematic bias.
\end{abstract}

\keywords{exoplanet, mass-radius}

\section{Introduction}

The \textit{Kepler} \citep{Borucki2010} mission revealed a rich exoplanet landscape showing that planets with short orbital periods (P$<$100 days) between the size of Earth and Neptune (1--4~$\rearth$) are ubiquitous in our galaxy \citep{Dressing2013,Fulton2018}. However, the stars \textit{Kepler} targeted are mostly too distant, and hence too faint, for precise RV follow up. In contrast, the Transiting Exoplanet Survey Satellite \citep[TESS,][]{Ricker2015} mission was designed to preferentially discover planets around brighter, closer host stars that are highly amenable to follow up efforts, including known planetary systems \citep{Kane2009,Kane2021}. The current era of precise RV mass measurements is beginning to reveal structure within the mass-radius distribution of small planets. For cool M dwarfs, \cite{Luque2022} found that the dichotomy of sub-Neptunes and super-Earths may indeed be a \textit{trichotomy} composed of rocky, icy, and gaseous planets. While the connection between these groups and specific formation scenarios is debated \citep{Rogers2023}, resolving nuances in this region of mass-radius space is a significant motivator to expand the sample of planets around FGK stars with precisely determined masses.

Understanding the processes that result in the observed planet population hinges not only on well-characterized planets, but also on the mitigation of bias and non-astrophysical scatter in datasets. This is especially important for the derivation of mass-radius relations. For example, publication bias \citep{Burt2018} can not only bias a mass-radius relation to higher masses but may also artificially decrease intrinsic scatter. Additionally, uniform analyses have the power to eliminate scatter that may be inherent in heterogeneous data sets such as those compiled from various literature sources. Previous derivations of mass-radius relations have found evidence for more scatter in planet masses than can be described by measurement error alone \citep{WeissMarcy2014,Wolfgang2016}. Some of this intrinsic scatter arises from the diversity of exoplanet compositions, but some may also be due to heterogeneous datasets. Mass-radius relations derived from homogeneous samples have the potential to more clearly distinguish between scatter caused by real compositional diversity as opposed to differences in analysis and target selection \citep{Teske2021}.

The plethora of planets now accessible with ground-based spectrographs has given rise to large collaborations that are able to pool many nights of telescope time together to follow up these targets and amass RV time series data. The TESS Follow-up Observing Program effort\footnote{https://tess.mit.edu/followup/} is a global collaboration encompassing many different efforts to confirm planet candidates including transit recovery, high contrast imaging, and RV follow up. Within RV follow up, large collaborations include the GAPS programme \citep{Covino2013}, the HARPS-N Rocky Planet Search \citep{Motalebi2015}, the CARMENES search for exoplanets around M dwarfs \citep{Reiners2018}, the Magellan-TESS survey \citep{Teske2021}, and many others. Another major RV follow-up effort, which is the focus of this work, is the TESS-Keck Survey (TKS).

TKS is an RV follow-up program that targeted TESS-discovered systems with Keck/HIRES and APF/Levy. The main goals of TKS, as well as our target selection function, are detailed in full by \citet[][hereafter TKS0]{TKS0}, with the primary goal being to measure 100 planet masses over a 3 year time span. Our target selection process\footnote{https://github.com/ashleychontos/sort-a-survey} identified TESS Objects of Interest (TOIs) that were valuable across both complementary and competing science cases. This resulted in a final target list of 108 transiting planet candidates in 86 systems\footnote{Targets in this work are sometimes referred to by their TKS ID which is the name used in the California Planet Search queue system. For reference, stars in the Henry Draper catalog are referred to by their numerical designation (e.g. HD 95072 $\rightarrow$ 95072) while TOIs are given a short-hand reference (e.g. TOI--1776 $\rightarrow$ T001776)  }. While many TKS publications have focused on the detailed characterization of individual systems, in this work we combines the homogeneous stellar properties of \cite{TKSXV} with a uniform analysis of all RV survey data to compile a catalog of planet masses, eccentricities, and densities. This includes a reanalysis of RV data for previously confirmed planets in addition to 32 unpublished TESS systems.

\par The paper is organized as follows. \S\ref{sec:data} describes our data collection for high resolution imaging and RV. \S\ref{sec:methods} details the algorithm we developed for homogeneously derived masses for all planets in our sample and we expand upon caveats to this algorithm in \S\ref{sec:caveats}. In \S\ref{sec:results}, we compare our measured masses with those previously published in TKS and in the wider literature. In \S\ref{sec:TKS_goals} we discuss how well our survey met its original goals, assess the contribution of APF to our survey, and in \S\ref{sec:exp_meas_mass} we review the ability of commonly used mass-radius relations to predict a planet's mass.

\section{Observations}\label{sec:data}

\subsection{Radial Velocities}\label{sec:RV}

All of our spectroscopic data used in this analysis were taken entirely by the High Resolution Echelle Spectrometer \citep[HIRES,][]{vogt94} on the Keck-I telescope at W. M. Keck Observatory, and the Levy Spectrometer located on the Automated Planet Finder \citep[APF,][]{vogt14} telescope. HIRES is an echelle spectrometer with a resolving power of $\sim$67,000 depending on the slit used and a wavelength range spanning from 374--970 nm. We extracted RVs from the spectroscopic data (reduced using the standard procedure described in \citealt{Howard2010}), using the iodine cell method outlined in \cite{butler96}. The APF is a robotic telescope operated at Lick Observatory \citep{APF}. The Levy spectrograph has a resolving power of 100,000, and also utilizes the same iodine cell method to extract precise RVs as for HIRES. The APF has the same wavelength span as HIRES; 374--970 nm.




The total survey includes 7,894 HIRES RVs taken over the course of 974 Keck observing nights when including archival RVs. Narrowing to only HIRES RVs taken during TKS observing programs, we collected 4,943 HIRES RVs over 301 observing nights. We also use 5,989 APF RVs taken over the course of 1,205 APF observing nights. Again, narrowing to only observations taken during TKS-affiliated observing semesters, we acquired 4,261 APF RVs over 710 observing nights. 


Observational baseline is a key component of RV surveys, as a long baseline allows for improved sensitivity \textcolor{black}{to long-period, non-transiting companions.} When utilizing archival HIRES RVs, our maximum survey baseline was 6,502 days, and our minimum was 59 days. The average RV baseline of the survey was 1,427 days. A more detailed breakdown by instrument is visible in Table~\ref{tab:obs_stats}. In addition, many of the stars in our survey have RVs from other instruments, but for this work we do not include non-TKS data.

In general, the TKS survey was broken down into a number of different science cases with different observation and cadence goals, and such science cases were the primary drivers in our observing strategy. Rather than detail all of the TKS science cases here, we refer the interested reader to \cite{TKS0}, which describes our strategies in detail.

\begin{deluxetable*}{lllll}
\label{tab:obs_stats}
\tabletypesize{\footnotesize}
\tablecaption{Observation Statistics}
\tablehead{\colhead{Statistic} &
\colhead{HIRES (w/ archival)} & \colhead{APF (w/ archival)} & \colhead{HIRES (TKS only)} & \colhead{APF (TKS only)}
}
\startdata
N$_{RV}$ & 7894 & 5989 & 4943 & 4261 \\
Nights Observed & 974 & 1205 & 301 & 710 \\
Longest Baseline (days) & 6502 & 3391  & 1076 & 1249 \\
Shortest Baseline (days) & 59 & 0 & 59 & 0 \\
Average Baseline (days) & 1427 & 965 & 828 & 745 \\
\enddata
\end{deluxetable*}

\subsection{High Resolution Imaging}\label{sec:HRI}

High resolution imaging (HRI) is an important step in validating planet candidates due to its ability to resolve additional stars not resolved by \textit{Gaia}. Visual binary companions, regardless of whether they are gravitationally bound or not, can dilute the TESS lightcurve leading to underestimated planet radii \citep{Ciardi2015}. In this section, we describe the observations taken of the systems that are newly validated in this work and discuss the presence of a companion to TOI-1443. Since many of our TOIs were observed by multiple instruments, in Table \ref{tab:hri_obs} we provide a summary of the imaging observations made.

\subsubsection{Keck-NIRC2}

NIRC2 is an adaptive optics enabled imaging camera on the on the 10 m Keck II telescope. Narrow angle mode was used, providing a pixel scale of around 0.01 \arcsec px$^{-1}$ and full field of view of about 10\arcsec. Observations were taken in either the Br-$\gamma$, K, or K' filters. Observations were taken using the natural guide star AO system in the standard three-point dither pattern to avoid the lower-left quadrant of the detector, which is typically noisier than the other three quadrants. The dither pattern has a step size of 3\arcsec. Each dither position was observed three times, with 0.5\arcsec~ positional offsets between each observation, for a total of nine frames.

\subsubsection{Palomar-PHARO}

PHARO is an adaptive optics enabled imaging camera on the 5.1 m Hale telescope with a pixel scale of 0.025 \arcsec px$^{-1}$ and a full field of view of around 25\arcsec. Observations were taken in the Br-$\gamma$ and acquired using the natural guide star AO system P3K \citep{Dekany2013} in the standard five-point \texttt{quincunx} dither pattern with steps of 5\arcsec. Each dither position
was observed three times, with 0.5\arcsec~ positional offsets between each observation, for a total of 15 frames.

\subsubsection{Shane-ShARCS}

ShARCS is an adaptive optics enabled camera on the Shane 3 m telescope at Lick Observatory \citep{2012SPIE.8447E..3GK, 2014SPIE.9148E..05G, 2014SPIE.9148E..3AM}. Observations were taken with the Shane adaptive optics system in natural guide star mode in order to search for nearby, unresolved stellar companions. We used a variety of filters including $K_s$ ($\lambda_0 = 2.150$ $\mu$m, $\Delta \lambda = 0.320$ $\mu$m) and $J$ ($\lambda_0 = 1.238$ $\mu$m, $\Delta \lambda = 0.271$ $\mu$m). Information about our observing strategy can be found in \citet{2020AJ....160..287S}. We reduced the data using the publicly available \texttt{SImMER} pipeline \citep{2022PASP..134l4501S}.\footnote{https://github.com/arjunsavel/SImMER}. Further information about these data can be found in Dressing et al. (in prep) and on ExoFOP.

\subsubsection{Gemini-NIRI}

NIRI \citep{Hodapp2003} is an apdative optics enabled imaging camera on the Gemini North 8 m telescope. In each case we observed using the Br$\gamma$ filter centered on 2.166$\micron$, and collected 9 individual frames, dithering the telescope in a grid-pattern between each frame. Integration times for individual frames were calculated based on the stellar magnitude, and were 4.4s, 6.6s and 4s respectively for TOI--1184, TOI--1443, and TOI--1451. To process the data, we first constructed a sky background by median-combining the dithered science frames. We then removed bad pixels, flat fielded, and subtracted the sky background. Cleaned frames were aligned by fitting the stellar PSF in each frame, and coadded. Sensitivity of each observation was calculated as a function of radius by injecting fake companions and scaling their background so they could be redetected at 5$\sigma$. For visual companions identified in the Gemini-NIRI data, we calculated astrometry relative to the host-star by fitting a Gaussian to each of the star and companion, and then applying the pixel scale of 0.021\arcsec/pixel. We performed aperture photometry on both the star and the visual companion in our reduced images, using a small aperture around the position determined in the previous section.

For each target, we reached contrast sensitivities of at least 5mag beyond $\sim$300mas, and at least 7mag in the background limited regime beyond $\sim$1\arcsec; images have a field of view of 26$\times$26\arcsec, though sensitivity is reduced in the outermost $\sim$3\arcsec where fewer dithered images are being combined.

\begin{figure}
    \centering
    \includegraphics[width=0.45\textwidth]{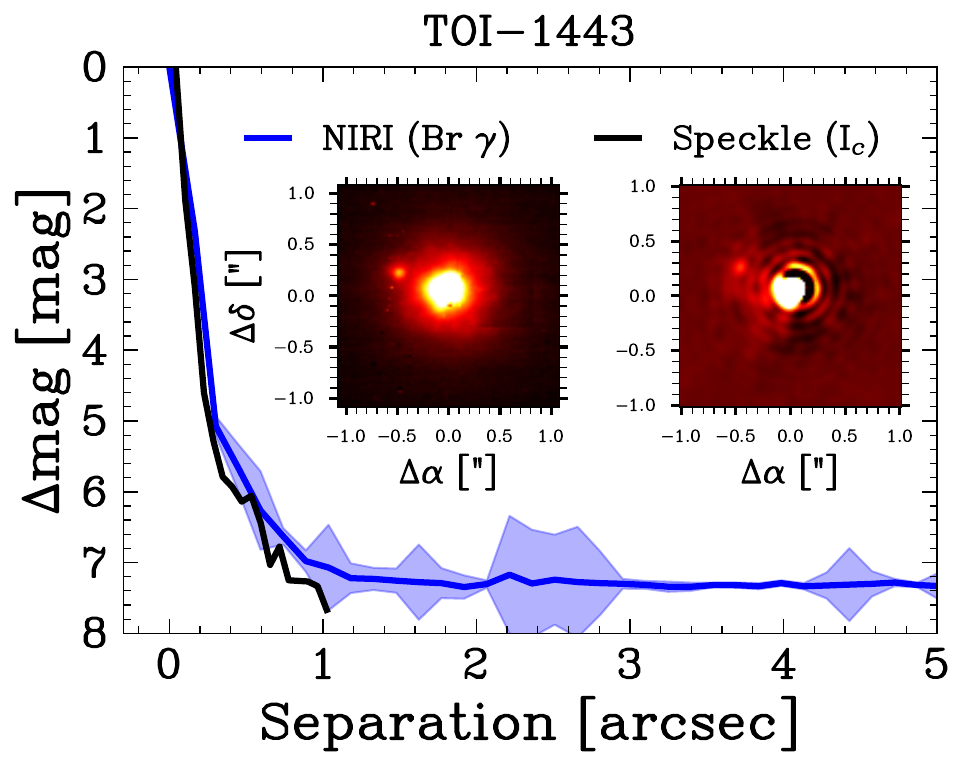}
    \caption{Contrast curves for TOI-1443 from Gemini-NIRI (blue) and SAI Speckle Polarimeter (black). The 1$\sigma$ error for the NIRI contrast curve is given as the blue envelope. A companion is observed for TOI-1443 at $\sim$0.5\arcsec that is not detected in \textit{Gaia} DR3.  }
    \label{fig:toi1443_contrast}
\end{figure}

\subsubsection{Gemini-'Alopkeke \& Zorro}

‘Alopeke and Zorro are high-resolution optical speckle instruments mounted on the Gemini North and South 8 m telescopes respectively. ‘Alopeke and Zorro are identical instruments which provide simultaneous speckle imaging in two bands, usually operated using narrow-band filters centered at 562nm and 832 nm\footnote {https://www.gemini.edu/sciops/instruments/alopeke-zorro/}. Speckle imaging consists of obtaining many thousands of very short exposures (typically 60 msec/frame at Gemini) which are then combined and subjected to Fourier analysis in our standard reduction pipeline \citep[see][]{Howell2011,Horch2012}. Output data products include a reconstructed image in each filter and determination of robust 5$\sigma$ contrast limits on companion detections in each bandpass. If a close (0.02-1.2\arcsec) companion source is discovered, its position angle, separation, and flux ratio are reported. The instruments are described in detail in \cite{Scott2021}.

\subsubsection{WIYN-NESSI}

The NESSI speckle imager \citep{Scott2019} is mounted on the 3.5 m WIYN telescope at Kitt Peak. NESSI simultaneously acquires data in two bands centered at 562\,nm and 832\,nm using high speed electron-multiplying CCDs (EMCCDs). We collected and reduced the data following the procedures described in \citet{Howell2011}. The resulting reconstructed images achieved contrast of $\Delta\mathrm{mag} \sim 6$ at a separation of 1\arcsec~ in the 832\,nm band.

\subsubsection{Speckle Polarimeter}

The Speckle Polarimeter \citep{Safonov2017} is mounted on the the 2.5 m telescope at the Caucasian Observatory of Sternberg Astronomical Institute (SAI) of Lomonosov Moscow State University. Electron Multiplying CCD Andor iXon 897 was used as a detector. For each target we accumulated 4000 frames with 30 ms exposure times. The detector has a pixel scale of $20.6$ mas pixel$^{-1}$, and the angular resolution was 89 mas. For all targets except TOI--1443 we did not detect any stellar companions.

\subsubsection{AstraLux}

The AstraLux instrument \citep{Hormuth2008} is mounted on the 2.2 m telescope of the Calar Alto Observatory (Almer\'ia, Spain). AstraLux uses the lucky-imaging technique, obtaining a large number of very short exposures (shorter than the coherence time of the atmospheric turbulence) that are subsequently selected and combined. The datacube is then processed by the automatic observatory pipeline \citep{Hormuth2008}, which besides doing the basic reduction of the individual frames, selects the 10\% with the highest Strehl ratio \citep{Strehl1902} and combines them into a final high-spatial resolution image. All images were obtained  in the SDSSz bandpass (peak of the instrument resolution) with an individual exposure time of 10 or 20\,ms (depending on the source and weather conditions) and a field-of-view windowed to $6\times6$ \arcsec. This final image is used to compute the sensitivity limits (also known as the contrast curve). Such limits are computed by using our own developed \texttt{astrasens} package\footnote{\url{https://github.com/jlillo/astrasens}} with the procedure described in \cite{Lillo-Box2012,Lillo-Box2014b}.

\subsubsection{A Companion to TOI--1443}

The only target in our sample where a diluting star is seen is TOI-1443, which was imaged with Gemini-NIRI and the SAI speckle polarimeter (Figure \ref{fig:toi1443_contrast}). TOI--1443 was imaged with the speckle polarimeter in four epochs: 2020--12--26, 2022--06--03, 2023--09--23, 2023--09--29 in the I$_c$ band. Each time the object was resolved, the binarity parameters were determined by approximation of the average power spectrum with the model of binary star \cite{Strakhov2023}. The separation and position angle of secondary companion for these epochs was found to be $518\pm6$~mas and $68.2^{\circ}\pm0.4$, $531\pm6$~mas and $69.3^{\circ}\pm0.3$, $537\pm5$~mas and $68.6^{\circ}\pm0.4$, $507\pm4$~mas and $68.2^{\circ}\pm0.4$. Total proper motion reported by \textit{Gaia} DR3 for TOI--1443 is 90.511~mas/yr. This would yield a total relative shift of 249.54~mas between first and last epoch, if the secondary companion was a background star. We conclude that this binary is a gravitationally bound system. We additionally note that, while orbital motion is statistically detected through these observations, they are not sufficient to provide meaningful orbital constraints.

From \citealt{Monet2003}, TOI--1443 has I$_{c}$ = 9.75 giving the primary an V-I$_{c}$ color of 1.08~mag. Interpolating Table 5 of \citealt{PecautMamajek2013}, this gives an effective temperature around 5300~K (spectral type of K0V), consistent with its spectroscopically derived parameters. The Speckle Polarimeter gives an average magnitude difference between the components of $\Delta$I$_c$ = 5.0 mag while NIRI gives $\Delta$K = 3.64$\pm$0.01 mag. From these values we obtain a K-I$_c$ color of 2.47 mag for the secondary component. Again interpolating Table 5 of \citealt{PecautMamajek2013} provides an approximate effective temperature of 3100~K, consistent with an M4.5V spectral type. A $\Delta$I$_{c}$ of 5.0 mag results in $\sim$0.5\% contamination, assuming the planet transits the primary. 

We also observed a significant RV trend of $0.029\pm0.0044$ m s$^{-1}$ d$^{-1}$. Using \texttt{ethraid} \citep{VanZandt2024}\footnote{https://github.com/jvanzand/ethraid}, we use the RV trend and the NIRI sensitivity curve to constrain the combinations of mass and semi-major axis of an object that could be responsible for the trend. We find that an M4.5V dwarf (M$_{*}\sim0.18$~M$_{\odot}$) at the observed angular separation is a plausible explanation for the observed trend in our RVs lending additional support to a bound scenario. 


\section{Methods}\label{sec:methods}

In this section, we describe the analysis of our survey data. In an effort to achieve homogeneity, we outline the standardized decision tree that informed our RV fits. We also describe the uniform set of priors used. In the few instances where RVs from Keck/HIRES and/or APF exist prior to the inception of our survey, we do include those in this analysis. This decision is motivated by a desire to release these data, and by the fact that the existence of RVs on a given target was a favorable metric in our target selection efforts \citep{TKS0}.

\begin{figure*}[t]
    \centering
    \includegraphics[width=0.85\textwidth]{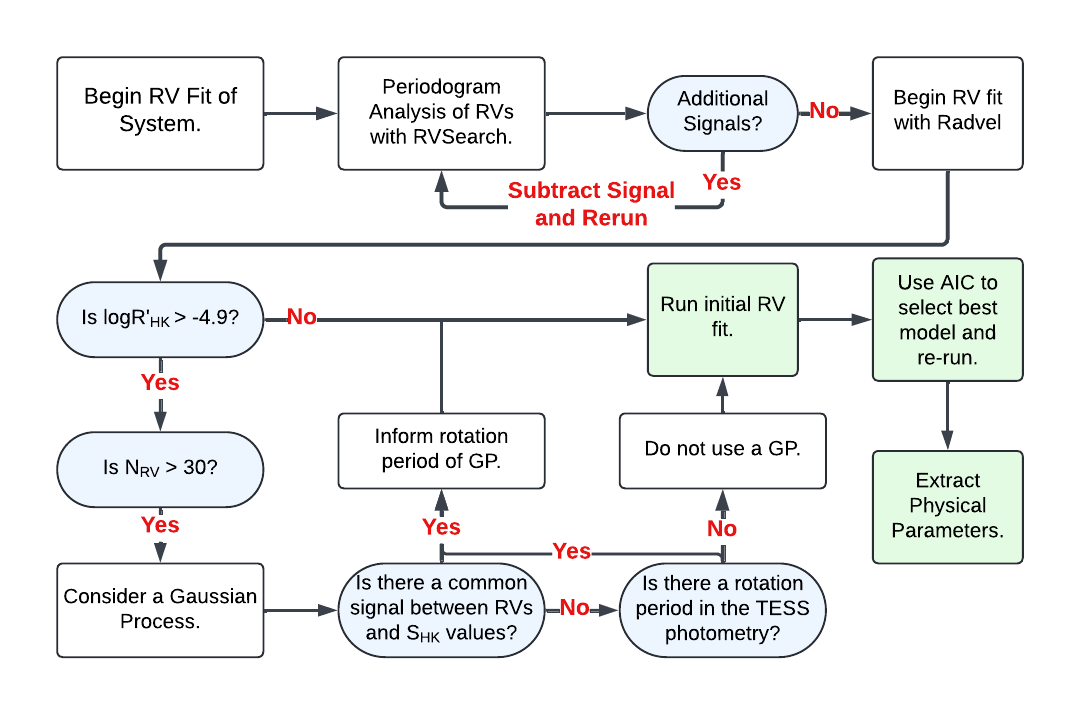}
    \caption{A flowchart of our algorithm for homogeneous determination of planet masses.}
    \label{fig:flowchart}
\end{figure*}

\subsection{Radial Velocity Fitting}\label{sec:fitting}

A schematic of our methodology is shown in Figure \ref{fig:flowchart}. We searched for signals in each RV data set using \texttt{RVSearch} \citep{CLS1_rvsearch}, which uses an iterative approach to fit candidate signals, as determined by False Alarm Probability, FAP > 1\%, and then compare against an N vs N-1 planet model. We seeded the search with parameters of known, transiting planets in each system taken from the TKS Systems Properties Catalog \citep{TKSXV}. 

Once all candidate signals within the RV dataset are identified by \texttt{RVSearch}, we begin our RV fit. Using \texttt{RadVel} \citep{Fulton2018}, we adopt a top-down approach whereby we assume a maximally complex RV model where all parameters (see below) are initially included. Free parameters included Keplerian signal parameters ($P_i$, $T_{0,i}$, $K_i$, $\sqrt{e_i}\sin{(\omega_i)}$, $\sqrt{e_i}\cos{(\omega_i)}$) for all known planets (transiting and non-transiting), and candidate signals, linear ($\dot{\gamma}$) and quadratic ($\ddot{\gamma}$) trend terms, and offset/jitter terms ($\gamma$, $\sigma$) specific to each instrument. 


We ran RadVel in the fitting basis described above and explored the posterior parameter space using Markov-chain Monte Carlo (MCMC) sampling \citep[\texttt{emcee},][]{foreman-mackey13}. This basis reparametrizes e (eccentricity) and $\omega$ (argument of periastron) into $\sqrt{e}\cos{(\omega)}$ and $\sqrt{e}\sin{(\omega)}$ for better MCMC performance.  The MCMC was run until the chains were well-mixed; a G-R statistic less than 1.001 \citep{Gelman1992, Ford2006} and the number of independent samples was greater than 1000 for all free parameters for at least five consecutive checks. We then determined which combination of model parameters was justified using the Akaike Information Criterion (AIC) with the preferred model being the one with the smallest AIC. In cases where the lowest model AIC values differed by $<2$, the model with the fewest free parameters was selected. One exception is when model selection rejects a signal of a known, transiting planet. In these cases we instead accept the model with lowest AIC score that still includes all transiting planets.

We aimed to uniformly derive mass estimates for as many planets as possible. For systems with high planet multiplicity and/or especially active host stars, our methodology may not return the best planet parameters. Nonetheless, the primary goal of this analysis is the homogeneous derivation of masses, not necessarily the best derivation of masses. Similarly, in cases where additional data from other instruments exists, a more precise mass could be determined. We discuss caveats to our uniform fits in \S \ref{sec:caveats} and compare our masses to literature values in \S\ref{sec:lit_comparison}.

\subsection{Selection of Priors}\label{sec:priors}

We implemented a homogeneous set of priors for every planet in the sample detailed in Table \ref{tab:priors}. Parameters such as trend, curvature, and jitter were given broad, uniform priors. For transiting planets, the priors applied to their period and time of conjunction were Gaussian, with centers and widths equal to the posteriors of the transit fits performed in \cite{TKSXV}. Non-transiting planets were given broad Jeffrey's priors \citep{Jeffreys1946} on their orbital period with the initial guess sourced from the \texttt{RVsearch} results. These planets were also given a broad, uniform prior on time of conjunction, with lower and upper bounds extending half of the orbital period. All other free parameters, including K-amplitude, \textit{K}, were unrestricted and allow for negative values. 


\subsection{Gaussian Processes}

A Gaussian process (GP) model can often be very effective at accounting for stellar activity from RV datasets \citep[e.g.,][]{haywood14, lopezmorales16}. GPs are flexible, however, and can overfit on sparse datasets \citep[e.g.,][]{blunt23}. A GP is used on a TKS systems if (1) $\log R^{\prime}_{HK} >$ -4.9, and (2) if N$_{o bs}$ $>$ 30. The first condition is to ensure that stellar activity modeling is indeed warranted, as $\log R^{\prime}_{HK}$ values are found to positively correlate with increased stellar activity \citep{Wright2005, IsaacsonFisher2010,hillenbrand15}. The second condition is intended to prevent overfitting while using a GP.

We used the QuasiPeriodic GP kernel to model activity \citep[][]{haywood14,lopezmorales16}. The covariance between observations taken at times $t_i$ and $t_j$ is given as
\begin{equation}
    \label{eqn:QP}
    \centering
    C_{i,j} = \eta_{1}^{2} * exp(\frac{-|t_{i} - t_{j}|^{2}}{\eta_{2}^{2}} - \frac{\sin^{2}(\frac{\pi|t_{i} - t_{j}|}{\eta_{3}})}{2\eta_{4}^{2}}).
\end{equation}

\noindent An advantage of the QuasiPeriodic GP kernel is the interpretability of its hyperparameters, as most correspond, at least approximately, with a physical parameter. $\eta_{1}$ is the GP amplitude in m s$^{-1}$, and corresponds to the amplitude of the stellar variability. $\eta_{2}$ is the exponential decay length, and approximately represents the lifetime of a stellar spot. $\eta_{3}$ is the recurrence timescale, which is usually well represented by the rotation period of a star. $\eta_{4}$ is the periodic scale length, which varies from 0 to 1 and approximately controls how periodic the structure of the GP signal is. 

Similar to planet parameters, we strove for homogeneous priors on GP hyperparameters (Table \ref{tab:priors}). The GP amplitude $\eta_{1}$ is not known a-priori, and we adopt a broad Jeffreys prior. We also use a broad Jeffreys prior for $\eta_{2}$, though we prevent the value from falling below the system rotation period, $\eta_{3}$, as recommended by \cite{kosiarek20}. $\eta_{3}$ is informed by either a periodogram of the photometry or S-values. Finally, we restrict $\eta_{4}$ using a Gaussian prior recommended by \cite{lopezmorales16}. This is to prevent overfitting with the GP, which can happen for small values of $\eta_{4}$.

Even when a GP is included in our maximally complex initial fit, if the GP is not favored in model comparison, it is not included in our final fit. The fitted values for our GP hyperparameters are given in Table \ref{tab:GPfitted}.

\subsection{Derivation of Physical Parameters}

The stellar and transit properties for this work are taken from \cite{TKSXV}. In summary, stellar spectroscopic properties ($T_{\text{{eff}}}$, [Fe/H], $\log{(g)}$) were derived from HIRES spectra using \texttt{SpecMatch} \citep{SMSynth,SMEmp} while M$_{*}$ and R$_{*}$ were obtained using \texttt{isoclassify} \citep{Huber2017,Berger2020,Berger2023} with the spectroscopic parameters used to inform input priors. Similarly, the transiting planet parameters are from the same study's analysis of TESS lightcurves. In \cite{TKSXV}, stellar parameter uncertainties do not account for systematic errors in model isochrones \citep{Tayar2022} but do provide a recommendation to inflate stellar parameter uncertainties. In the following derivation of physical properties, we make the recommended corrections.

We derive the parameters listed in Table \ref{tab:derived} by multiplying the MCMC chains from each fit. For the planet mass we use the equation:

\begin{equation}
    \label{eqn:msini}
    \centering
    \frac{\text{m}_p\sin{(i)}}{\text{M}_{J}} = \frac{\text{K}}{\text{K}_0} \sqrt{1-e^2} \left(\frac{\text{M}_*}{\text{M}_{\odot}}\right)^{2/3} \left(\frac{P}{1 \text{yr}}\right)^{1/3}
\end{equation}

\noindent where $\text{K}_0$ = 28.4329$\text{m s}^{-1}$. The semi-major axis, $a$, is calculated using Kepler's 3rd law in the form:

\begin{equation}
    \label{eqn:smax}
    \centering
    a = \left( \frac{P^2 ~G ~\text{M}_{Total}}{4\pi^2} \right)^{1/3} 
\end{equation}

The majority of the planets in our sample are transiting allowing us to derive the true mass with knowledge of the planet's inclination. We calculate the inclination of the transiting planets using\footnote{The precise definition of impact parameter includes a factor of $(1-e^2)/(1+e\sin{\omega})$, but given that transiting planets have near-zero $\cos{i}$ this factor usually has a negligible contribution to the derived mass.}:

\begin{equation}
    \label{eqn:impact_param}
    \centering
    b = \frac{a}{\text{R}_*} \cos{(i)}
\end{equation}

\noindent where $a$ is taken from Equation \ref{eqn:smax} and both R$_*$ and $b$ are from \cite{TKSXV}. For R$_*$, we assume Gaussian distributed uncertainties, however, the posterior distribution for the impact parameter in transit analyses is typically non-Gaussian. To reflect this, we use the MCMC chains from \cite{TKSXV}\footnote{The chains for each transit fit are available at \url{https://github.com/mason-macdougall/tks_system_properties}} and use \texttt{sklearn}'s Kernel Density Estimation function to construct the impact parameter's distribution. We sample the resulting distribution to obtain a chain of samples for $b$ of equal length to those used in our \texttt{RadVel} analyses. The chains for $b$, R$_*$, and $a$ are used to derive $i$. For each derived parameter in Table \ref{tab:derived} we list the median value with the 15.6\% and 84.1\% percentiles as uncertainties. 

\begin{figure*}[p]
    \centering
    \includegraphics[width=\textwidth]{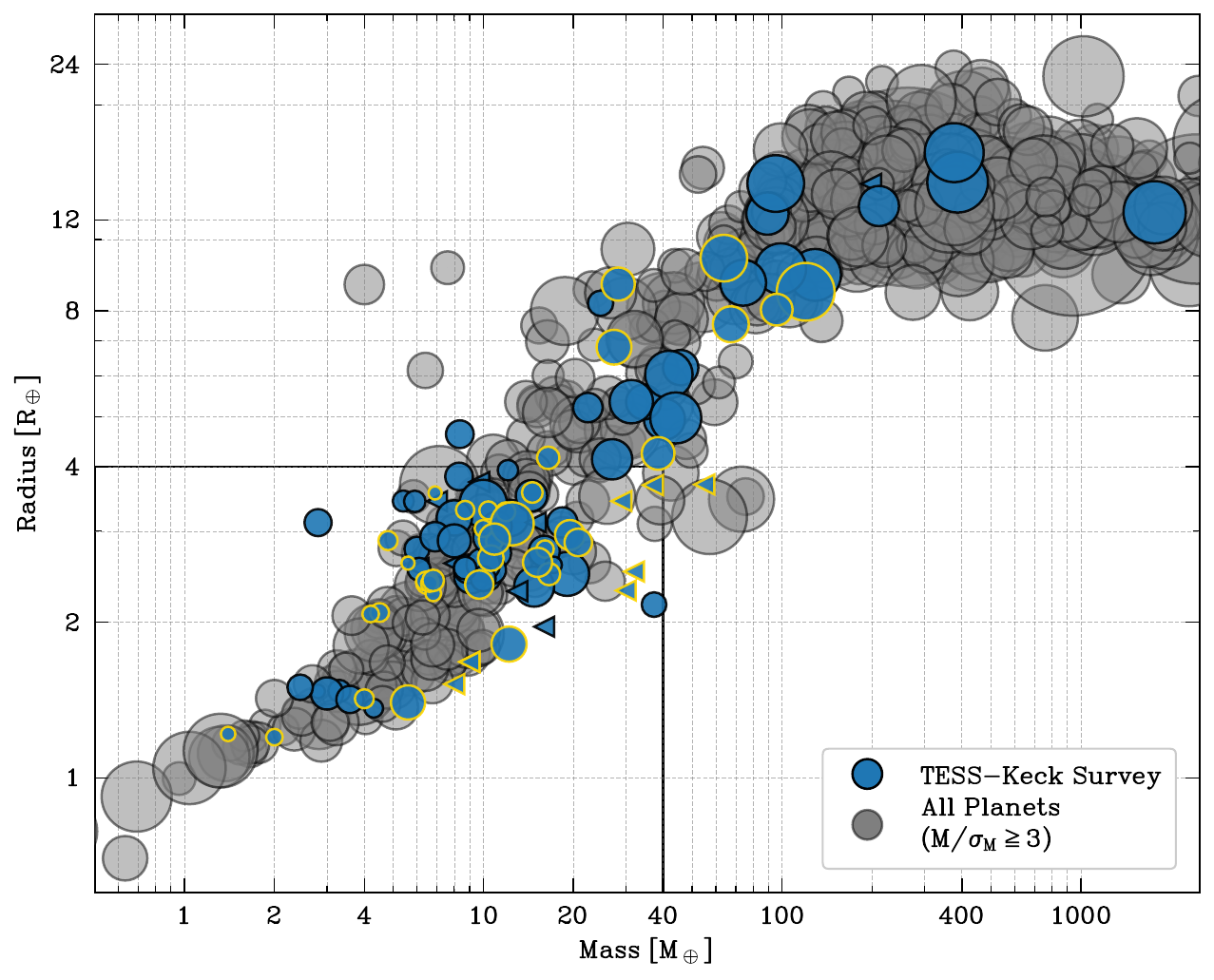}
    \caption{Mass-radius diagram for the full TKS sample (blue) in comparison to the all known exoplanets with a $\geq3\sigma$ mass. Markers are scaled according to the mass precision (size $\propto M/\sigma_{M}$) while triangles represent 3 $\sigma$ upper limits. Points outlined in gold represent newly confirmed and validated planets in our sample. The region outlined in black represents the region of M-R space shown in more detail in Figure \ref{fig:small_mr}.}
    \label{fig:big_mr}
\end{figure*}

\begin{deluxetable*}{llll}
\label{tab:priors}
\tabletypesize{\footnotesize}
\tablecaption{Survey Priors}
\tablehead{\colhead{~~~Parameter Name} &
\colhead{Prior} & \colhead{Units} & \colhead{Description}
}
\startdata
\sidehead{\textbf{Transiting Planets:}}
~~~P$_{orb}$ & $\mathcal{N}^{a}(\mu_{M23}^{*}, sd_{M23})$ & days & Period \\
~~~T$_{c}$ & $\mathcal{N}(\mu_{M23}, sd_{M23})$ & BJD (days) & Time of Inferior Conjunction \\ 
~~~$\sqrt{e} \cos\omega $ & $\mathcal{U}^{b}(-1, 1)$  &  ...  & Eccentricity Reparametrization$^{d}$ \\
~~~$\sqrt{e} \sin\omega $ & $\mathcal{U}(-1, 1)$ & ...  & Eccentricity Reparametrization \\
~~~K & $\mathcal{U}(-\infty,+\infty)$ & m s$^{-1}$  & Velocity Semi-amplitude \\
\sidehead{\textbf{Non-Transiting Planets:}}
~~~P$_{orb}$ & $\mathcal{J}^{c}(1, \rm{Observation Baseline})$ & days & Period \\
~~~T$_{c}$ & $\mathcal{U}(T_{c} - \frac{P_{orb}}{2}, T_{c} + \frac{P_{orb}}{2})$ & BJD (days) & Time of Inferior Conjunction \\ 
~~~$\sqrt{e} \cos\omega $ & $\mathcal{U}(-1, 1)$  &  ...  & Eccentricity Reparametrization \\
~~~$\sqrt{e} \sin\omega $ & $\mathcal{U}(-1, 1)$ & ...  & Eccentricity Reparametrization \\
~~~K & $\mathcal{U}(-\infty,+\infty)$ & m s$^{-1}$   & Velocity Semi-amplitude \\
\sidehead{\textbf{Trend and Curvature:}}
~~~$  \dot{\gamma}$ & $\mathcal{U}(-\infty, +\infty)$  &  m s$^{-1}$ d$^{-1}$  & Linear Trend \\
~~~$ \ddot{\gamma}$ & $\mathcal{U}(-\infty, +\infty)$ & m s$^{-1}$ d$^{-2}$  & Quadratic Curvature \\
\sidehead{\textbf{GP Hyperparameters}}
~~~$\eta_{1}$ & $\mathcal{J}(0.1, 100)$ & m s$^{-1}$  & GP Amplitude \\
~~~$\eta_{2}$ & $\mathcal{J}(\eta_{3}, 10000)$ & days   & Exponential Scale Length \\
~~~$\eta_{3}$ & $\mathcal{N}(\mu_{GLS}, sd_{GLS})$  & days   & Periodic Term \\
~~~$\eta_{4}$ & $\mathcal{N}(0.5,0.05)$   & ...  & Periodic Scale Length \\
\sidehead{\textbf{Instrumental Parameters}}
~~~$\gamma_{\rm{HIRES}}$ & $\mathcal{U}(-\infty,+\infty)$ & m s$^{-1}$  & HIRES offset \\
~~~$\gamma_{\rm{APF}}$ & $\mathcal{U}(-\infty,+\infty)$ &  m s$^{-1}$ & APF offset \\
~~~$\sigma_{\rm{HIRES}}$ & $\mathcal{U}(0.0,+\infty)$ &  m s$^{-1}$    & Instrumental Jitter, HIRES \\
~~~$\sigma_{\rm{APF}}$ & $\mathcal{U}(0.0,+\infty)$ &  m s$^{-1}$   & Instrumental Jitter, APF \\
\enddata

\tablenotetext{a}{$\mathcal{N}$ is a normal prior with $\mathcal{N}$(mean, standard deviation)}
\tablenotetext{b}{$\mathcal{U}$ is a uniform prior with $\mathcal{U}$(lower,upper)}
\tablenotetext{c}{$\mathcal{J}$ is a Jeffreys prior with $\mathcal{J}$(lower,upper)}
\tablenotetext{d}{Reparameterization of eccentricity is motivated by \citet{Anderson2011} and \citet{Eastman2013}. Here, $\omega$ is the argument of periapsis of the star’s orbit in a left-handed coordinate system.}
\tablenotetext{*}{$\mu_{M23}$ and $sd_{M23}$ refers to the mean and standard deviation of the posterior taken from \cite{TKSXV}.}
\end{deluxetable*}

\section{Caveats To a Homogeneous RV Fit}
\label{sec:caveats}

We developed our fitting methodology to be applicable to most systems. However there were some instances where we departed from the steps stated in \S \ref{sec:methods}. In the interest of reproducibility, here we clarify decisions made for three systems.

\subsection{TOI--1444}\label{TOI1444}
TOI--1444 hosts a confirmed ultra-short period planet (b) at 0.47 days along with a candidate non-transiting companion (c) at $\sim$ 16 days \citep{TKSX}. Since the publication of this system, we collected 20 RVs in addition to the existing 56; however, these additional observations did not significantly increase in the detection significance for this signal. Curiously, when both the `b' and `c' components are removed, \texttt{RVSearch} reveals a strong peak at 89 days. When only removing the `b' component, neither the 16 day or 89 day signals are confidently detected. According to our methodology, the 89 day signal would be included as a candidate signal, but since its detection is dependent on the removal of a signal that has yet to be definitively confirmed, we decide to exclude it from our fits. 89 days is also close to the 90 day alias seen in HIRES data \citep{Rosenthal2021}.

\subsection{TOI--1726}\label{TOI1726}

TOI--1726 is a young Sun-like star that hosts two super-Earths of comparable size \citep{2023AA...672A.126D} with a rotation period of $\sim$6.4 days that is clearly seen in periodograms of both our RVs and our activity indicators. This rotation period happens to be close to the orbital period of TOI--1726 b at 7.1 days which, combined with the low expected semi-amplitude for both planets, motivated us to stop observing this target well before it hit its 100 observations quota. 
According to our methodology, we placed an informative prior on the periodic component of the GP informed by the rotation period of the star resulting in a mass constraint for the c planet of 17.3$\pm9.9 \mearth$ which is in agreement with the previous publication of this system. However, we obtain a mass for the `b' planet of $37\pm9.6 \mearth$ which is significantly higher than both the estimate from \cite{2023AA...672A.126D} and the masses of all other sub-Neptunes of similar size.\footnote{\cite{Hadden2014}'s transit timing analysis of \textit{Kepler} planets found sub-Neptunes comparable in mass to our estimate for TOI--1726 c. However they are still consistent with the highest RV-determined sub-Neptune mass within their uncertainties.} Given the similarity between the rotation and orbital period we suspect the GP might be over-fitting and possibly inflating the RV semi-amplitude for the `b' planet to some degree. We keep this result in the catalog for completeness but caution accepting it at face value.

\subsection{TOI--1759}\label{TOI1759}

TOI--1759 is an M0V dwarf with moderate levels of activity ($\log{R'_{HK}}=-4.69$) that hosts a $\sim 3 ~\rearth$ planet \citep{Espinoza2022,Martioli2022}. A stellar rotation period of 35.65 days was identified through SPIRou polarized spectra. Though this signal was not clearly seen in either the CARMENES or SPIRou RVs, it is prominent in the HIRES RVs. This could be due to the different band passes of the instruments and the expected spot temperatures. Additionally, this signal is absent in both our $S_{HK}$ timeseries and the TESS photometry. Given our methodology for building RV fits, this stellar signal would have been treated as Keplerian with a blind GP applied. In this case, we chose to place a Gaussian prior on the rotation term in the GP. While we are only able obtain an upper limit for the mass, a reanalysis of this system using additional new data sets is underway (Polanski et. al \textit{in prep}).

\begin{figure*}[t]
    \centering
    \includegraphics[width=\textwidth]{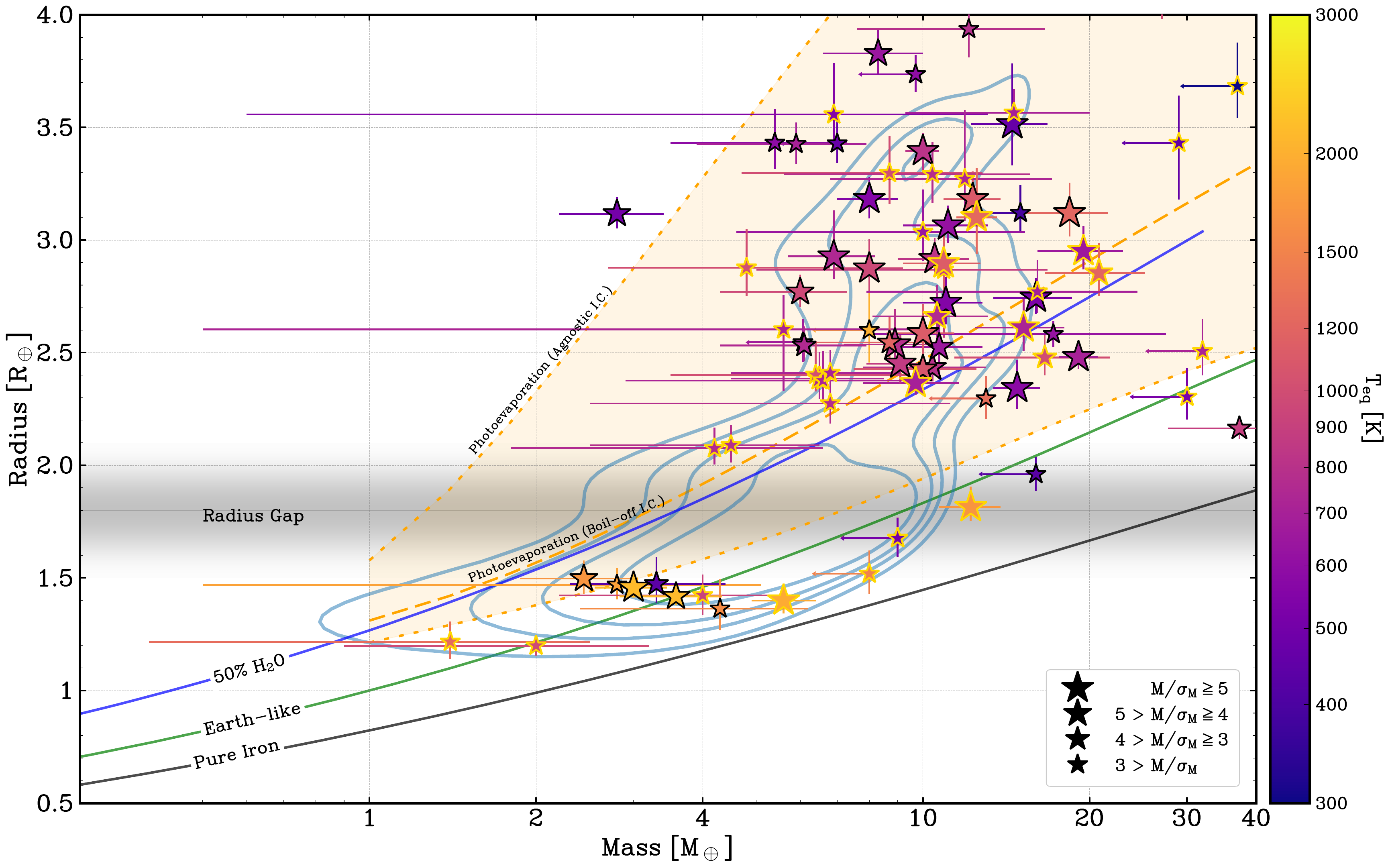}
    \caption{The mass-radius space for planets below 4~\rearth~ in the TKS sample. Points are color-coded based on the planets' equilibrium temperature while the size is scaled according their mass significance: larger points having a more precise mass measurement. Points outlined in gold are planets where their mass constraints are newly published in this work. Blue contours represent a Gaussian Kernel Density Estimate (KDE) of all known exoplanets around FGK stars. We show the composition tracks for pure iron, an Earth-like composition, and a 50/50 water-silicate mix \citep{Zeng2019}. In addition, we include the photo-evaporation mass-radius distributions from \cite{Rogers2023}.}
    \label{fig:small_mr}
\end{figure*}

\section{Results}\label{sec:results}

We report mass estimates or upper limits for a total of 126 planets and candidate signals across 86 planetary systems. 112 of these planets transit their host star. In Figure \ref{fig:big_mr} we show our full sample in comparison to all known exoplanets with a significant mass measurement,\footnote{From the NASA Exoplanet Archive as of Sept. 2023} while in Figure \ref{fig:small_mr} we show a detailed view of the mass-radius space for small planets. In Table \ref{tab:fitted} we report values for all fitted parameters while Table \ref{tab:derived} lists the derived mass, density and semi-major axis for each candidate along with the previously published mass estimate (where available).

In this section we compare our results to previously published measurements, present newly confirmed planetary systems, and statistically validate planets that did not reach a mass detection significance threshold of 3$\sigma$ (see \S\ref{sec:statval}).

\subsection{Comparison with Confirmed Planets}\label{sec:lit_comparison}

\begin{figure}
    \centering
    \includegraphics[width=0.48\textwidth]{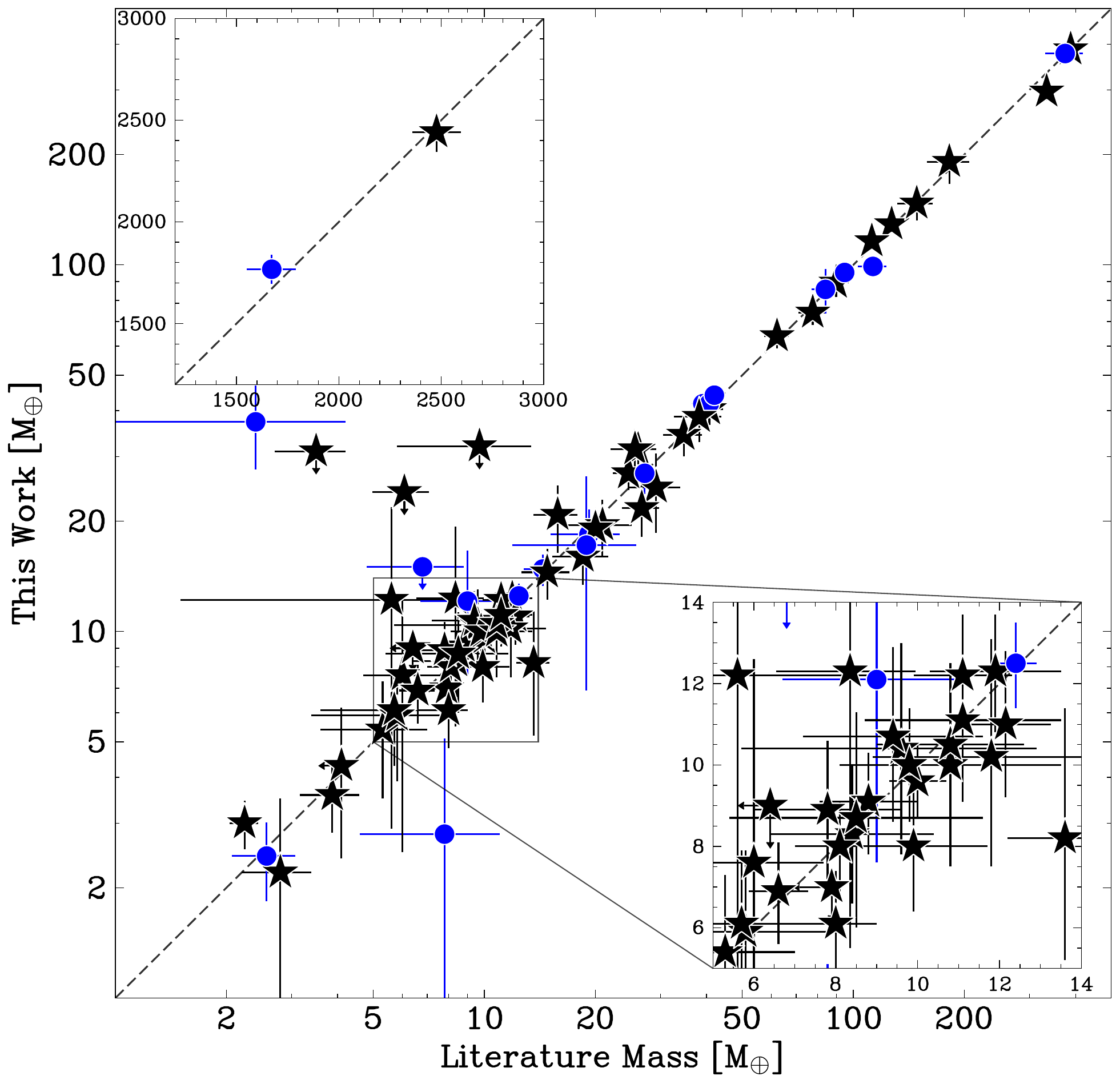}
    \caption{A comparison of mass measurements for the 62 planets in our sample for which masses had been published previously. Black stars indicate the literature value comes from a TKS publication while blue circles indicate a mass publication outside of our survey. The insets show the comparison for objects exceeding 500\mearth (upper left) and a zoomed-in view of the dense region between 5 and 14 \mearth (lower right).}
    \label{fig:mass_comparison}
\end{figure}

Figure \ref{fig:mass_comparison} compares our derived masses with literature masses for 62 planets in 33 previously confirmed TESS systems. For the majority of planets, we find that our approach produces consistent results with most measurements within 1$\sigma$.

Of the 11 planets where we were only able to determine an upper limit, 5 of them were in the TOI--1136 system where both multiplicity and stellar activity conspired to require a much more comprehensive analysis (\citealp[Beard et al. in review]{2023AJ....165...33D}). Although expected, this highlights the difficulty in homogeneous mass measurements for multiplanet systems which is exacerbated by high levels of activity. Many of our mass measurements with the largest measurement error ($M/\sigma_{M}<3\sigma$) in Figure \ref{fig:mass_comparison} come from such systems (e.g TOI--561, HD 191939, TOI--1136). In two cases, HD 25463 c and HD 12572 c, only an upper limit was obtained in both this work and in their dedicated TKS publication despite a more in-depth joint RV-transit fit \citep{TKSXVI},  suggesting these planets require additional data. Conversely, for 24 previously published planets, we are able to obtain a more precise mass measurement than previous literature values, with an average improvement in the mean mass uncertainty of $~$20\%.

\subsection{New RV Confirmation of TOIs}\label{sec:rv_confirm}

We achieve a mass significance of $\geq 3\sigma$ for 15 previously unconfirmed candidate planets. Given the vetting performed on these targets in \citet{TKS0}, we confirm these as bona fide planets. While a periodogram analysis of the RV data for each of these systems all reveal some amount of power at the candidate's period, 8 had a false alarm probability less than 1\%. For the remaining 7 systems, we take the additional step of statistically validating them (\S \ref{sec:statval}). Individual descriptions of these newly confirmed planetary systems are given in this section and our periodogram analyses and RV fits can be found in Appendix \ref{sec:rv_fits}. 

\subsubsection{TOI--260 b}

TOI--260 b (HIP 1532) is a super-Earth (1.47$^{+0.12}_{-0.08}$ R$_{\oplus}$) orbiting a K type star (4050$\pm109$ K) with an orbital period of $\sim$13.5 days. Our S-index timeseries contains a stellar activity signal at $\sim$37 days which is also seen in our RVs; however, we note that this is significantly different from the rotation period of $\sim$15 days obtained through TESS photometry by \cite{Howard2021}. An RV fit (Fig \ref{fig:HIP1532_rv_panel}) with a GP included allows us to modestly constrain the mass of the transiting planet to 3.3$\pm1.1$ M$_{\oplus}$, placing it between the Earth-like and 50/50 water world composition curves, although it is consistent with an Earth-like composition within the mass error. With an RV semi-amplitude of 1.3 m s$^{-1}$, a mass measurement for this planet lies at the extreme end of HIRES' capabilities, and the system would benefit from additional observations with higher precision spectrographs. 

\subsubsection{TOI--1173 b}

TOI--1173 b is a Saturn-sized (9.02$^{+0.16}_{-0.15}$ R$_{\oplus}$) planet orbiting a G type star (5414$\pm125$ K) with an orbital period of $\sim$7.1 days. We were able to obtain a 7$\sigma$ mass constraint of $28.3 \pm 4.0$ M$_{\oplus}$ (Fig \ref{fig:T001173_rv_panel}) with 28 RVs, giving TOI--1173 b a bulk density of $0.21 \pm 0.04$ \gcc~ which is a third of that of Saturn. TOI-1173 has an estimated TSM of 182, making it a valuable target for transmission spectroscopy.

\subsubsection{TOI--1184 b}

TOI--1184 b is a sub-Neptune (2.41$^{+0.10}_{-0.14}$ R$_{\oplus}$) orbiting a K type star (4616$\pm104$ K) with an orbital period of $\sim$5.7 days. Our RV fits (Fig \ref{fig:T001184_rv_panel}) were able to constrain the planet mass to 6.8$\pm2.3$ M$_{\oplus}$, making it one of the least dense sub-Neptunes in the TKS sample. 

\subsubsection{TOI--1194 b}

TOI--1194 b is a Saturn sized (8.72$^{+0.16}_{-0.15}$ R$_{\oplus}$) planet orbiting a G type star (5393$\pm124$ K) with an orbital period of $\sim$2.3 days. We were able to obtain a 30$\sigma$ mass constraint of 120$\pm6$ M$_{\oplus}$ (Fig \ref{fig:T001194_rv_panel}) with 32 RVs corresponding to a density slightly higher than that of Saturn. The shorter period of TOI--1194 b places it at the upper edge of the Neptune desert and its high density could be the result of past or ongoing atmospheric erosion. 

\subsubsection{TOI--1248 b}

TOI--1248 b is a sub-Saturn sized (6.81$^{+0.12}_{-0.12}$ R$_{\oplus}$) planet orbiting a K type star (5205$\pm120$ K) with an orbital period of $\sim$4.4 days. Similar to TOI--1173 b, our RVs (Fig \ref{fig:T001248_rv_panel}) constrain the mass to 27.4$\pm3.6$ \mearth, giving it a density of 0.47$\pm0.1$ \gcc.

\subsubsection{TOI--1279 b}

TOI--1279 b is a sub-Neptune (2.66$^{+0.14}_{-0.11}$ R$_{\oplus}$) orbiting a G type star (5457$\pm127$ K) with an orbital period of $\sim$9.6 days. We obtain a 4$\sigma$ mass constraint of 10.6$\pm2.5$ \mearth~ (Fig \ref{fig:T001279_rv_panel}) placing this planet in a region of the mass-radius diagram where many of the sub-Neptunes in our sample are concentrated. We also measure a significant downward RV trend of -0.0079$\pm$0.0020 m s$^{-1}$ d$^{-1}$ that does not correlate with activity indicators, suggesting the possible presence of a long period companion. Additionally, TOI--1279 b is the only single-TOI system where \cite{TKSXV} detect significant transit timing variations.

\subsubsection{TOI--1451 b}

TOI--1451 b is a sub-Neptune (2.61$^{+0.13}_{-0.10}$ R$_{\oplus}$) orbiting a G type star (5800$\pm139$ K) with an orbital period of $\sim$16.5 days. We measure a 5$\sigma$ mass of 15.2$\pm2.8$ \mearth~ (Fig \ref{fig:T001451_rv_panel}) placing TOI--1451 b in a region of the mass-radius diagram where many of the sub-Neptunes in our sample are concentrated (see Fig. \ref{fig:small_mr}). Our periodogram analysis of the RVs after subtraction of the transiting planet also reveals a strong peak at $\sim$90 days that is not seen in the activity indicators. We treat this signal with caution since multiple $\sim$90 day aliases were seen in the California Legacy Survey data \citep{CLS1_rvsearch}. A periodogram of the window function for all RV data reveals no strong peak near 90 days, however, when considering the window function of \textit{only} HIRES data, multiple peaks at integer fractions of a year are seen, including at 90 days. Because of this, we consider the 90 day signal to be an alias in the HIRES data.

\subsubsection{TOI--1472 b}

TOI--1472 b is a Neptune analog (4.16$^{+0.16}_{-0.13}$ R$_{\oplus}$) orbiting a K type star with a period of $\sim$6.4 days. Our RV fits constrain the mass to 16.5$\pm5.1$ \mearth~ (Fig. \ref{fig:T001472_rv_panel}), giving it a mass also comparable to that of Neptune. 

\subsubsection{HD 156141 b (TOI--1742 b)}

HD 156141 b (TOI--1742 b) is a sub-Neptune (2.37$^{+0.06}_{-0.05}$ R$_{\oplus}$) orbiting a G type star (5814$\pm137$ K) with an orbital period of $\sim$21.3 days. We obtain a nearly 5.5$\sigma$ mass constraint of 9.7$\pm1.9$ \mearth~ (Fig. \ref{fig:156141_rv_panel}), placing this planet in a region of the mass-radius diagram where many of the sub-Neptunes in our sample are concentrated. Our RVs also constrain the orbit of the planet to a modest eccentricity of 0.3$\pm0.1$, making this the second highest eccentricity small (R$<4\rearth$) planet in the TKS sample after TOI--1751 b. We also detect a significant amount of curvature in our RV time series that correlates with our activity indicators. 

\subsubsection{TOI--1753 b}

TOI-1753 b is a sub-Neptune (2.34$^{+0.08}_{-0.06}$ R$_{\oplus}$) orbiting a G type star (5620$\pm134$ K) with an orbital period of $\sim$5.3 days. Our RV fits (Fig. \ref{fig:T001753_rv_panel}) constrain the mass of the transiting planet to 16.6$\pm5.2$ M$_{\oplus}$. It falls in the relatively common population of sub-Neptunes, though its mass and radius, consistent with either a large H/He envelope, or some water enrichment (see Fig. \ref{fig:small_mr}). A more precise mass would be required to disambiguate different planetary compositions.

\subsubsection{TOI--1775 b}

TOI--1775 b is a Saturn sized (8.05$^{+0.14}_{-0.13}$ R$_{\oplus}$) planet orbiting a  late G/early K type star (5283$\pm116$ K) with an orbital period of $\sim10.2$ days. We are obtain over a 6$\sigma$ (Fig. \ref{fig:T001775_rv_panel}) mass constraint of 96$\pm15$ \mearth~, giving TOI--1775 b a mass consistent with that of Saturn. 

\subsubsection{HD 77946 b (TOI--1778 b)}

HD 77946 b (TOI--1778 b) is a sub-Neptune (2.9$^{+0.19}_{-0.14}\rearth$) planet orbiting an F type star (6000$\pm147$ K) with an orbital period of $\sim$10.2 days. We obtain a nearly 6.5$\sigma$ (Fig. \ref{fig:77946_rv_panel}) mass constraint of 10.9$\pm1.7$ \mearth, placing this planet in a region of the mass-radius diagram where many of the sub-Neptunes in our sample are concentrated (see Fig. \ref{fig:small_mr}). Our RVs obtain a 3$\sigma$ constraint on the orbital eccentricity of 0.21$\pm0.07$, also placing this among highest eccentricity small (R$<4\rearth$) planets in the TKS sample. We also find evidence of curvature in our RV time series which does not correlate with our activity indices, suggesting the potential presence of a long period companion.

\subsubsection{TOI--1798 b \& c}

TOI--1798 is a K type star (5106$\pm115$ K) that hosts two transiting planets, the first being TOI--1798 b, a 2.76$^{+0.13}_{-0.1}\rearth$ sub-Neptune with a period of $\sim$8.0 days. RV fits (Fig \ref{fig:T001798_rv_panel}) constrain the mass to 6.5$\pm2$ M$_{\oplus}$. 

The second planet is TOI--1798 c, a 1.4$^{+0.06}_{-0.07}\rearth$ super-Earth with a period of $\sim$0.44 days, making this another ultra-short period planet in the TKS sample. We were able to obtain an 8$\sigma$ mass of 5.6$\pm0.7$ \mearth~, giving this planet a high density of 11.5$\pm2.0$ \gcc. \cite{TKSX} compiled all known USPs with a mass at the time and observed a correlation between planet mass and disk surface density of solid material ($10^{\text{[Fe/H]}}\times M_*$), suggesting that planet mass increases with the availability of planetary building blocks within 1 AU. Estimating the surface density of solids available in the protoplanetary disk as $10^{\text{[Fe/H]}}\times M_*$, we find TOI--1798 c is placed comfortably within the population of USPs studied in \cite{TKSX}. TOI--1798 is the third system in the TKS sample to host an inner USP accompanied by a longer period, sub-Neptune-mass, companion.

\subsubsection{TOI--1823 b}

TOI-1823 b is a sub-Saturn sized (7.54$^{+0.33}_{-0.25}$ R$_{\oplus}$) planet orbiting a K type star (4926$\pm113$ K) with an orbital period of $\sim$38.8 days. We obtain an 8$\sigma$ mass constraint of 67.4$\pm8.2$ M$_{\oplus}$ (Fig. \ref{fig:TIC142381532_rv_panel}) with 32 RVs, giving it a density slightly higher than that of Saturn.

\subsection{Statistical Validation of TOIs}\label{sec:statval}

For the 30 systems that either did not reach $\geq 3\sigma$ mass significance or did not have a clear and independent periodogram detection, we used \texttt{TRICERATOPS} \citep{Giacalone2021AJ} to calculate the false positive and nearby false positive probabilities (FPP/NFPP) and statistically validate these targets. Statistical validation attempts to rule out astrophysical false positive scenarios such as eclipsing binaries. While statistical validation has been employed since the \textit{Kepler} era \citep{Borucki2012B,Crossfield2016} with packages such as \texttt{BLENDER} \citep{Torres2004} and \texttt{VESPA} \citep{Morton2015}, we opt to use \texttt{TRICERATOPS} since it is actively maintained and designed specifically for use on TESS candidates \citep{Morton2023}.

We began by pulling the TOI's short cadence lightcurve from the MAST database using \texttt{Lightkurve} \citep{lightkurve} and then use transit information from the TKS Systems Properties catalog \citep{TKSXV} to phase fold the data. We bin the timeseries data down to 100 bins in the interest of computational speed.\footnote{https://triceratops.readthedocs.io/en/latest/tutorials/example.html} The apertures used in the SPOC pipeline are pulled and used as input to \texttt{TRICERATOPS}. High resolution imaging data is also incorporated; however, \texttt{TRICERATOPS} can only run with a single HCI dataset at a time so we chose the imaging set with the best available contrast (\S \ref{sec:HRI}).

When calculating the NFPP, we remove any NFPP candidates flagged by the TESS Input Catalog \citep[TIC,][]{TIC} as `artifact' to avoid biasing the NFPP value. If the TIC is missing stellar parameters for NFPP candidates, we use the candidate's Bp-Rp color to estimate $T_{\text{eff}}$, $M_{*}$, and $R_{*}$ from \cite{PecautMamajek2013}.

To obtain an accurate estimate of the FPP and NFPP, we run \texttt{TRICERATOPS} 20 times per candidate planet and report the average and standard deviation in Table \ref{tab:statval}. Typically, for a planet to be validated it must satisfy FPP $<$ 0.015 and NFPP $<$ 0.001, In this work, we compute the average and standard deviation of the FPP and NFPP from the 20 runs and  require a candidate to clear those thresholds to 3$\sigma$ to be considered validated. In total we graduate 17 candidates to validated planets, while six targets, TOI--1436.01, 1438.01, 1438.02, 1269.02, 1716.01, 2114.01 did not reach the requisite threshold for validation and remain planetary candidates. That said, 1436.01 and 2114.01 have low enough FPP/NFPP to be considered likely planets \citep{Giacalone2021AJ}. In all cases, the most likely scenario was a planet transiting the target star followed by a planet transiting the primary star but with unresolved bound companions or unresolved background stars. Two targets, TOI--1279.01 and TOI--1742.01, have FPPs of 26\% and 1.8\%, respectively, which would place them in the likely planet category. However, given the significant mass obtained for both of these planets (4$\sigma$ and 5$\sigma$ constraints, respectively), we consider them confirmed. Additionally, the low NFPP across all of our targets suggests that none of the planets we targeted are likely to be nearby false positives.

\renewcommand{\arraystretch}{1.5}
\startlongtable
\begin{deluxetable*}{c c l c c r}
\tablecaption{\texttt{TRICERATOPS} Results \label{tab:statval}}
\tablehead{
  \colhead{TOI} & 
  \colhead{No. Sectors} & 
  \colhead{Imaging Used} &
  \colhead{FPP} &
  \colhead{NFPP} &
  \colhead{Disposition}}
\startdata
260.01 & 2 & Keck/NIRC2 Br$\gamma$ & 0.0003$\pm$0.0003 & $<10^{-4}$ & Confirmed Planet \\
1184.01 & 26 & Gemini/NIRI K & 0.0001$\pm$0.0001 & $<10^{-4}$ & Confirmed Planet \\
1279.01* & 6 & Palomar/PHARO Br$\gamma$ & 0.2631$\pm$0.0855 & $<10^{-4}$ & Confirmed Planet\\
1472.01 & 3 & Keck/NIRC2 Br$\gamma$ & 0.0001$\pm$0.0001 & $<10^{-4}$ & Confirmed Planet \\
1742.01* & 26 & Keck/NIRC2 Br$\gamma$ & 0.0184$\pm$0.0792 & $<10^{-4}$ & Confirmed Planet \\
1753.01 & 26 & Palomar/PHARO Br$\gamma$ & 0.0005$\pm$0.0002 & $<10^{-4}$ & Confirmed Planet \\
1798.01 & 5 & Gemini/'Alopeke 832 nm & 0.0003$\pm$0.0001 & $<10^{-4}$ & Confirmed Planet \\
\hline 
 \hline
1180.01 & 11 & WIYN/NESSI 832 nm & 0.0035$\pm$0.0032 & $<10^{-4}$ & Validated Planet \\
1244.01 & 26 & Keck/NIRC2 K & $<10^{-4}$ & $<10^{-4}$ & Validated Planet \\
1249.01 & 7 & Gemini/'Alopeke 832 nm & 0.0083$\pm$0.0016 & 0.0007$\pm$0.0001 & Validated Planet \\
1269.01 & 29 & Palomar/PHARO Br$\gamma$ & 0.0013$\pm$0.0004 & $<10^{-4}$ & Validated Planet \\
1269.02 & 29 & Palomar/PHARO Br$\gamma$ & 0.0091$\pm$0.0012 & 0.0064$\pm$0.001 & Candidate \\
1438.01 & 26 & Gemini/'Alopeke 832 nm & 0.0047$\pm$0.0027 & 0.0016$\pm$0.0014 & Candidate \\
1691.01 & 5 & Gemini/'Alopeke 832 nm & 0.0001$\pm$0.0003 & $<10^{-4}$ & Validated Planet \\
1716.01 & 3 & Gemini/'Alopeke 832 nm & 0.0018$\pm$0.0002 & 0.001$\pm$0.0001 & Candidate \\
1723.01 & 3 & Shane/ShARCS K-short & 0.0006$\pm$0.0001 & $<10^{-4}$ & Validated Planet \\
1758.01 & 7 & Gemini/'Alopeke 832 nm & 0.0002$\pm$0.0002 & $<10^{-4}$ & Validated Planet \\
1794.01 & 5 & Palomar/PHARO H-continuous & 0.0002$\pm$0.0001 & $<10^{-4}$ & Validated Planet \\
1799.01 & 2 & Palomar/PHARO Br$\gamma$ & 0.001$\pm$0.0001 & $<10^{-4}$ & Validated Planet \\
2128.01 & 3 & Palomar/PHARO H-continuous & $<10^{-4}$ & $<10^{-4}$ & Validated Planet \\
\hline 
 \hline
1174.01 & 7 & Gemini/'Alopeke 832 nm & 0.0012$\pm$0.0004 & $<10^{-4}$ & Validated Planet \\
1347.02 & 27 & Keck/NIRC2 K' & 0.0097$\pm$0.0007 & $<10^{-4}$ & Validated Planet \\
1436.01 & 6 & Keck/NIRC2 Br$\gamma$ & 0.063$\pm$0.002 & $<10^{-4}$ & Likely Planet \\
1438.02 & 26 & Gemini/'Alopeke 832 nm & 0.0067$\pm$0.0017 & 0.0045$\pm$0.0014 & Candidate \\
1443.01 & 21 & Gemini/NIRI K & 0.0003$\pm$0.0002 & $<10^{-4}$ & Validated Planet \\
1467.01 & 3 & Keck/NIRC2 K & $<10^{-4}$ & $<10^{-4}$ & Validated Planet \\
1669.01 & 8 & Gemini/'Alopeke 832 nm & 0.0021$\pm$0.0004 & $<10^{-4}$ & Validated Planet \\
1776.01 & 2 & Palomar/PHARO Br$\gamma$ & 0.0029$\pm$0.0003 & $<10^{-4}$ & Validated Planet \\
2088.01 & 28 & WIYN/NESSI 832 nm & 0.003$\pm$0.004 & $<10^{-4}$ & Validated Planet \\
2114.01 & 2 & Palomar/PHARO Br$\gamma$ & 0.019$\pm$0.022 & $<10^{-4}$ & Likely Planet \\
\enddata
\tablecomments{The break in the table separates TOIs with a $\geq3\sigma$ mass measurement (top) those with a $1-3\sigma$ mass measurement (middle) and those with only an upper limit (bottom). Validated planets satisfy FPP $\leq$ 0.015 and NFPP $\leq$ 0.001. Likely planets satisfy FPP $\leq$ 0.5 and NFPP $\leq$ 0.001. * Indicates a target that would be considered a likely planet but is still confirmed through a significant mass measurement.}
\end{deluxetable*}
\renewcommand{\arraystretch}{1.0}


\section{Survey Performance}\label{sec:TKS_goals}

The overarching goal of TKS was to measure precise (5$\sigma$) planet masses \citep{TKS0}. We use this section to assess the success of TKS as well as reflect on how well we were able to our reach observing quotas. We additionally discuss how effective we were at allocating resources to each individual science case. While detailed information on each science case (SC) can be found in \citet{TKS0}, we provide a brief description here: SC1 is dedicated to measuring bulk compositions, SC2 is dedicated to probing system architectures, SC3 is dedicated to measuring masses of possible future atmospheric targets, and SC4 is dedicated to measuring masses of planets orbiting evolved stars.

\subsection{Observation and Cadence Goals}\label{cadence_goals}

\begin{figure*}[t]
    \centering
    \includegraphics[width=0.8\textwidth]{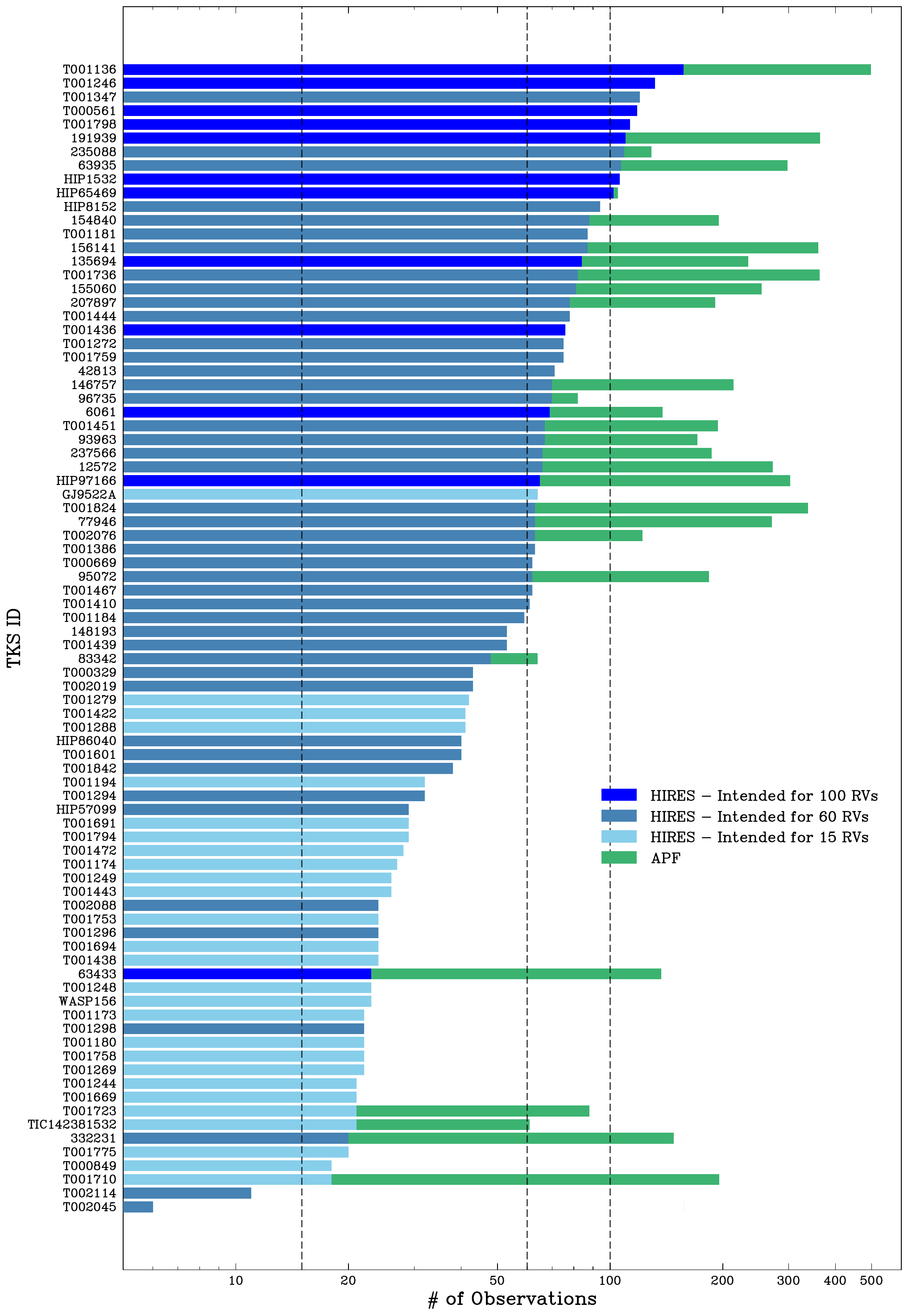}
    \caption{The number of observations acquired for each target by instrument. Vertical dashed lines are at the 15, 60, and 100 observation thresholds set for the several TKS science cases.}
    \label{fig:Nobs_breakdown}
\end{figure*}

\begin{figure}[t!]
    \centering
    \includegraphics[width=0.47\textwidth]{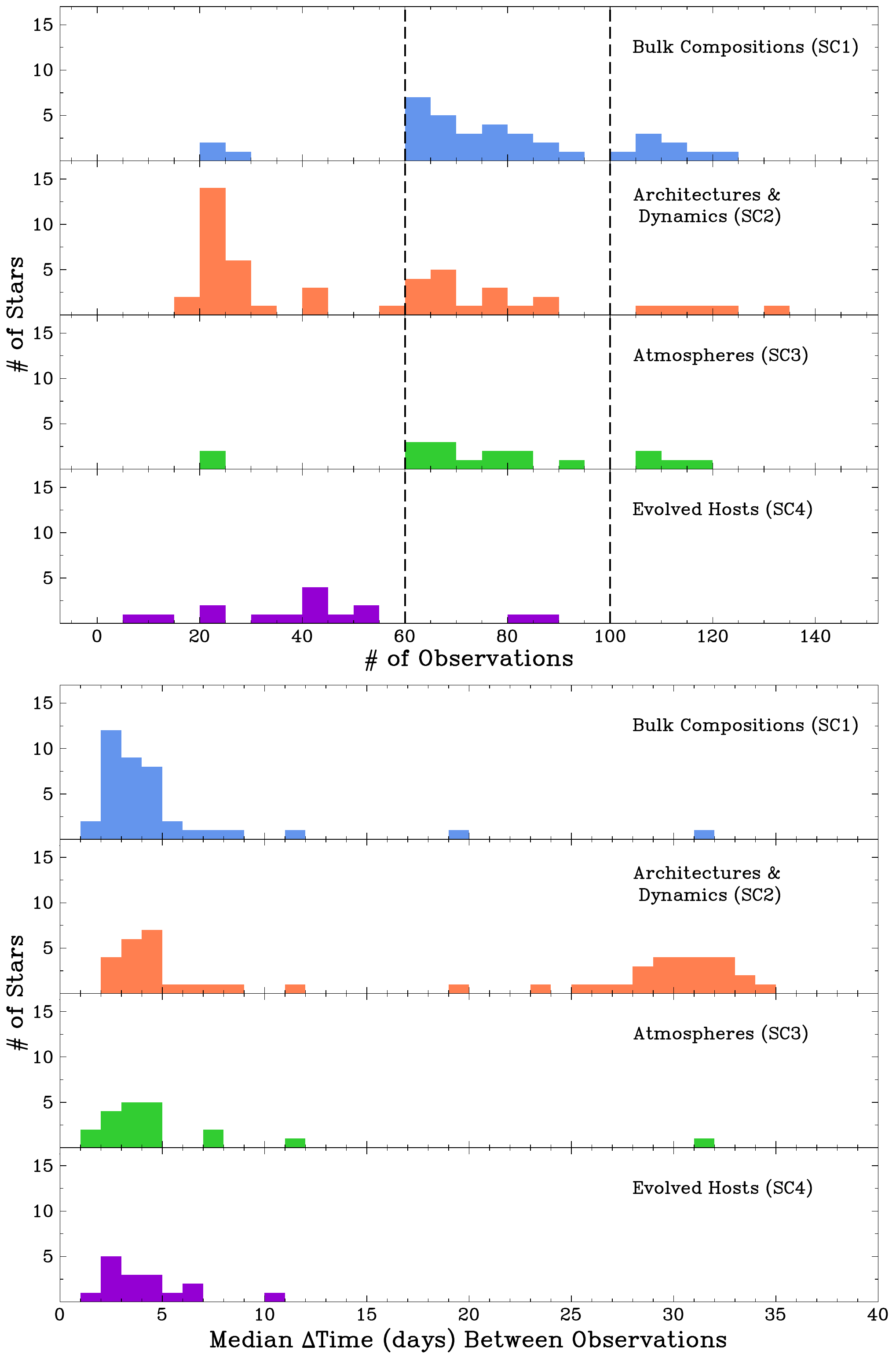}
    \caption{\textbf{Top:} The number of observations obtained broken down by science case. The vertical dashed lines represent the $N_{\text{obs}}$ targets set in \cite{TKS0}. \textbf{Bottom:} The median time difference between observations (cadence) for our targets broken down by science case. }
    \label{fig:obs_cadence_plot}
\end{figure}

We intended to get a maximum of 60 RVs for the majority of targets in our survey. However, three notable exceptions include: those designated as interesting multi-planet systems (100 RVs), those orbiting particularly active hosts (100 RVs), and those in the Distant Giant survey (15 RVs).Figure \ref{fig:Nobs_breakdown} shows a breakdown of number of observations per target in the survey and Figure \ref{fig:obs_cadence_plot} (top) shows the number of observations broken down by science case. 

In total, 62 stars received \textit{more} observations than their stated goal. Of the 22 targets that did not meet their observing quota, 14 of the targets had surpassed a $\geq5\sigma$ mass and therefore, no more observations were needed. At the extreme ends, there were 8 stars that missed their stated goal of observations by greater than 40 observations. TOI--1136 is an activity target, which was intended to get 100 observations. Instead, we opted to pool all the time  allocated to the stellar activity case (Technical Outcomes B; TB) into this high interest target, and so we got 157 total observations. Out of 86 TOIs, only 3 were dropped\footnote{Targets were dropped for various reasons including higher than expected rotational broadening and stellar rotation periods that coincided with planet orbital periods.}: HD 63433 (TOI--1726, \S\ref{TOI1726}), TOI--2045, and TOI--2114. Other targets, HD 63935 (TOI--509) and HD 235088 (TOI--1430), received more than 100 observations, despite being initially designated 60, due to increased interest in the target: HD 63935 later presented a second transiting planet and HD 235088 was a high priority atmospheres target which displayed increased stellar activity. 

To determine our attained cadence, we computed the time difference for every observation of a given target (eliminating differences less than 0.85 days to avoid intra-night cadence), and took the median time difference. In Figure \ref{fig:obs_cadence_plot} (bottom) we break down the observational cadence of targets by their SC. Note that many targets are represented in multiple science cases. We find that the Bulk Compositions, Atmospheres, and Evolved Hosts have median cadences between 3 and 4 days, while the System Architectures saw a spread of cadences. The 3rd panel from the bottom of the lower plot in Figure \ref{fig:obs_cadence_plot} shows the lower RV cadence required by the Distant Giants program ($\sim$30 days between observations) which is part of the Architectures SC. The distribution of cadences achieved for different science cases shows what is possible with a multi-year RV survey and may be used in simulating observations for future TKS-like programs.

\subsection{Mass Significance Achieved}

\begin{figure}[t!]
    \centering
    \includegraphics[width=0.48\textwidth]{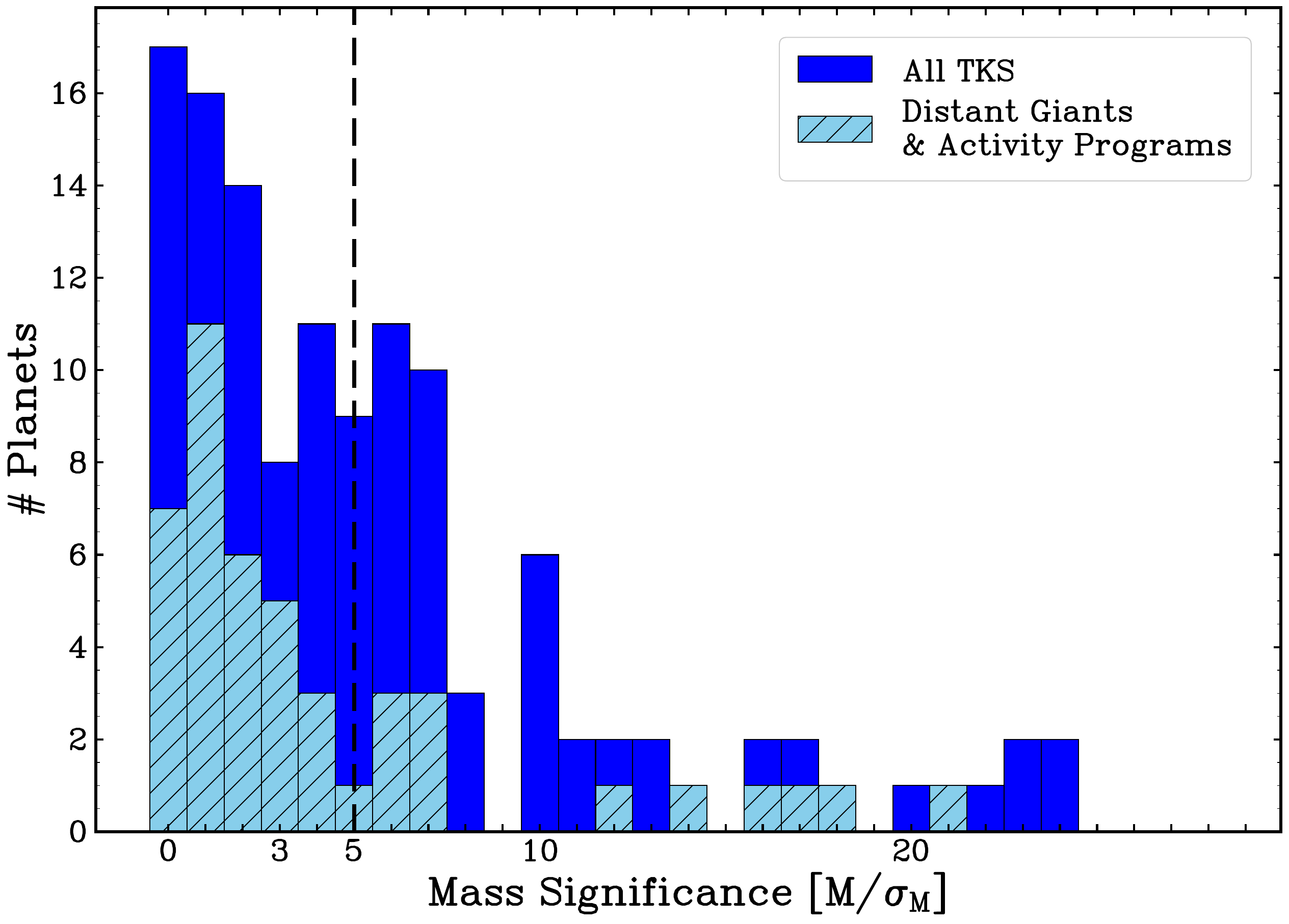}
    \caption{A histogram of the mass significance of each planet in our sample (blue). We highlight TOIs that were part of either the Distant Giants (SC2A) or Stellar Activity (TB) science cases, i.e programs where the transiting planet was not of primary interest, in the light blue, hashed histogram.}
    \label{fig:massSig}
\end{figure}

It is similarly straight-forward to assess our performance of measured mass significance. We measure at least a $5\sigma$ mass for 58 planets. This increases to 77 planets for $\geq3\sigma$ masses; see Figure \ref{fig:massSig} for a histogram of mass significances of our survey. Breaking this down a bit further, if we filter by planets with masses below 30 $M_{\oplus}$, then we have measured 30 planets to $5\sigma$ mass or better and 44 planets to $3\sigma$ mass. Through these numbers, given the over-arching goal of TKS of measuring many precise planet masses, we believe our survey has been very successful.

We can go further by acknowledging that our survey was driven by some science cases that either were not interested in the transiting TOIs (Distant Giants, SC2A) or intentionally made it challenging to measure a precise mass (Stellar Activity, TB). In Figure \ref{fig:massSig} we highlight this group of targets and find that many of our "under-performing" targets are in either of these categories. For TOIs with radii  $<4~\rearth$, we were able to reach at least a 3$\sigma$ mass for 38 of them. To put this number in context, the baseline science requirement of the TESS Follow-up Program (TFOP) was to measure masses for 50 TESS-discovered planets less than 4 $\rearth$ radii\footnote{https://tess.mit.edu/followup/}. Our survey alone was nearly able to achieve this goal.

\subsection{APF as Survey Support}\label{APF_support}

The use of the Automated Planet Finder in our survey was intended to provide additional RVs to bright (V <10) targets. To that end, we conducted complementary APF observations of 28 TOIs, which account for nearly half the RV measurements presented in this work. In order to assess APF's contribution to the goals of the survey, we select a subset of the 28 TOIs, and re-analyze these systems \textit{excluding} APF data. This subset includes massive targets for which a 5$\sigma$ mass was obtained without reaching the requisite 60 observations, targets that had their observations halted at 60 HIRES data points, and targets that received considerable attention from APF, making up $\sim$2/3 of their datasets. 

\begin{figure}[t]
    \centering
    \includegraphics[width=0.48\textwidth]{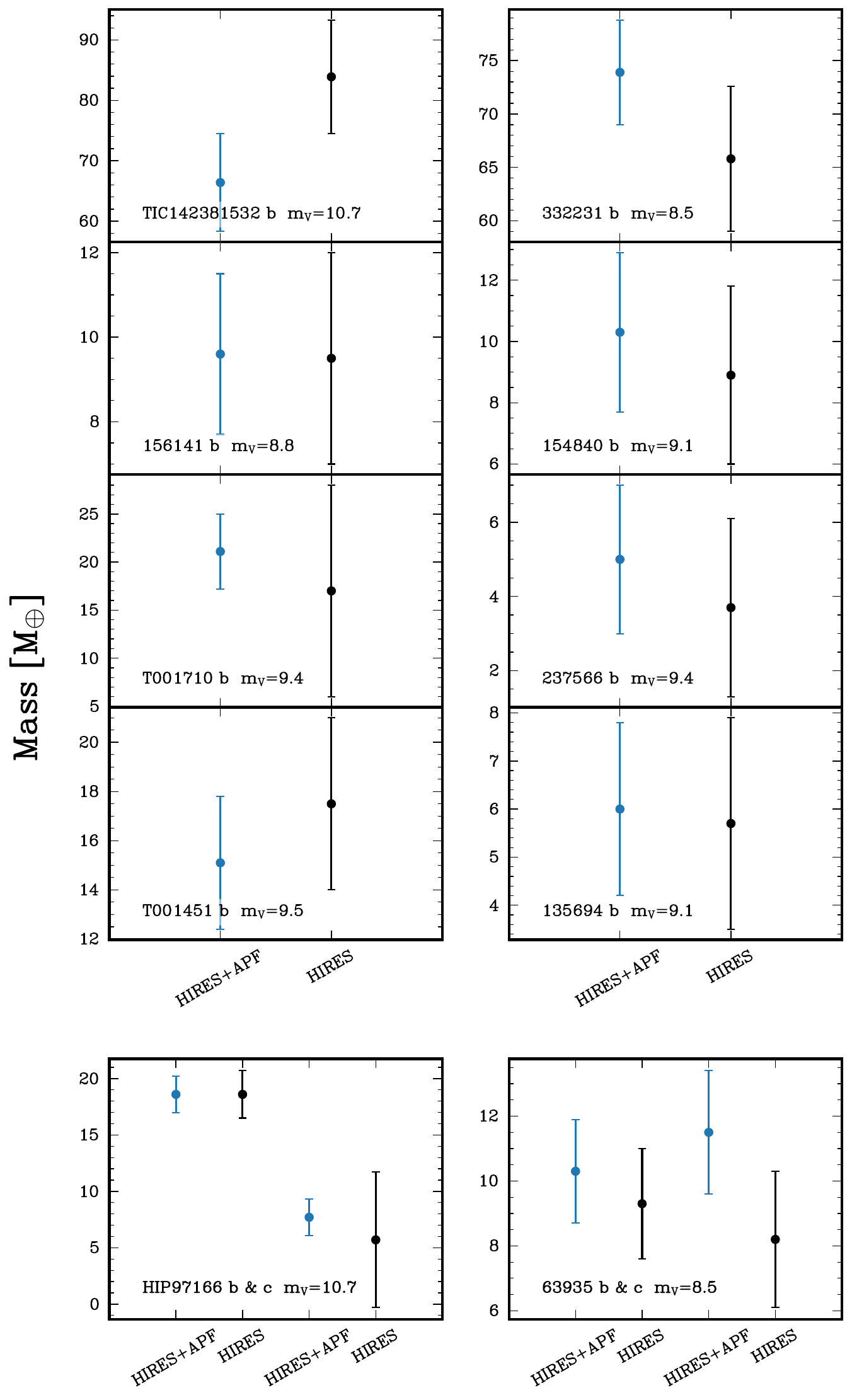}
    \caption{Comparisons of the mass constraints for a sample of systems observed with both HIRES and APF. Blue errorbars denote the mass obtained with both datasets while black is without APF data. Generally, APF contributed only slightly to the mass significance for each system, with the exception of TOI--1710 b and HIP97166 c.}
    \label{fig:sans_APF}
\end{figure}

A straightforward test of APF's contribution to the survey is its effect on mass precision. In Figure \ref{fig:sans_APF} we show the mass constraints for the TOIs reanalyzed without APF data. Masses measured from the full RV dataset were generally larger than those obtained with only HIRES data, though the difference is often statistically insignificant. For two systems, TOI--1710 b and HIP97166 c, APF had a significant impact on the mass significance achieved. With TOI--1710 b, a 3$\sigma$ mass was only obtained due to APF RVs, while without APF the non-transiting HIP97166 c would not have been detected in a periodogram of HIRES-only data. In most cases, targets saw slight reductions in their mass error after the inclusion of APF RVs.


For the magnitude range of these stars (V$\sim$8.5-11), APF is a comparably lower precision instrument, with average per-datum uncertainties being $\sim$2 times greater than the RV errors attained with HIRES. This means that mass constraints will generally be dominated by HIRES data. However, APF's contribution should not be assessed on mass precision alone. The higher cadence of APF observations (Table \ref{tab:obs_stats}) allowed for more uniform phase coverage, and for HD 332231 b, HIP 97166 b, and HD 156141 b this resulted in significantly improved constraints on the eccentricity of these planets, with all three cases seeing a reduction in the eccentricity uncertainty by a factor of two.

TOI--1736, HD 146757, and HD 191939 are also systems that benefited greatly from the inclusion of APF data. In TOI--1736's case we were able to pick up the acceleration of the system due to an outer companion through the initial 60 observations allocated to the single-TOI system. Additional observations made with APF were able to extend the baseline and critically catch two passages of the companion through periastron \citep[][Figure 25]{TKSXVI}. APF provided continual cadence, which wasn't feasible with HIRES, resulting strong constraints on both the orbital period and orbital eccentricity of TOI--1736 b's companion. 

HD 146757 \citep{TKSXVIII} benefited from APF data through the identification of an alias in the HIRES data. This case highlights APF's strength as a sanity check when searching for long period, massive companions with limited primary survey instrument data.

Finally, APF observations of HD 191939 during March through May 2020 (22 observations), at a time when few other observatories were taking data due to COVID-19 shutdowns, were crucial for early detection of planet e, the 101 day Saturn-mass planet \citep{TKSIX}. Without these observations, the signal may not have been recognized until much later, and possibly at an aliased period. Subsequent APF data continued to fill the phase space of the companion at a cadence higher than would have been possible with HIRES alone. HD 146757 benefited from APF data through the identification of an alias in the HIRES data. 

APF's contribution to the goals of TKS should not be understated. Our understanding of the architecture of exoplanet systems hinges on both a high degree of phase coverage to determine the eccentricity, and long baselines to detect companions. For surveys with limited time allocations, these criteria create opposing interests. Supplementing the observations of the primary survey instrument with those made with instruments with less demand not only helps to allocate more time to targets that require a greater density of higher precision data, but also serves to extend the scientific productivity of previous generation instruments. Future TKS-like surveys should seriously consider using APF-like instruments to extend their observational baseline and obtain denser phase coverage for eccentric planets. Additionally, support instruments can also be used to conduct observations ahead of the primary survey to identify systems that may be more interesting to certain science cases (e.g. Distant Giants) than would be expected based on the presence of a transiting planet alone.

\section{Expected vs. Measured Masses}\label{sec:exp_meas_mass}

\begin{figure*}[th]
    \centering
    \includegraphics[width=\textwidth]{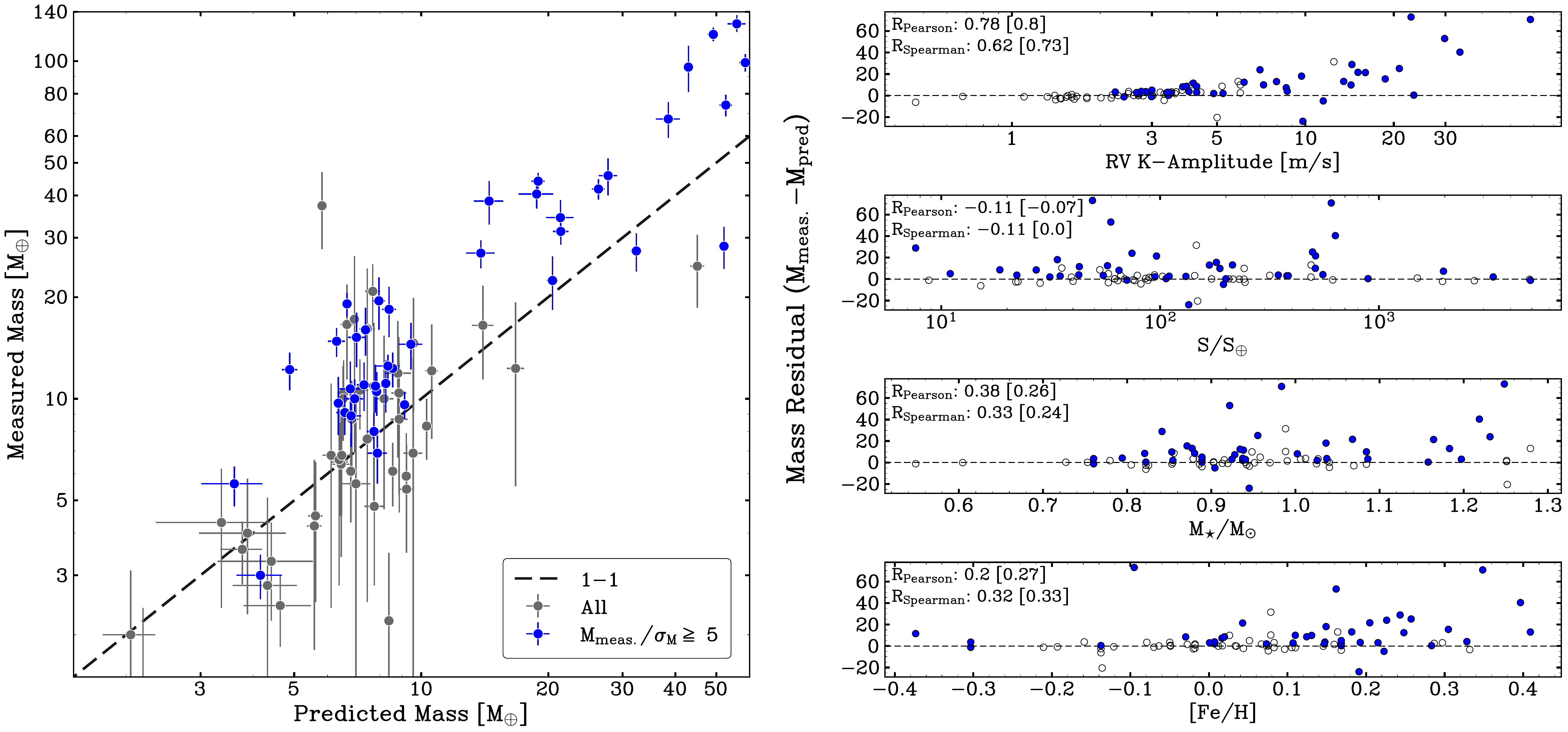}
    \caption{\textbf{Left:} A comparison of RV-measured masses to those predicted by the empirical mass-radius relations \citealt{WeissMarcy2014} and \citealt{ChenKipping2017}. Blue points highlight planets that achieved a 5$\sigma$ mass significance while the unfilled point represent the full sample. \textbf{Right:} The mass residual (measured minus predicted) as a function of stellar and planetary parameters (from top): RV semi-amplitude, planetary insolation flux, stellar mass, and stellar metallicity. Pearson and Spearman correlation coefficients are given for both the 5$\sigma$ sample and the full sample (in brackets). }
    \label{fig:meas_vs_pred_mass}
\end{figure*}

In \cite{TKS0}, the expected RV semi-amplitude for each planet in our sample was calculated using theoretical masses derived from mass-radius relations. The relations used were \cite{WeissMarcy2014} (WM14) for planets with R < $4 \rearth$ and \cite{ChenKipping2017} (CK17) for planets with $4\leq$ R $< 11.3~\rearth$, both of which are empirically derived. Figure \ref{fig:meas_vs_pred_mass} shows the  masses we measure compared to their expected values, with the error in the predicted mass being propagated from the radius uncertainty. It can immediately be seen that across the mass range sampled there is significant scatter around the one-to-one mass comparison line. This scatter is not surprising: we should not expect simple power law mass-radius relationships to be able to capture the compositional diversity of planets at a given radius.

Deriving a new mass-radius relation from the TKS sample is beyond the scope of this paper: we are using a limited sample of planets and not using all the RV data available for these systems. But, here we examine how our measured masses compare to what was predicted by these commonly used relations. We restrict this discussion to planets below $11.3~\rearth$ since beyond this threshold the theoretical mass was assumed to be a Jupiter mass. 

\subsection{Weiss and Marcy 2014}

In their analysis, WM14 report an RMS scatter of 2.7 \mearth~ and 4.3 \mearth~ for planets with R < $1.5~\rearth$ and $1.5\leq$ R $< 4~\rearth$, respectively, and $\chi ^2$ values that cannot be explained by mass measurement error alone. For the TKS sample, we find the RMS scatter to be comparable: 1.2 \mearth~ and 5.3 \mearth~ for the same radius bins (noting that our sample contains far fewer planets with R < $1.5 ~\rearth$). 

We then check if the WM14 M-R relation appropriately predicted the planet masses, i.e. if the mass residual is centered around zero. For planets with $1.5\leq$ R $< 4~\rearth$, we find that a slight bias towards under-predicted masses of $\sim10$\% while masses predicted from WM14 for the smallest planets (R < $1.5~\rearth$) are in agreement with what we measure from RVs. Exploring this further, we also recalculate mass predictions using the modified version of WM14 that only includes planet masses derived through RVs (as opposed to masses derived from RVs \textit{and} transit timing variations) and find that this largely mitigates the difference. This suggests that the WM14 M-R relations, which were based primarily on \textit{Kepler} planets, may still be relied on to approximate the masses of planets coming from TESS.

\subsection{Chen and Kipping 2017}

For planets with masses predicted from CK17 ($4\leq$ R $< 11.3~\rearth$) we report a higher RMS scatter of 32 \mearth~, but also find this M-R relation produces the largest mass discrepancies, with an average percent difference of 24\%. Considering planets in this category had an average measurement uncertainty of $11\%$ this represents an appreciable difference.
We also use the complete CK17 M-R relation to include planets below $4~\rearth$ to check if this underestimation persists for smaller worlds. This extends the Neptunian relation down to $\sim$1.24 \rearth~ at which point it transitions to the Terran world relation. The predicted masses from these relations show a percentage difference comparable to the RV+TTV WM14 relation with the exception that this underestimation in mass is more prevalent for the smallest planets. Below 1.5 \rearth,~ the mean percentage difference increases to 28\%, although we note that we have few planets in this radius space and the average measurement uncertainty for these planets is 37\%.

\subsection{Comparison to Non-TKS TESS Planets}

We also perform a similar analysis for both M-R relations with the 169 TESS-discovered exoplanets with a mass measurement\footnote{From the NASA Exoplanet Archive as of Sept. 2023}. For WM14, we find a similar RMS scatter and, again, a slight positive bias in the mass residual that disappears once the RV-only relation is used.

With CK17, the underestimation of mass persists for the greater TESS sample. Between 1.5 and 4 \rearth, the percent difference nearly doubles from 13\% in just the TKS sample to 20\% for all TESS-discovered planets. Beyond 4 \rearth~, the difference is comparable to the TKS sample at 27\% suggesting that on average CK17 underestimates planet masses by around a quarter and, in extreme cases, by one half. However, we do note that below R $\sim1.5~\rearth$, the mass residual reduces to 0.5 \mearth~ for the entire TESS-discovered sample. In contrast to WM14, CK17 has fewer ``Neptunian'' worlds with the bulk of their sample having radii larger than $\sim10~\rearth$. Therefore, this sample may not be appropriately describing the lower mass population of planets, which comprises the majority of worlds discovered to date.

\subsection{Sources of M-R Relation Bias and Implications}

Empirical M-R relations tend to rely on planets that are well-characterized in both radius and mass space. This is especially true for CK17 who instituted a 3$\sigma$ mass significance threshold for their sample thereby not taking into account selection effects or survey bias. Historically, more massive planets have been easier to reach a high mass significance, and subsequent M-R relations based on these samples may be biased to higher masses \citep[known as publication bias,][]{Burt2018,Montet2018}. 

Comparing measured masses to their predicted values, we see the opposite trend in parts of the TKS sample. Restricting ourselves to planets that reached a $5\sigma$ mass (blue points in Fig. \ref{fig:mass_comparison}), we see that they are mainly the higher mass planets for a given radius bin showing that M-R relations based on ``well-characterized'' samples will indeed be biased towards higher mass. \cite{WeissMarcy2014} mitigated this bias by allowing for ``negative'' masses. It is possible that if we were able to obtain greater mass significance on the remaining under-performing planets, the average mass residual for the well-characterized sample would move towards zero for the WM14 relation.

We may also expect the affect of publication bias to grow with decreasing RV semi-amplitude as the noise floor of the instrument is approached. The right panel of Figure \ref{fig:mass_comparison} shows the mass residual of our sample as a function of select planetary and stellar parameters. Interestingly, the mass residual increases with larger RV semi-amplitude. HIRES is more than capable of measuring semi-amplitudes above 5 m s$^{-1}$ and planet masses in this range, primarily predicted with CK17, should not be affected by a stopping criteria based on mass significance. This implies that the CK17 M-R relation, at least for planets between 1.5 and $\sim11$ \rearth~, is not necessarily affected by publication bias.

Another explanation is that the CK17 relations may simply be outdated. The number of planets below 100 \mearth~ has doubled since CK17's study providing a far more diverse data set from which to derive relationships between planet properties. The vast majority of CK17's sample were Jovian worlds with few planets between 1 and 10 \rearth. A consequence of this, perhaps resulting from the 3$\sigma$ threshold criteria, is that the transition point between Terran and Neptunian worlds is placed at 2 \mearth. In a recent reassessment of the M-R relation, \citealt{Baron2023} find that a transition nearer to $\sim8$\mearth~ more appropriately describes the current sample of exoplanets. A higher mass transition point results in a steeper M-R slope predicting higher masses for a given radius. While \citealt{Baron2023} also institute a quality cut on planet mass and radius, their analysis still benefits from the significant increase in well-characterized small planets.

Similarly, \citealt{Edmondson2023} also find discrepancies between measurements and CK17's predictions. When used to predict planetary radius, a scenario that will become increasingly important with the discovery of more non-transiting planets, CK17 tends to \textit{overestimate} planet radius in agreement with our underestimate in mass.

M-R relations serve an important role in not only helping us to understand the population of exoplanets we are presented with today but also for highlighting targets that are particularly valuable for future follow up. The transit spectroscopy metric \citep[TSM,][]{Kempton2018} estimates the expected signal-to-noise ratio of an atmosphere signal obtained with JWST, and it depends on the mass of the planet. Studies that use M-R relations to calculate the TSM can be crucial for bringing planets with high potential to be well-characterized to the forefront \citep{Hord2023}. However, care should be employed with which M-R relations to use, and these empirical trends should not be treated as a substitute for obtaining actual mass measurements. In particular, TSMs calculated with masses predicted from CK17 may be higher than expected, especially for planets larger than $\sim1.5~\rearth$, ~producing exaggerated scale heights and overestimated TSM values.

\section{Conclusions}

In this work we present a uniform analysis of 9,204 RV measurements taken as part of the TESS-Keck Survey. Ultimately, we obtain masses or mass upper limits on a total of 112 transiting and 15 non-transiting planets marking the culmination of 3 years of observations from Keck/HIRES and the Automated Planet Finder. We conclude with our results and main findings:

\begin{itemize}
    \item The TESS-Keck Survey has generated a wealth of RV data that, with this paper, is now public (\S \ref{sec:RV}). This includes 5,110 Keck/HIRES RVs and activity indicators in addition to 4,534 RVs from the Automated Planet Finder covering an average baseline of nearly 4 years (inclusive of archival data). 
    
    \item We uniformly fit all the data in our survey (\S \ref{sec:methods}) to produce the largest homogeneously analyzed RV dataset of TESS-discovered planets to date. We also incorporate updated stellar and planetary parameters from the TKS System Properties Catalog \citep{TKSXV}.
    
    \item We confirm or validate, through RV measurements and statistical validation, 32 new planets of varying size, adding to the many previously published systems in our survey (\S \ref{sec:rv_confirm} \& \ref{sec:statval}). This ensures that future mass-radius relations derived with the TKS sample can incorporate mass upper limits on bona fide planets.
    
    \item We assess the performance of our survey (\S \ref{sec:TKS_goals}) and find that we were able to achieve our observing goals \citep{TKS0} for the majority of our targets. In total, we were able to achieve a $\geq5\sigma$ mass significance for 58 planets in our sample, increasing to 77 planets at $\geq3\sigma$ significance. We measured masses for 38 small planets below 4 \rearth to 3$\sigma$, nearly achieving the TESS Follow-up Program's base science requirement of 50 small planets with masses in a single survey.
    
    \item We gauged the benefit of using the Automated Planet Finder as survey support (\S \ref{APF_support}). APF data  played a crucial role in filling out the phase space for many planets. This resulted in longer baselines for these systems, more precise constraints in the eccentricity of 3 TOIs, and the identification of an observing alias that would have been mistaken as a distant companion. 
    
    \item We compare our masses to those predicted using empirical mass-radius relations (\S \ref{sec:exp_meas_mass}). We find that the \citet{WeissMarcy2014} mass-radius relationship under-predicts the planet masses when compared to our data set. However this discrepancy largely disappears when using the version of their MR relation that excludes masses derived from TTVs. Similarly, masses from \cite{ChenKipping2017} seem to be underestimated by factors of 1/4 to 1/2 compared to our measurements suggesting this mass-radius relation may be outdated and is not able to accurately describe the planet population we observe today.
\end{itemize}

The derived masses for our sample can be found in Table \ref{tab:derived} and the fitted parameters for each system in Table \ref{tab:fitted}. This work represents another set of uniformly analyzed planet masses for TESS-discovered planets, making this the second work to do so \citep{Teske2021}. Our survey contributes to the legacy of HIRES as a prolific instrument for planet detection and mass measurement as we enter an era of sub-m/s RV precision pushing down to smaller planet masses and longer orbital periods.

\renewcommand{\arraystretch}{1.2}
\startlongtable
\begin{deluxetable*}{c c c c c c c c}
\tablecaption{Masses and Derived Parameters \label{tab:derived}}
\tablehead{
  \colhead{TOI} & 
  \colhead{TKS ID} & 
  \colhead{Mass} &
  \colhead{Mass Provenance} &
  \colhead{Density} &
  \colhead{a} &
  \colhead{Literature Mass} &
  \colhead{Reference} \\
  \colhead{} &
  \colhead{} &
  \colhead{(\mearth)} &
  \colhead{} &
  \colhead{(\gcc)} &
  \colhead{(AU)} &
  \colhead{(\mearth)} &
  \colhead{} 
}
\startdata
260.01 & HIP1532 & $3.3^{+1.1}_{-1.0}$ & Mass & $5.6^{+2.02}_{-2.83}$ & $0.091^{+0.002}_{-0.002}$ & - & - \\
266.01 & HIP8152 & $8.9^{+1.7}_{-1.7}$ & Mass & $2.97^{+0.81}_{-1.08}$ & $0.093^{+0.002}_{-0.002}$ & $7.8^{+1.8}_{-1.8}$ & \cite{2023AJ....166..153A}* \\
266.02 & HIP8152 & $10.7^{+2.1}_{-2.2}$ & Mass & $3.63^{+1.02}_{-1.38}$ & $0.139^{+0.003}_{-0.002}$ & $9.4^{+2.2}_{-2.2}$ & \cite{2023AJ....166..153A}* \\
329.01 & T000329 & $40.4^{+3.8}_{-3.7}$ & Mass & $1.84^{+0.41}_{-0.55}$ & $0.0638^{+0.0012}_{-0.0013}$ & $40.7^{+3.7}_{-3.6}$ & \cite{2024arXiv240207893C}* \\
465.01 & WASP156 & $41.8^{+2.9}_{-2.9}$ & Mass & $1.05^{+0.18}_{-0.23}$ & $0.0458^{+0.0008}_{-0.0009}$ & $40.7^{+3.2}_{-2.9}$ & \cite{2018AA...610A..63D} \\
469.01 & 42813 & $5.9^{+2.0}_{-2.0}$ & Mass & $0.8^{+0.29}_{-0.35}$ & $0.11^{+0.002}_{-0.002}$ & $5.8^{+2.4}_{-2.4}$ & \cite{2023AJ....166..153A}* \\
480.01 & 39688 & $20.8^{+4.4}_{-4.2}$ & Mass & $4.91^{+1.3}_{-1.65}$ & $0.077^{+0.001}_{-0.001}$ & $15.8^{+2.2}_{-2.1}$ & \cite{2024arXiv240207893C}* \\
509.01 & 63935 & $10.5^{+1.6}_{-1.5}$ & Mass & $2.32^{+0.5}_{-0.65}$ & $0.083^{+0.002}_{-0.002}$ & $10.8^{+1.8}_{-1.8}$ & \cite{2021AJ....162..215S}* \\
509.02 & 63935 & $11.1^{+2.0}_{-1.9}$ & Mass & $2.11^{+0.49}_{-0.62}$ & $0.148^{+0.003}_{-0.003}$ & $11.1^{+2.4}_{-2.4}$ & \cite{2021AJ....162..215S}* \\
554.01 & 25463 & $8.7^{+2.7}_{-2.6}$ & Mass & $2.87^{+0.95}_{-1.2}$ & $0.076^{+0.001}_{-0.001}$ & $8.5^{+3.1}_{-3.1}$ & \cite{2023AJ....166..153A}* \\
554.02 & 25463 & $4.3^{+1.9}_{-1.9}$ & Mass & $9.09^{+4.32}_{-5.87}$ & $0.0443^{+0.0008}_{-0.0008}$ & $<4.1$ & \cite{2023AJ....166..153A}* \\
561.01 & T000561 & $3.0^{+0.45}_{-0.45}$ & Mass & $5.3^{+1.2}_{-1.51}$ & $0.0104^{+0.0002}_{-0.0002}$ & $2.24^{+0.2}_{-0.2}$ & \cite{2023AJ....165...88B}* \\
561.02 & T000561 & $6.9^{+1.3}_{-1.2}$ & Mass & $1.5^{+0.39}_{-0.51}$ & $0.087^{+0.002}_{-0.002}$ & $6.6^{+0.73}_{-0.73}$ & \cite{2023AJ....165...88B}* \\
561.03 & T000561 & $11.0^{+1.8}_{-1.8}$ & Mass & $2.97^{+0.7}_{-0.91}$ & $0.156^{+0.003}_{-0.003}$ & $12.15^{+1.1}_{-1.1}$ & \cite{2023AJ....165...88B}* \\
561.04 & T000561 & $8.2^{+3.0}_{-3.2}$ & Mass & - & $0.327^{+0.008}_{-0.008}$ & $13.6^{+1.4}_{-1.4}$ & \cite{2023AJ....165...88B}* \\
669.01 & T000669 & $10.0^{+1.4}_{-1.4}$ & Mass & $3.17^{+0.73}_{-0.95}$ & $0.048^{+0.001}_{-0.001}$ & $9.8^{+1.5}_{-1.5}$ & \cite{2023AJ....166..153A}* \\
849.01 & T000849 & $41.8^{+2.5}_{-2.4}$ & Mass & $5.6^{+0.76}_{-0.92}$ & $0.0155^{+0.0003}_{-0.0003}$ & $39.09^{+2.66}_{-2.55}$ & \cite{2020Natur.583...39A} \\
1136.01 & T001136 & $7.6^{+5.1}_{-5.0}$ & Mass & $1.93^{+1.29}_{-1.44}$ & $0.067^{+0.001}_{-0.001}$ & $6.0^{+1.3}_{-1.7}$ & \cite{2024AJ....167...70B}* \\
1136.02 & T001136 & $12.3^{+6.8}_{-7.0}$ & Mass & $0.68^{+0.38}_{-0.43}$ & $0.107^{+0.002}_{-0.002}$ & $8.35^{+1.8}_{-1.6}$ & \cite{2024AJ....167...70B}* \\
1136.03 & T001136 & $<24.0$ & Mass & $<8.0$ & $0.14^{+0.002}_{-0.002}$ & $6.07^{+1.09}_{-1.01}$ & \cite{2024AJ....167...70B}* \\
1136.04 & T001136 & $<32.0$ & Mass & $<3.0$ & $0.175^{+0.003}_{-0.003}$ & $9.7^{+3.9}_{-3.7}$ & \cite{2024AJ....167...70B}* \\
1136.05 & T001136 & $<31.0$ & Mass & $<25.0$ & $0.0514^{+0.0009}_{-0.0009}$ & $3.5^{+0.8}_{-0.7}$ & \cite{2024AJ....167...70B}* \\
1136.06 & T001136 & $12.2^{+9.3}_{-9.6}$ & Mass & $4.11^{+3.18}_{-3.37}$ & $0.23^{+0.004}_{-0.004}$ & $5.6^{+4.1}_{-3.2}$ & \cite{2024AJ....167...70B}* \\
1173.01 & T001173 & $28.3^{+4.1}_{-4.0}$ & Mass & $0.21^{+0.04}_{-0.05}$ & $0.071^{+0.001}_{-0.001}$ & - & - \\
1174.01 & T001174 & $<32.0$ & Mass & $<11.0$ & $0.08^{+0.001}_{-0.001}$ & - & - \\
1180.01 & T001180 & $10.0^{+5.3}_{-5.4}$ & Mass & $1.93^{+1.07}_{-1.25}$ & $0.082^{+0.002}_{-0.002}$ & - & - \\
1181.01 & T001181 & $377.0^{+17.0}_{-17.0}$ & Mass & $0.49^{+0.08}_{-0.11}$ & $0.0364^{+0.0006}_{-0.007}$ & $375.0^{+44.0}_{-44.0}$ & \cite{2022MNRAS.513.5955K} \\
1184.01 & T001184 & $6.8^{+2.3}_{-2.3}$ & Mass & $2.63^{+0.98}_{-1.24}$ & $0.056^{+0.001}_{-0.001}$ & - & - \\
1194.01 & T001194 & $120.0^{+5.6}_{-5.7}$ & Mass & $0.99^{+0.16}_{-0.21}$ & $0.034^{+0.0006}_{-0.0006}$ & - & - \\
1244.01 & T001244 & $6.6^{+3.8}_{-3.7}$ & Mass & $2.66^{+1.52}_{-1.78}$ & $0.061^{+0.001}_{-0.001}$ & - & - \\
1246.01 & T001246 & $5.4^{+1.9}_{-1.9}$ & Mass & $0.73^{+0.28}_{-0.33}$ & $0.132^{+0.003}_{-0.003}$ & $5.3^{+1.7}_{-1.7}$ & \cite{2022BAAS...54e.281T}* \\
1246.02 & T001246 & $8.0^{+1.2}_{-1.1}$ & Mass & $1.84^{+0.42}_{-0.54}$ & $0.049^{+0.001}_{-0.001}$ & $8.1^{+1.1}_{-1.1}$ & \cite{2022BAAS...54e.281T}* \\
1246.03 & T001246 & $9.1^{+1.3}_{-1.2}$ & Mass & $3.39^{+0.78}_{-1.02}$ & $0.061^{+0.001}_{-0.001}$ & $8.8^{+1.2}_{-1.2}$ & \cite{2022BAAS...54e.281T}* \\
1246.04 & T001246 & $14.5^{+2.3}_{-2.3}$ & Mass & $1.83^{+0.47}_{-0.65}$ & $0.212^{+0.004}_{-0.004}$ & $14.8^{+2.2}_{-2.2}$ & \cite{2022BAAS...54e.281T}* \\
1246.05 & T001246 & $31.4^{+4.5}_{-4.3}$ & m$_p$$\sin{(i)}$ & - & $0.389^{+0.008}_{-0.008}$ & $25.6^{+3.6}_{-3.6}$ & \cite{2022BAAS...54e.281T}* \\
1247.01 & 135694 & $6.1^{+1.8}_{-1.8}$ & Mass & $2.05^{+0.68}_{-0.8}$ & $0.12^{+0.002}_{-0.002}$ & $5.7^{+2.1}_{-2.1}$ & \cite{2023AJ....166..153A}* \\
1248.01 & T001248 & $27.4^{+3.6}_{-3.5}$ & Mass & $0.47^{+0.1}_{-0.12}$ & $0.051^{+0.001}_{-0.001}$ & - & - \\
1249.01 & T001249 & $11.9^{+5.2}_{-5.1}$ & Mass & $1.84^{+0.85}_{-1.11}$ & $0.109^{+0.002}_{-0.002}$ & - & - \\
1255.01 & HIP97166 & $19.1^{+1.6}_{-1.5}$ & Mass & $6.86^{+1.26}_{-1.58}$ & $0.091^{+0.002}_{-0.002}$ & $20.0^{+1.5}_{-1.5}$ & \cite{2021AJ....162..265M}* \\
1255.02 & HIP97166 & $8.0^{+1.6}_{-1.6}$ & m$_p$$\sin{(i)}$ & - & $0.124^{+0.002}_{-0.002}$ & $9.9^{+1.8}_{-1.8}$ & \cite{2021AJ....162..265M}* \\
1269.01 & T001269 & $6.4^{+3.0}_{-2.9}$ & Mass & $2.53^{+1.22}_{-1.42}$ & $0.0496^{+0.0009}_{-0.001}$ & - & - \\
1269.02 & T001269 & $6.8^{+4.4}_{-4.3}$ & Mass & $3.15^{+2.05}_{-2.29}$ & $0.083^{+0.001}_{-0.002}$ & - & - \\
1272.01 & T001272 & $27.0^{+2.7}_{-2.5}$ & Mass & $2.09^{+0.44}_{-0.59}$ & $0.0417^{+0.0007}_{-0.0007}$ & $24.6^{+2.3}_{-2.3}$ & \cite{2022AJ....164...97M}* \\
1272.02 & T001272 & $21.7^{+3.6}_{-3.6}$ & m$_p$$\sin{(i)}$ & - & $0.079^{+0.001}_{-0.001}$ & $26.7^{+3.1}_{-3.1}$ & \cite{2022AJ....164...97M}* \\
1279.01 & T001279 & $10.6^{+2.5}_{-2.5}$ & Mass & $3.05^{+0.88}_{-1.14}$ & $0.085^{+0.002}_{-0.002}$ & - & - \\
1288.01 & T001288 & $44.1^{+2.7}_{-2.6}$ & Mass & $1.97^{+0.34}_{-0.44}$ & $0.0374^{+0.0007}_{-0.0008}$ & $42.0^{+3.0}_{-3.0}$ & \cite{2023MNRAS.519.5637K} \\
1288.02 & T001288 & $85.7^{+12.1}_{-11.7}$ & m$_p$$\sin{(i)}$ & - & $1.07^{+0.03}_{-0.03}$ & $84.0^{+7.0}_{-7.0}$ & \cite{2023MNRAS.519.5637K} \\
1294.01 & T001294 & $63.9^{+4.7}_{-4.5}$ & Mass & $0.34^{+0.06}_{-0.08}$ & $0.051^{+0.001}_{-0.001}$ & $62.1^{+4.8}_{-4.6}$ & \cite{2024arXiv240207893C}* \\
1294.02 & T001294 & $147.0^{+15.0}_{-14.0}$ & m$_p$$\sin{(i)}$ & - & $0.61^{+0.02}_{-0.02}$ & $148.5^{+17.0}_{-15.8}$ & \cite{2024arXiv240207893C}* \\
1296.01 & T001296 & $95.3^{+4.7}_{-4.6}$ & Mass & $0.19^{+0.03}_{-0.04}$ & $0.051^{+0.001}_{-0.001}$ & $94.7^{+1.4}_{-1.4}$ & \cite{2021AA...653A.147M} \\
1298.01 & T001298 & $99.0^{+6.0}_{-6.0}$ & Mass & $0.6^{+0.1}_{-0.13}$ & $0.057^{+0.001}_{-0.001}$ & $113.0^{+10.0}_{-10.0}$ & \cite{2021AA...653A.147M} \\
1339.01 & 191939 & $9.6^{+0.9}_{-0.9}$ & Mass & $1.34^{+0.25}_{-0.33}$ & $0.079^{+0.002}_{-0.002}$ & $10.0^{+0.7}_{-0.7}$ & \cite{2023AA...669A..40O}* \\
1339.02 & 191939 & $6.1^{+1.3}_{-1.3}$ & Mass & $1.03^{+0.27}_{-0.34}$ & $0.171^{+0.003}_{-0.003}$ & $8.0^{+1.0}_{-1.0}$ & \cite{2023AA...669A..40O}* \\
1339.03 & 191939 & $2.2^{+1.3}_{-1.3}$ & Mass & $0.4^{+0.23}_{-0.26}$ & $0.209^{+0.004}_{-0.004}$ & $2.8^{+0.6}_{-0.6}$ & \cite{2023AA...669A..40O}* \\
1339.04 & 191939 & $116.0^{+5.0}_{-5.0}$ & m$_p$$\sin{(i)}$ & - & $0.4^{+0.008}_{-0.008}$ & $112.2^{+4.0}_{-4.0}$ & \cite{2023AA...669A..40O}* \\
1339.05 & 191939 & - & m$_p$$\sin{(i)}$ & - & - & $<660.0$ & \cite{2023AA...669A..40O}* \\
1339.06 & 191939 & - & m$_p$$\sin{(i)}$ & - & - & $<13.5$ & \cite{2023AA...669A..40O}* \\
1347.01 & T001347 & $12.2^{+1.6}_{-1.5}$ & Mass & $11.17^{+2.46}_{-3.2}$ & $0.0171^{+0.0003}_{-0.0003}$ & $11.1^{+1.2}_{-1.2}$ & \cite{2024AJ....167..153R}* \\
1347.02 & T001347 & $<9.0$ & Mass & $<11.0$ & $0.0546^{+0.0009}_{-0.001}$ & $<6.4$ & \cite{2024AJ....167..153R}* \\
1386.01 & T001386 & $45.8^{+5.8}_{-5.7}$ & Mass & $1.04^{+0.21}_{-0.27}$ & $0.173^{+0.003}_{-0.003}$ & - & - \\
1386.02 & T001386 & $75.2^{+15.3}_{-15.0}$ & m$_p$$\sin{(i)}$ & - & $0.75^{+0.03}_{-0.02}$ & - & - \\
1410.01 & T001410 & $12.5^{+1.1}_{-1.0}$ & Mass & $2.3^{+0.53}_{-0.74}$ & $0.0207^{+0.0004}_{-0.0004}$ & $12.4^{+0.5}_{-0.5}$ & Livingston et al. \textit{in review} \\
1410.02 & T001410 & $27.0^{+3.3}_{-3.2}$ & m$_p$$\sin{(i)}$ & - & $0.239^{+0.004}_{-0.005}$ & $27.2^{+1.7}_{-1.7}$ & Livingston et al. \textit{in review} \\
1411.01 & GJ9522A & $2.0^{+1.2}_{-1.1}$ & Mass & $6.36^{+3.71}_{-4.14}$ & $0.02122^{+0.0004}_{-0.0004}$ & - & - \\
1422.01 & T001422 & $12.1^{+4.5}_{-4.5}$ & Mass & $1.08^{+0.42}_{-0.5}$ & $0.108^{+0.002}_{-0.002}$ & $9.0^{+2.3}_{-2.0}$ & \cite{2022AA...667A...8N} \\
1430.01 & 235088 & $4.2^{+2.4}_{-2.4}$ & Mass & $2.58^{+1.47}_{-1.66}$ & $0.071^{+0.001}_{-0.001}$ & - & - \\
1436.01 & T001436 & $<8.0$ & Mass & $<12.0$ & $0.0164^{+0.0003}_{-0.0003}$ & - & - \\
1437.01 & 154840 & $10.4^{+2.6}_{-2.6}$ & Mass & $3.91^{+1.11}_{-1.4}$ & $0.139^{+0.003}_{-0.003}$ & $9.6^{+3.9}_{-3.3}$ & Pidhorodetska et al. \textit{in review}* \\
1438.01 & T001438 & $10.9^{+5.9}_{-5.9}$ & Mass & $2.47^{+1.4}_{-1.85}$ & $0.056^{+0.001}_{-0.001}$ & - & - \\
1438.02 & T001438 & $5.6^{+5.0}_{-5.1}$ & Mass & $1.7^{+1.51}_{-1.85}$ & $0.084^{+0.002}_{-0.002}$ & - & - \\
1439.01 & T001439 & $38.5^{+5.7}_{-5.6}$ & Mass & $2.76^{+0.64}_{-0.85}$ & $0.192^{+0.004}_{-0.004}$ & $38.2^{+5.6}_{-5.7}$ & \cite{2024arXiv240207893C}* \\
1443.01 & T001443 & $<30.0$ & Mass & $<14.0$ & $0.147^{+0.003}_{-0.003}$ & - & - \\
1444.01 & T001444 & $3.58^{+0.76}_{-0.75}$ & Mass & $6.84^{+1.78}_{-2.27}$ & $0.0116^{+0.0002}_{-0.0002}$ & $3.87^{+0.71}_{-0.71}$ & \cite{2021AJ....162...62D}* \\
1444.02 & T001444 & $10.2^{+2.7}_{-2.9}$ & m$_p$$\sin{(i)}$ & - & $0.123^{+0.002}_{-0.002}$ & $11.8^{+2.9}_{-2.9}$ & \cite{2021AJ....162...62D}* \\
1451.01 & T001451 & $15.2^{+2.8}_{-2.8}$ & Mass & $4.67^{+1.18}_{-1.55}$ & $0.127^{+0.003}_{-0.003}$ & - & - \\
1456.01 & 332231 & $74.1^{+5.6}_{-5.4}$ & Mass & $0.55^{+0.1}_{-0.12}$ & $0.145^{+0.003}_{-0.003}$ & $77.6^{+6.7}_{-6.7}$ & \cite{2020AJ....159..241D}* \\
1467.01 & T001467 & $<9.0$ & Mass & $<11.0$ & $0.0495^{+0.0009}_{-0.0009}$ & - & - \\
1471.01 & 12572 & $8.3^{+1.7}_{-1.7}$ & Mass & $0.81^{+0.21}_{-0.25}$ & $0.145^{+0.003}_{-0.003}$ & $8.4^{+2.0}_{-2.0}$ & \cite{2023AJ....166..153A}* \\
1471.02 & 12572 & $<7.0$ & Mass & $<1.0$ & $0.269^{+0.005}_{-0.005}$ & $<7.9$ & \cite{2023AJ....166..153A}* \\
1472.01 & T001472 & $16.5^{+5.1}_{-5.1}$ & Mass & $1.24^{+0.41}_{-0.51}$ & $0.065^{+0.001}_{-0.001}$ & - & - \\
1473.01 & 6061 & $10.0^{+2.5}_{-2.5}$ & Mass & $3.82^{+1.1}_{-1.39}$ & $0.06^{+0.001}_{-0.001}$ & $10.8^{+2.7}_{-2.7}$ & \cite{2023AJ....166..153A}* \\
1601.01 & T001601 & $387.0^{+16.0}_{-17.0}$ & Mass & $0.74^{+0.14}_{-0.18}$ & $0.069^{+0.001}_{-0.001}$ & $388.0^{+10.0}_{-10.0}$ & \cite{2024arXiv240207893C}* \\
1611.01 & 207897 & $14.8^{+1.5}_{-1.4}$ & Mass & $6.29^{+1.32}_{-1.78}$ & $0.117^{+0.002}_{-0.002}$ & $14.4^{+1.6}_{-1.6}$ & \cite{2022AA...658A.176H} \\
1669.01 & T001669 & $<13.0$ & Mass & $<6.0$ & $0.0376^{+0.0007}_{-0.0007}$ & - & - \\
1669.02 & T001669 & $191.0^{+25.0}_{-24.0}$ & m$_p$$\sin{(i)}$ & - & $1.27^{+0.03}_{-0.03}$ & $182.0^{+24.0}_{-24.0}$ & \cite{2023AJ....165...60V}* \\
1691.01 & T001691 & $14.6^{+5.4}_{-5.3}$ & Mass & $1.76^{+0.68}_{-0.8}$ & $0.126^{+0.003}_{-0.003}$ & - & - \\
1694.01 & T001694 & $31.3^{+2.7}_{-2.5}$ & Mass & $1.13^{+0.21}_{-0.26}$ & $0.045^{+0.0008}_{-0.0008}$ & $26.1^{+1.1}_{-1.1}$ & \cite{2023AJ....165...60V}* \\
1694.02 & T001694 & $297.0^{+16.0}_{-16.0}$ & m$_p$$\sin{(i)}$ & - & $0.1^{+0.02}_{-0.02}$ & $334.0^{+16.0}_{-16.0}$ & \cite{2023AJ....165...60V}* \\
1710.01 & T001710 & $22.4^{+4.1}_{-4.0}$ & Mass & $0.87^{+0.2}_{-0.25}$ & $0.166^{+0.003}_{-0.0013}$ & - & - \\
1716.01 & 237566 & $4.8^{+2.0}_{-2.1}$ & Mass & $1.09^{+0.49}_{-0.59}$ & $0.081^{+0.002}_{-0.002}$ & - & - \\
1723.01 & T001723 & $10.4^{+5.2}_{-4.8}$ & Mass & $1.6^{+0.79}_{-0.91}$ & $0.114^{+0.002}_{-0.002}$ & - & - \\
1726.01 & 63433 & $37.3^{+9.6}_{-9.6}$ & Mass & $20.15^{+5.89}_{-7.15}$ & $0.072^{+0.001}_{-0.001}$ & $2.4^{+3.0}_{-1.8}$ & \cite{2023AA...672A.126D} \\
1726.02 & 63433 & $17.2^{+10.3}_{-9.3}$ & Mass & $5.43^{+3.13}_{-3.47}$ & $0.146^{+0.003}_{-0.003}$ & $18.9^{+7.0}_{-6.9}$ & \cite{2023AA...672A.126D} \\
1736.01 & T001736 & $12.3^{+1.5}_{-1.4}$ & Mass & $2.09^{+0.43}_{-0.56}$ & $0.073^{+0.001}_{-0.001}$ & $11.9^{+1.6}_{-1.6}$ & \cite{2023AJ....166..153A}* \\
1736.02 & T001736 & $2441.0^{+97.0}_{-99.0}$ & m$_p$$\sin{(i)}$ & - & $1.37^{+0.03}_{-0.03}$ & $2477.0^{+118.0}_{-118.0}$ & \cite{2023AJ....166..153A}* \\
1742.01 & 156141 & $9.7^{+1.9}_{-1.9}$ & Mass & $4.0^{+0.98}_{-1.21}$ & $0.154^{+0.003}_{-0.003}$ & - & - \\
1751.01 & 146757 & $19.5^{+3.5}_{-3.4}$ & Mass & $4.13^{+0.99}_{-1.26}$ & $0.215^{+0.004}_{-0.004}$ & $20.9^{+4.5}_{-4.2}$ & \cite{TKSXVIII}* \\
1753.01 & T001753 & $16.6^{+5.2}_{-5.2}$ & Mass & $5.95^{+2.06}_{-2.49}$ & $0.059^{+0.001}_{-0.001}$ & - & - \\
1758.01 & T001758 & $6.9^{+6.2}_{-6.3}$ & Mass & $0.83^{+0.75}_{-0.83}$ & $0.138^{+0.003}_{-0.003}$ & - & - \\
1759.01 & T001759 & $<15.0$ & Mass & $<3.0$ & $0.115^{+0.002}_{-0.002}$ & $6.8^{+2.0}_{-2.0}$ & \cite{2022AA...660A..86M} \\
1775.01 & T001775 & $96.0^{+15.0}_{-15.0}$ & Mass & $1.01^{+0.22}_{-0.27}$ & $0.08^{+0.002}_{-0.002}$ & - & - \\
1776.01 & 95072 & $1.4^{+1.1}_{-1.0}$ & Mass & $4.13^{+3.13}_{-3.79}$ & $0.0378^{+0.0007}_{-0.0008}$ & - & - \\
1778.01 & 77946 & $10.9^{+1.8}_{-1.7}$ & Mass & $2.44^{+0.61}_{-0.83}$ & $0.073^{+0.001}_{-0.01}$ & - & - \\
1794.01 & T001794 & $8.7^{+4.1}_{-4.0}$ & Mass & $1.31^{+0.62}_{-0.76}$ & $0.082^{+0.002}_{-0.002}$ & - & - \\
1797.01 & 93963 & $18.4^{+3.2}_{-3.1}$ & Mass & $3.31^{+0.8}_{-1.01}$ & $0.04726^{+0.0008}_{-0.0009}$ & $19.2^{+4.1}_{-4.1}$ & \cite{2022AA...667A...1S} \\
1797.02 & 93963 & $2.8^{+2.3}_{-2.3}$ & Mass & $4.81^{+3.91}_{-4.36}$ & $0.02061^{+0.0004}_{-0.0004}$ & $7.8^{+3.2}_{-3.2}$ & \cite{2022AA...667A...1S} \\
1798.01 & T001798 & $6.5^{+2.0}_{-2.0}$ & Mass & $2.61^{+0.9}_{-1.11}$ & $0.074^{+0.001}_{-0.001}$ & - & - \\
1798.02 & T001798 & $5.6^{+0.8}_{-0.7}$ & Mass & $11.17^{+2.5}_{-3.39}$ & $0.0107^{+0.0002}_{-0.0002}$ & - & - \\
1799.01 & 96735 & $4.0^{+1.7}_{-1.8}$ & Mass & $7.51^{+3.45}_{-4.4}$ & $0.071^{+0.001}_{-0.001}$ & - & - \\
1801.01 & HIP57099 & $<16.0$ & Mass & $<12.0$ & $0.075^{+0.001}_{-0.001}$ & - & - \\
1807.01 & HIP65469 & $2.44^{+0.6}_{-0.57}$ & Mass & $3.98^{+1.18}_{-1.53}$ & $0.0122^{+0.0002}_{-0.0002}$ & $2.57^{+0.5}_{-0.5}$ & \cite{2022AA...664A.163N} \\
1823.01 & TIC142381532 & $67.4^{+8.1}_{-8.3}$ & Mass & $0.86^{+0.18}_{-0.24}$ & $0.212^{+0.004}_{-0.004}$ & - & - \\
1824.01 & T001824 & $16.0^{+2.6}_{-2.6}$ & Mass & $4.24^{+0.96}_{-1.21}$ & $0.151^{+0.003}_{-0.003}$ & $18.5^{+3.2}_{-3.2}$ & Lange et al. \textit{in prep}* \\
1836.01 & 148193 & $24.7^{+6.1}_{-5.9}$ & Mass & $0.24^{+0.07}_{-0.08}$ & $0.157^{+0.003}_{-0.003}$ & $29.2^{+4.7}_{-4.8}$ & \cite{2024arXiv240207893C}* \\
1836.02 & 148193 & $<8.0$ & Mass & $<3.0$ & $0.03089^{+0.0005}_{-0.0005}$ & - & - \\
1842.01 & T001842 & $89.4^{+7.8}_{-7.8}$ & Mass & $0.26^{+0.05}_{-0.06}$ & $0.1^{+0.002}_{-0.002}$ & $89.8^{+7.4}_{-7.4}$ & \cite{2024arXiv240207893C}* \\
1898.01 & 83342 & $129.0^{+7.0}_{-8.0}$ & Mass & $0.85^{+0.15}_{-0.2}$ & $0.269^{+0.005}_{-0.005}$ & $127.0^{+7.0}_{-7.0}$ & \cite{2024arXiv240207893C}* \\
2019.01 & T002019 & $34.4^{+4.4}_{-4.4}$ & Mass & $1.25^{+0.28}_{-0.37}$ & $0.128^{+0.003}_{-0.003}$ & $34.7^{+4.2}_{-4.1}$ & \cite{2024arXiv240207893C}* \\
2045.01 & T002045 & $211.0^{+21.0}_{-20.0}$ & Mass & $0.55^{+0.11}_{-0.14}$ & $0.093^{+0.002}_{-0.002}$ & - & - \\
2076.01 & T002076 & $16.1^{+8.3}_{-8.2}$ & Mass & $4.1^{+2.14}_{-2.52}$ & $0.088^{+0.002}_{-0.002}$ & - & - \\
2076.02 & T002076 & $<55.0$ & Mass & $<6.0$ & $0.142^{+0.003}_{-0.003}$ & - & - \\
2076.03 & T002076 & $<29.0$ & Mass & $<4.0$ & $0.199^{+0.004}_{-0.004}$ & - & - \\
2088.01 & T002088 & $<37.0$ & Mass & $<2.0$ & $0.472^{+0.009}_{-0.009}$ & - & - \\
2114.01 & T002114 & $<199.0$ & Mass & $<0.0$ & $0.075^{+0.001}_{-0.001}$ & - & - \\
2128.01 & 155060 & $4.5^{+2.0}_{-2.0}$ & Mass & $2.68^{+1.24}_{-1.48}$ & $0.127^{+0.003}_{-0.003}$ & - & - \\
2145.01 & HIP86040 & $1767.0^{+71.0}_{-72.0}$ & Mass & $5.06^{+0.86}_{-1.11}$ & $0.111^{+0.002}_{-0.002}$ & $1672.0^{+121.0}_{-118.0}$ & \cite{2023MNRAS.521.2765R} \\
\enddata
\end{deluxetable*}
\tablecomments{* Denotes a TKS publication. Upper limitd are given as 3$\sigma$ upper limits. Note that mass marked as m$_p$$\sin{(i)}$ lack an inclination measurement and should be treated as a lower limit. }
\renewcommand{\arraystretch}{1.0} 

\section{Acknowledgements}

The authors thank the anonymous referee whose thorough review greatly increased the quality of this publication. We also thank the time assignment committees of the University of California, the California Institute of Technology, NASA, and the University of Hawaii for supporting the TESS-Keck Survey with observing time at Keck Observatory and on the Automated Planet Finder.  We thank NASA for funding associated with our Key Strategic Mission Support project.  We gratefully acknowledge the efforts and dedication of the Keck Observatory staff for support of HIRES and remote observing.  We recognize and acknowledge the cultural role and reverence that the summit of Maunakea has within the indigenous Hawaiian community. We are deeply grateful to have the opportunity to conduct observations from this mountain.  We thank Ken and Gloria Levy, who supported the construction of the Levy Spectrometer on the Automated Planet Finder. We thank the University of California and Google for supporting Lick Observatory and the UCO staff for their dedicated work scheduling and operating the telescopes of Lick Observatory.  

We thank Johanna Teske for her encouragement and for providing helpful advice on the initial scope of this project.

This research is partially funded through the Caltech-IPAC Visiting Graduate Research Fellowship (VGRF).

This research is based on data collected by the TESS mission. Funding for the TESS mission is provided by the NASA Explorer Program.

This research has made use of the NASA Exoplanet Archive, which is operated by the California Institute of Technology, under contract with the National Aeronautics and Space Administration under the Exoplanet Exploration Program.

This research was carried out in part at the Jet Propulsion Laboratory, California Institute of Technology, under a contract with the National Aeronautics and Space Administration (80NM0018D0004).

This work was partially support by the Future Investigators in NASA Earth and Space Science and Technology (FINESST) program Grant No. 80NSSC22K1754.

J.V.Z. acknowledges support from the Future Investigators in NASA Earth and Space Science and Technology (FINESST) grant 80NSSC22K1606.

E.A.P. acknowledges the support of the Alfred P. Sloan Foundation. 

L.M.W. is supported by the Beatrice Watson Parrent Fellowship and NASA ADAP Grant 80NSSC19K0597. 

D.H. acknowledges support from the Alfred P. Sloan Foundation, the National Aeronautics and Space Administration (80NSSC21K0652) and the Australian Research Council (FT200100871).

I.J.M.C. acknowledges support from the NSF through grant AST-1824644.

A.B. is supported by the NSF Graduate Research Fellowship, grant No. DGE 1745301.

R.A.R. is supported by the NSF Graduate Research Fellowship, grant No. DGE 1745301.

C. D. D. acknowledges the support of the Hellman Family Faculty Fund, the Alfred P. Sloan Foundation, the David \& Lucile Packard Foundation, and the National Aeronautics and Space Administration via the TESS Guest Investigator Program (80NSSC18K1583).  

J.M.A.M. is supported by the NSF Graduate Research Fellowship, grant No. DGE-1842400. J.M.A.M. also acknowledges the LSSTC Data Science Fellowship Program, which is funded by LSSTC, NSF Cybertraining Grant No. 1829740, the Brinson Foundation, and the Moore Foundation; his participation in the program has benefited this work.

M.R. acknowledges support from Heising-Simons Foundation Grant \#2023-4478

\facilities{Keck:I(HIRES), Lick:APF(Levy), TESS, Exoplanet Archive} 

\software{\texttt{RadVel} \citep{Fulton2018}, \texttt{Lightkurve} \citep{lightkurve}, \texttt{triceratops} \citep{Giacalone2021AJ}}

\appendix

\section{Radial Velocity Fits}\label{sec:rv_fits}

In this appendix, we show the periodogram analysis of our RV, activity indicators, and window function for each of the systems we were able to confirm through mass measurements (\S\ref{sec:rv_confirm}). We also show the fits to the RV data points. 

\begin{figure*}
     \centering
     \includegraphics[width=0.9\textwidth]{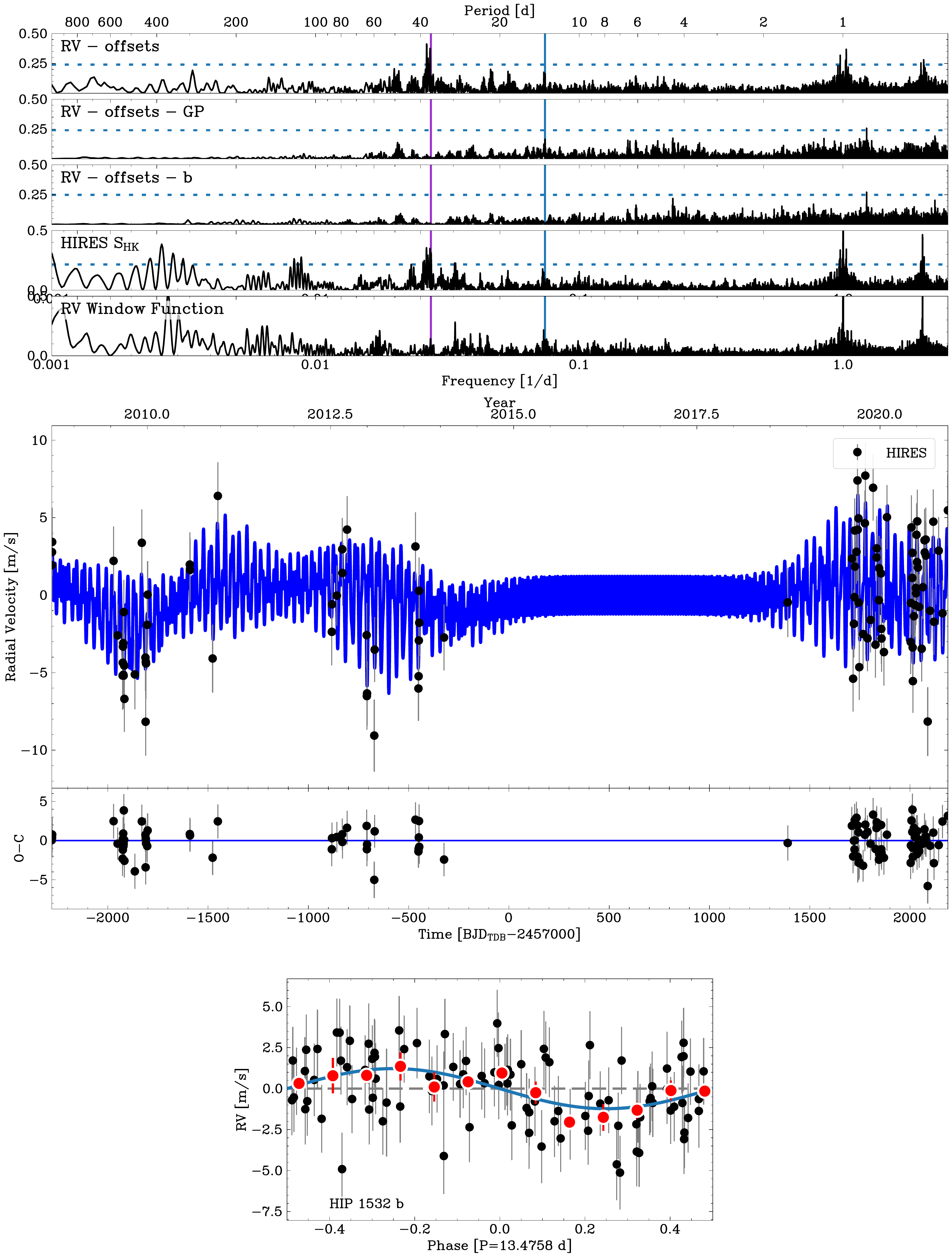}
     \caption{\textbf{Top:} Periodogram analysis of all RVs for HIP 1532 b in our dataset first subtracting any offsets and the Gaussian process, then removing the best fit planet signal (first three panels). We also show a periodogram of the S-indices and the RV window function (last two panels). The planet period is given is shown with the blue vertical line while the rotation period of the star is in purple. The 1\% false alarm probability (FAP) is the horizontal blue dashed line. \textbf{Middle:} The best fit Keplerian orbital plus Gaussian process model (blue line). We add in quadrature the jitter terms in Table \ref{tab:fitted} with the measurement errors for each RV. Residuals to the 1-planet fit are show in the bottom panel. \textbf{Bottom:} RVs phase-folded to the orbital ephemeris of planet b.}
     \label{fig:HIP1532_rv_panel}
\end{figure*}

\begin{figure*}
     \centering
     \includegraphics[width=0.9\textwidth]{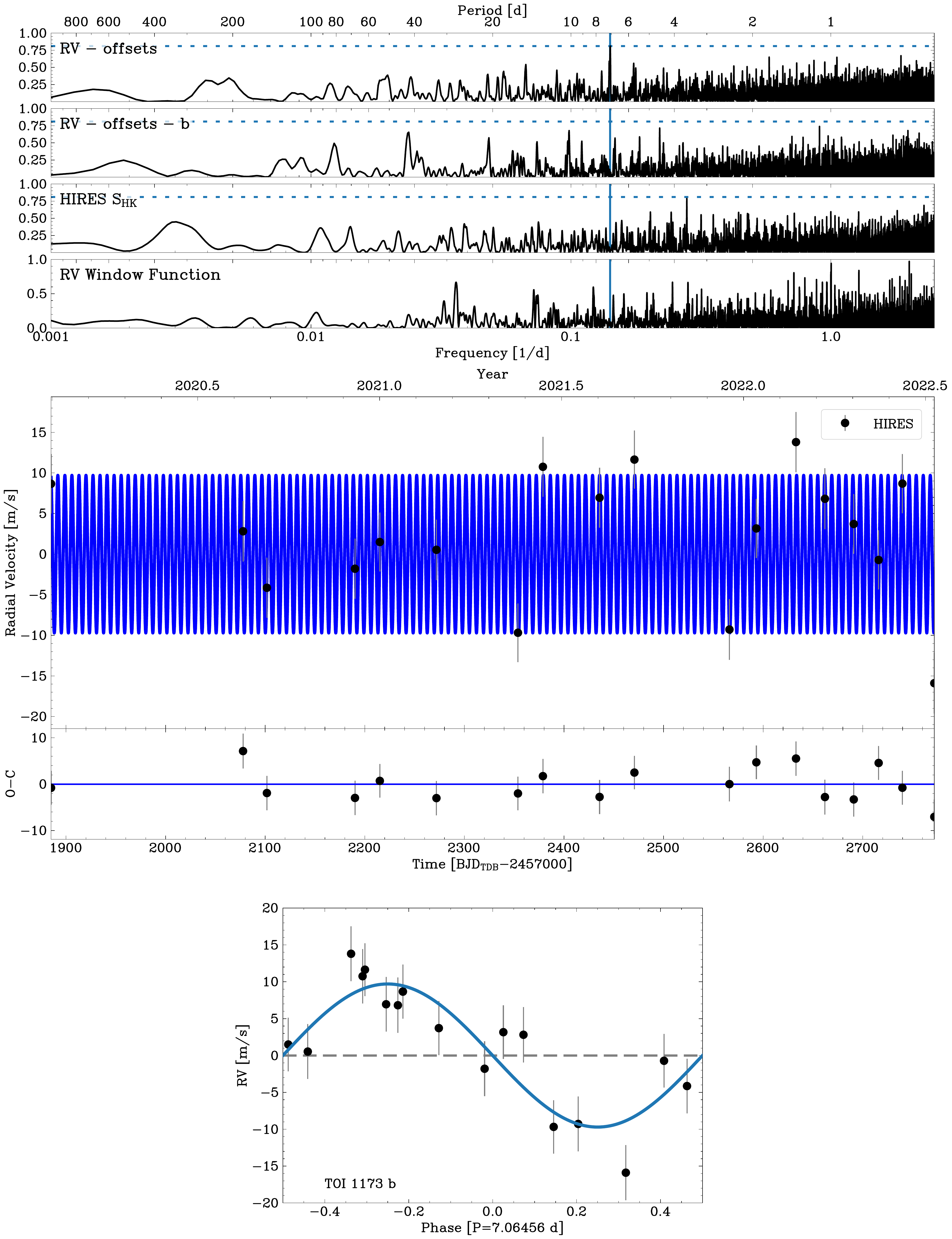}
     \caption{\textbf{Top:} Periodogram analysis of all RVs for TOI--1173 b in our dataset first subtracting any offsets, then removing the best fit planet signal (first two panels). We also show a periodogram of the S-indices and the RV window function (last two panels). The planet period is given is shown with the blue vertical line while the 1\% false alarm probability (FAP) is the horizontal blue dashed line. \textbf{Middle:} The best fit Keplerian orbital model (blue line). We add in quadrature the jitter terms in Table \ref{tab:fitted} with the measurement errors for each RV. Residuals to the 1-planet fit are show in the bottom panel. \textbf{Bottom:} RVs phase-folded to the orbital ephemeris of planet b.}
     \label{fig:T001173_rv_panel}
\end{figure*}

\begin{figure*}
     \centering
     \includegraphics[width=0.9\textwidth]{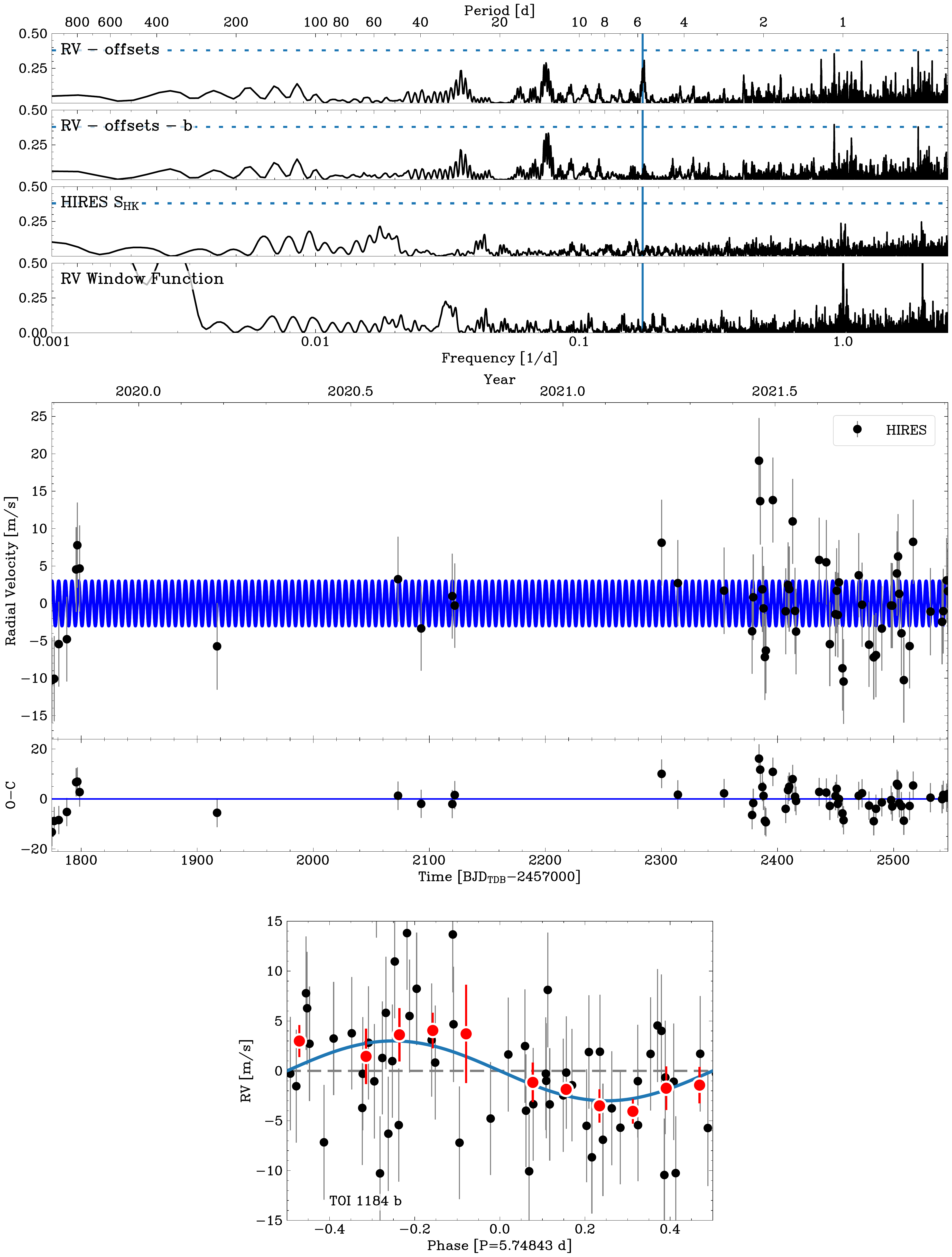}
     \caption{\textbf{Top:} Periodogram analysis of all RVs for TOI--1184 b in our dataset first subtracting any offsets, then removing the best fit planet signal (first two panels). We also show a periodogram of the S-indices and the RV window function (last two panels). The planet period is given is shown with the blue vertical line while the 1\% false alarm probability (FAP) is the horizontal blue dashed line. \textbf{Middle:} The best fit Keplerian orbital model (blue line). We add in quadrature the jitter terms in Table \ref{tab:fitted} with the measurement errors for each RV. Residuals to the 1-planet fit are show in the bottom panel. \textbf{Bottom:} RVs phase-folded to the orbital ephemeris of planet b.}
     \label{fig:T001184_rv_panel}
\end{figure*}

\begin{figure*}
     \centering
     \includegraphics[width=0.9\textwidth]{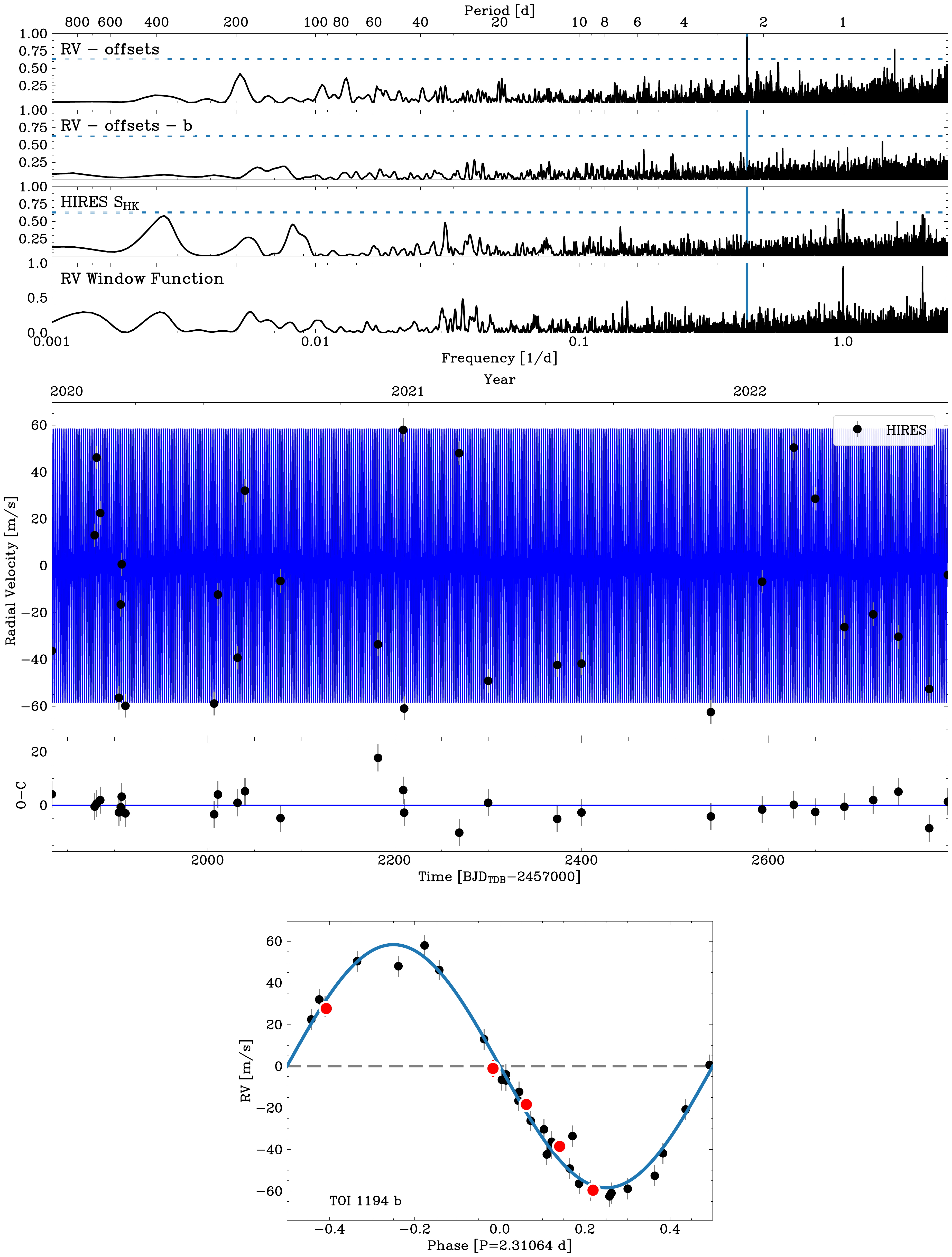}
     \caption{\textbf{Top:} Periodogram analysis of all RVs for TOI--1194 b in our dataset first subtracting any offsets, then removing the best fit planet signal (first two panels). We also show a periodogram of the S-indices and the RV window function (last two panels). The planet period is given is shown with the blue vertical line while the 1\% false alarm probability (FAP) is the horizontal blue dashed line. \textbf{Middle:} The best fit Keplerian orbital model (blue line). We add in quadrature the jitter terms in Table \ref{tab:fitted} with the measurement errors for each RV. Residuals to the 1-planet fit are show in the bottom panel. \textbf{Bottom:} RVs phase-folded to the orbital ephemeris of planet b.}
     \label{fig:T001194_rv_panel}
\end{figure*}

\begin{figure*}
     \centering
     \includegraphics[width=0.9\textwidth]{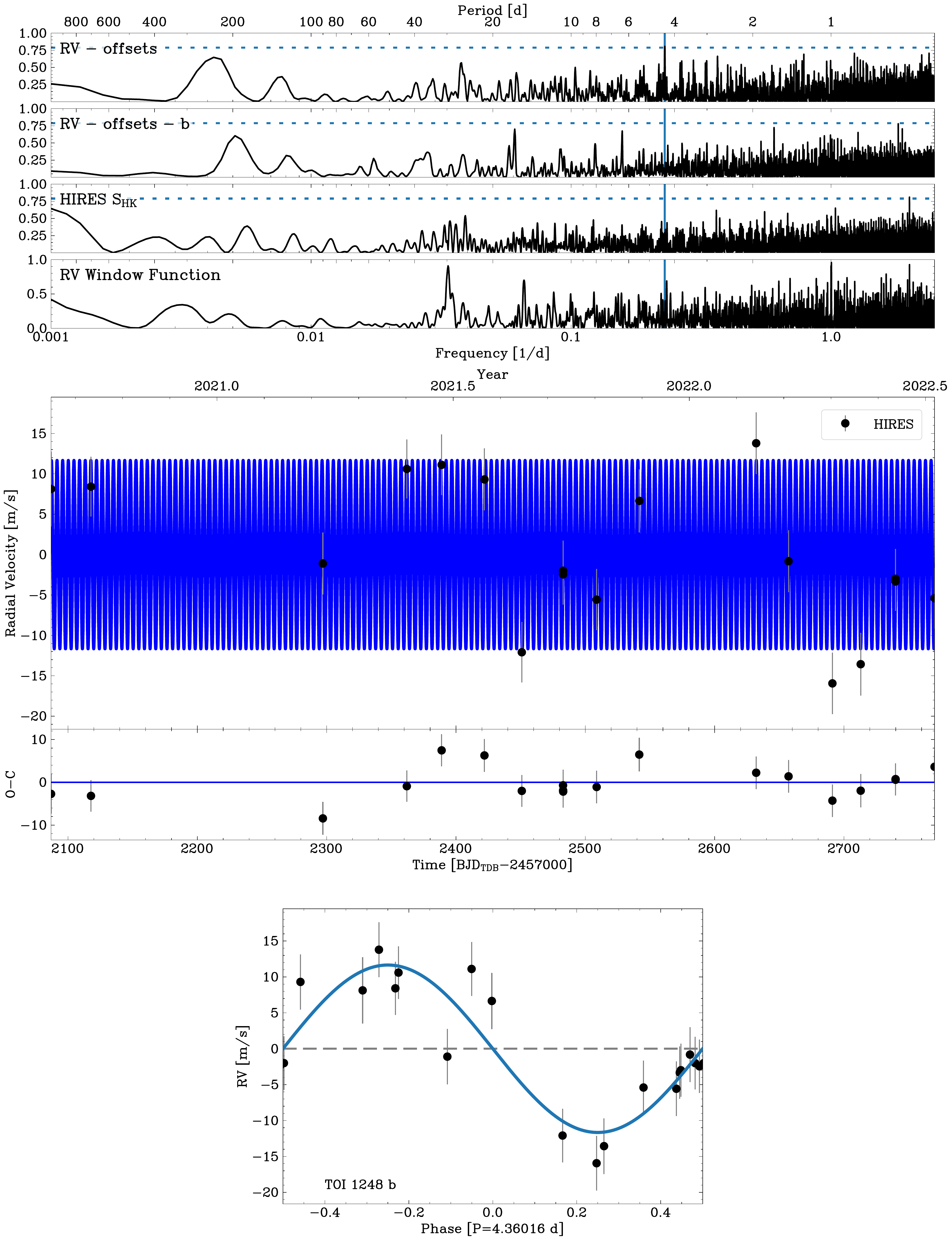}
     \caption{\textbf{Top:} Periodogram analysis of all RVs for TOI--1248 b in our dataset first subtracting any offsets, then removing the best fit planet signal (first two panels). We also show a periodogram of the S-indices and the RV window function (last two panels). The planet period is given is shown with the blue vertical line while the 1\% false alarm probability (FAP) is the horizontal blue dashed line. \textbf{Middle:} The best fit Keplerian orbital model (blue line). We add in quadrature the jitter terms in Table \ref{tab:fitted} with the measurement errors for each RV. Residuals to the 1-planet fit are show in the bottom panel. \textbf{Bottom:} RVs phase-folded to the orbital ephemeris of planet b.}
     \label{fig:T001248_rv_panel}
\end{figure*}

\begin{figure*}
     \centering
     \includegraphics[width=0.9\textwidth]{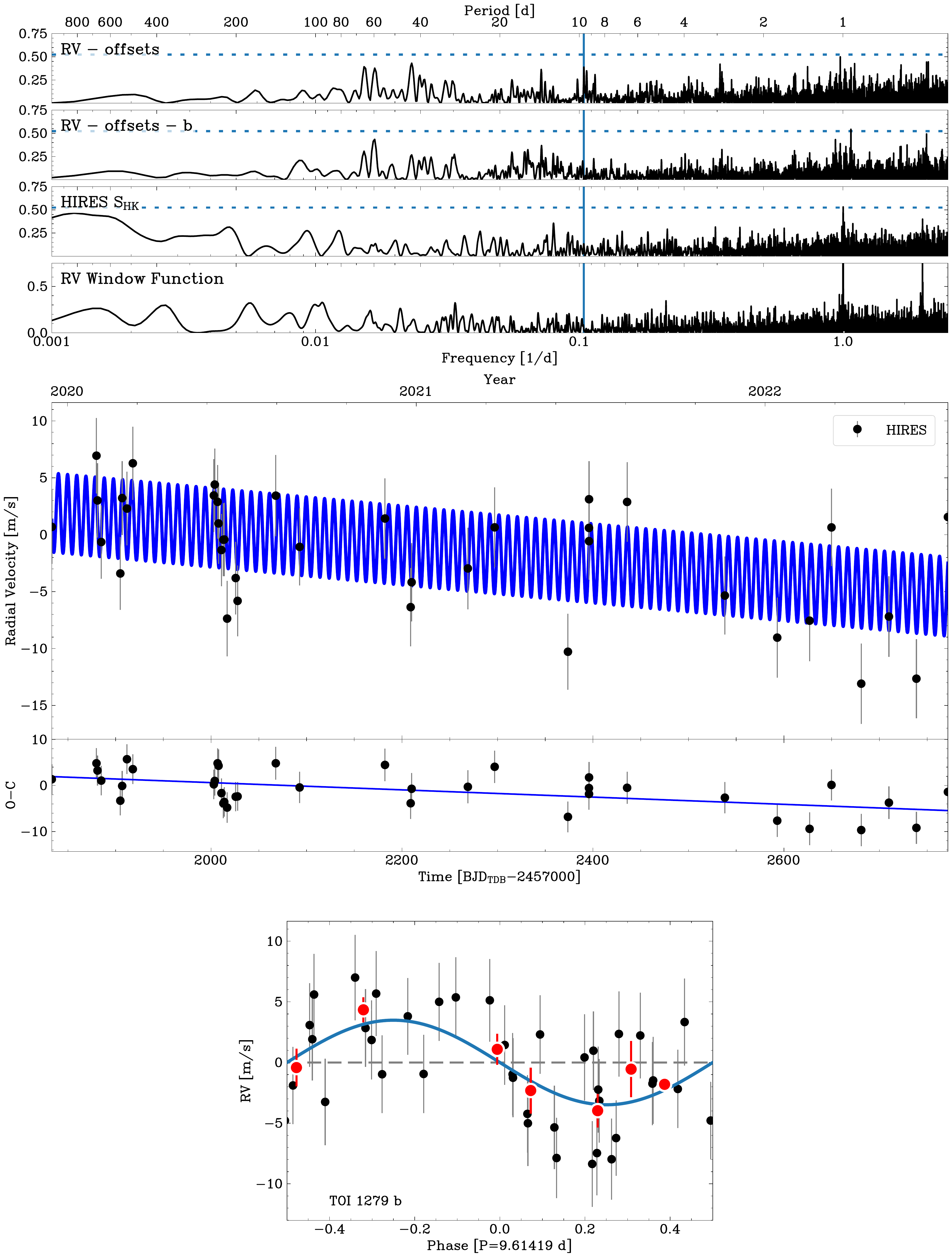}
     \caption{\textbf{Top:} Periodogram analysis of all RVs for TOI--1279 b in our dataset first subtracting any offsets, then removing the best fit planet signal (first two panels). We also show a periodogram of the S-indices and the RV window function (last two panels). The planet period is given is shown with the blue vertical line while the 1\% false alarm probability (FAP) is the horizontal blue dashed line. \textbf{Middle:} The best fit Keplerian orbital model (blue line). We add in quadrature the jitter terms in Table \ref{tab:fitted} with the measurement errors for each RV. Residuals to the 1-planet fit are show in the bottom panel. \textbf{Bottom:} RVs phase-folded to the orbital ephemeris of planet b.}
     \label{fig:T001279_rv_panel}
\end{figure*}


\begin{figure*}
     \centering
     \includegraphics[width=0.9\textwidth]{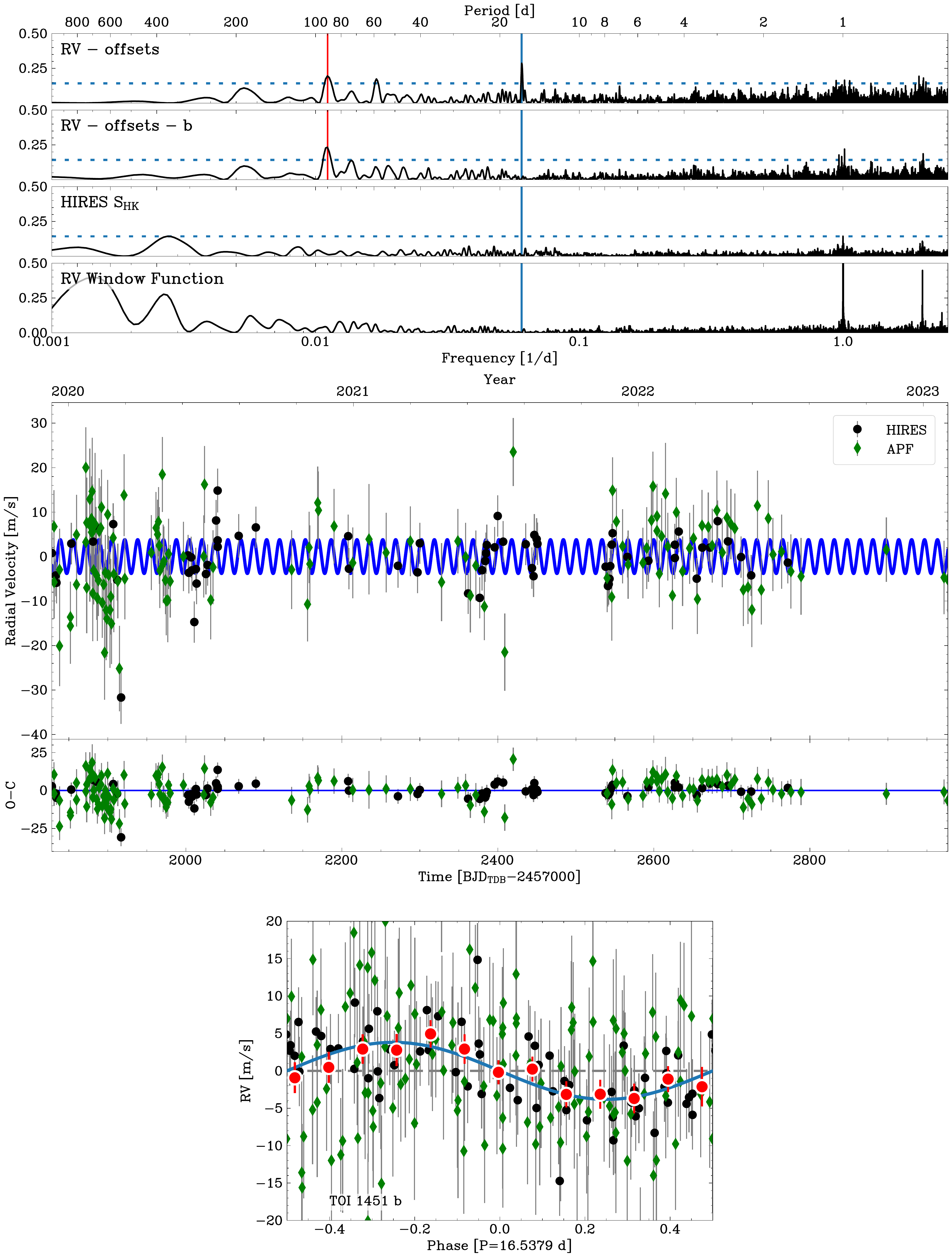}
     \caption{\textbf{Top:} Periodogram analysis of all RVs for TOI 1451 b in our dataset first subtracting any offsets, then removing the best fit planet signal (first two panels). We also show a periodogram of the S-indices and the RV window function (last two panels). The planet period is given is shown with the blue vertical line while the 1\% false alarm probability (FAP) is the horizontal blue dashed line. The vertical red line denotes the location of a known HIRES alias at $\sim90$ days. \textbf{Middle:} The best fit Keplerian orbital model (blue line). We add in quadrature the jitter terms in Table \ref{tab:fitted} with the measurement errors for each RV. Residuals to the 1-planet fit are show in the bottom panel. \textbf{Bottom:} RVs phase-folded to the orbital ephemeris of planet b.}
     \label{fig:T001451_rv_panel}
\end{figure*}

\begin{figure*}
     \centering
     \includegraphics[width=0.9\textwidth]{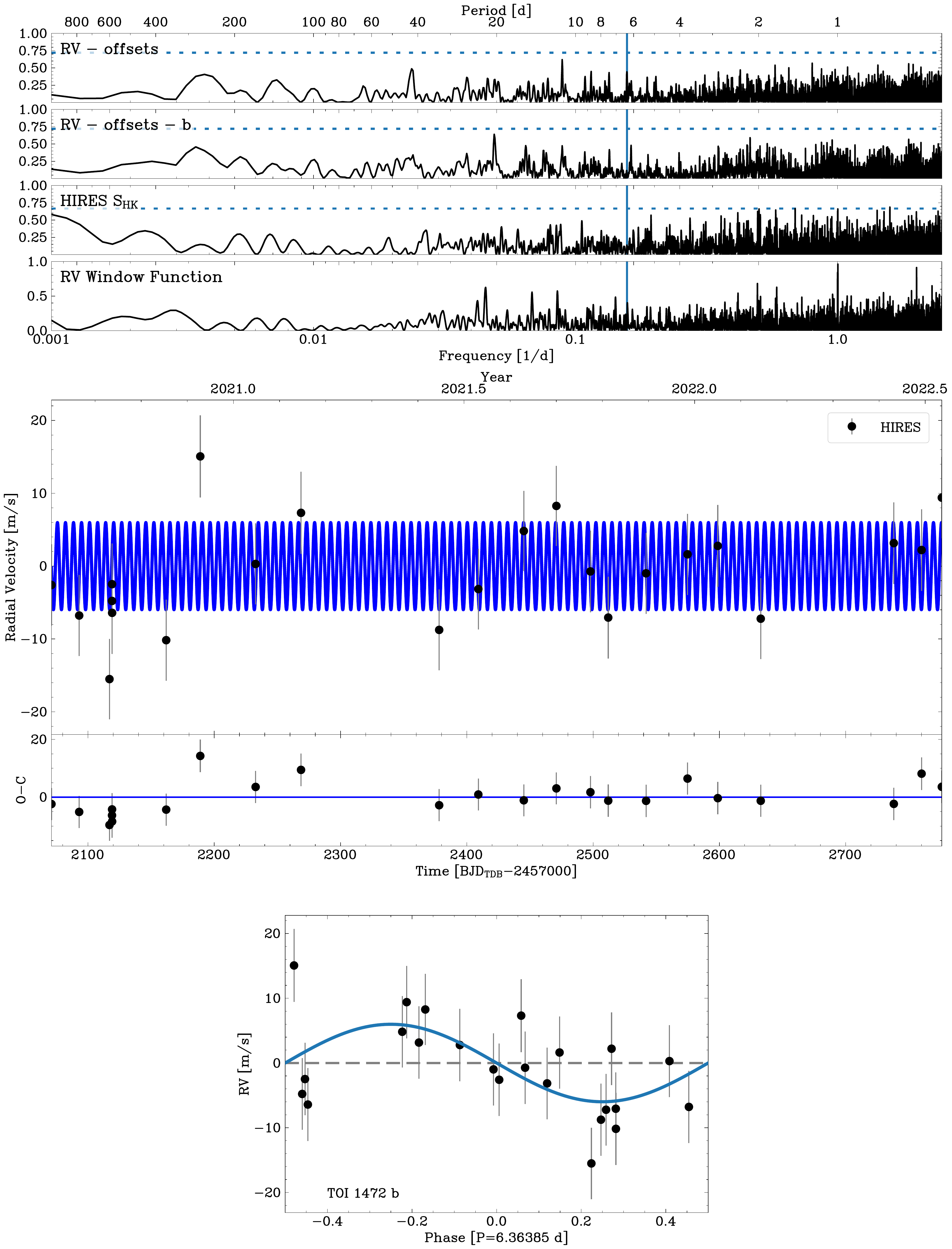}
     \caption{\textbf{Top:} Periodogram analysis of all RVs for TOI--1472 b in our dataset first subtracting any offsets, then removing the best fit planet signal (first two panels). We also show a periodogram of the S-indices and the RV window function (last two panels). The planet period is given is shown with the blue vertical line while the 1\% false alarm probability (FAP) is the horizontal blue dashed line. \textbf{Middle:} The best fit Keplerian orbital model (blue line). We add in quadrature the jitter terms in Table \ref{tab:fitted} with the measurement errors for each RV. Residuals to the 1-planet fit are show in the bottom panel. \textbf{Bottom:} RVs phase-folded to the orbital ephemeris of planet b.}
     \label{fig:T001472_rv_panel}
\end{figure*}

\begin{figure*}
     \centering
     \includegraphics[width=0.9\textwidth]{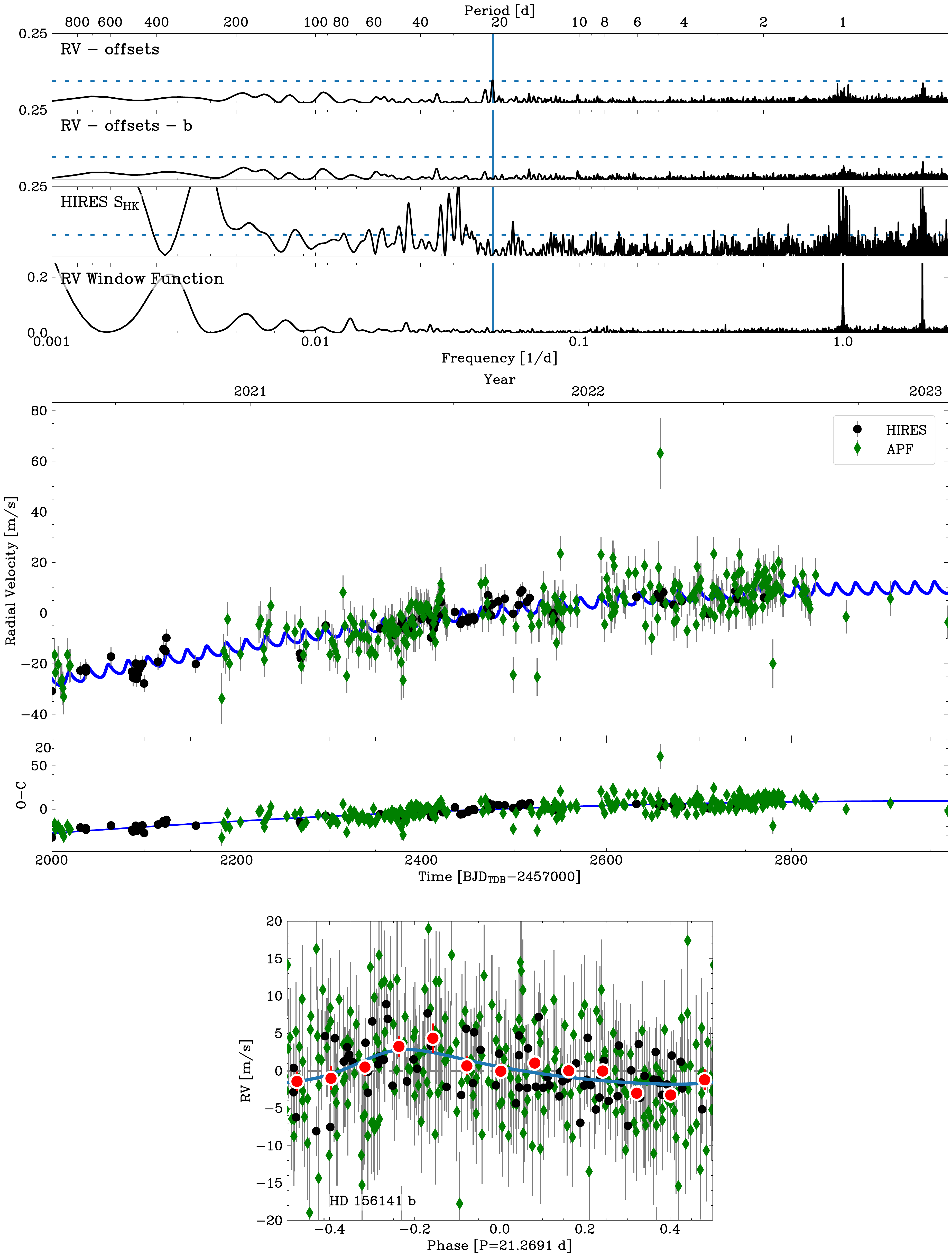}
     \caption{\textbf{Top:} Periodogram analysis of all RVs for HD 156141 b in our dataset first subtracting any offsets, then removing the best fit planet signal (first two panels). We also show a periodogram of the S-indices and the RV window function (last two panels). The planet period is given is shown with the blue vertical line while the 1\% false alarm probability (FAP) is the horizontal blue dashed line. \textbf{Middle:} The best fit Keplerian orbital model (blue line). We add in quadrature the jitter terms in Table \ref{tab:fitted} with the measurement errors for each RV. Residuals to the 1-planet fit are show in the bottom panel. \textbf{Bottom:} RVs phase-folded to the orbital ephemeris of planet b.}
     \label{fig:156141_rv_panel}
\end{figure*}

\begin{figure*}
     \centering
     \includegraphics[width=0.9\textwidth]{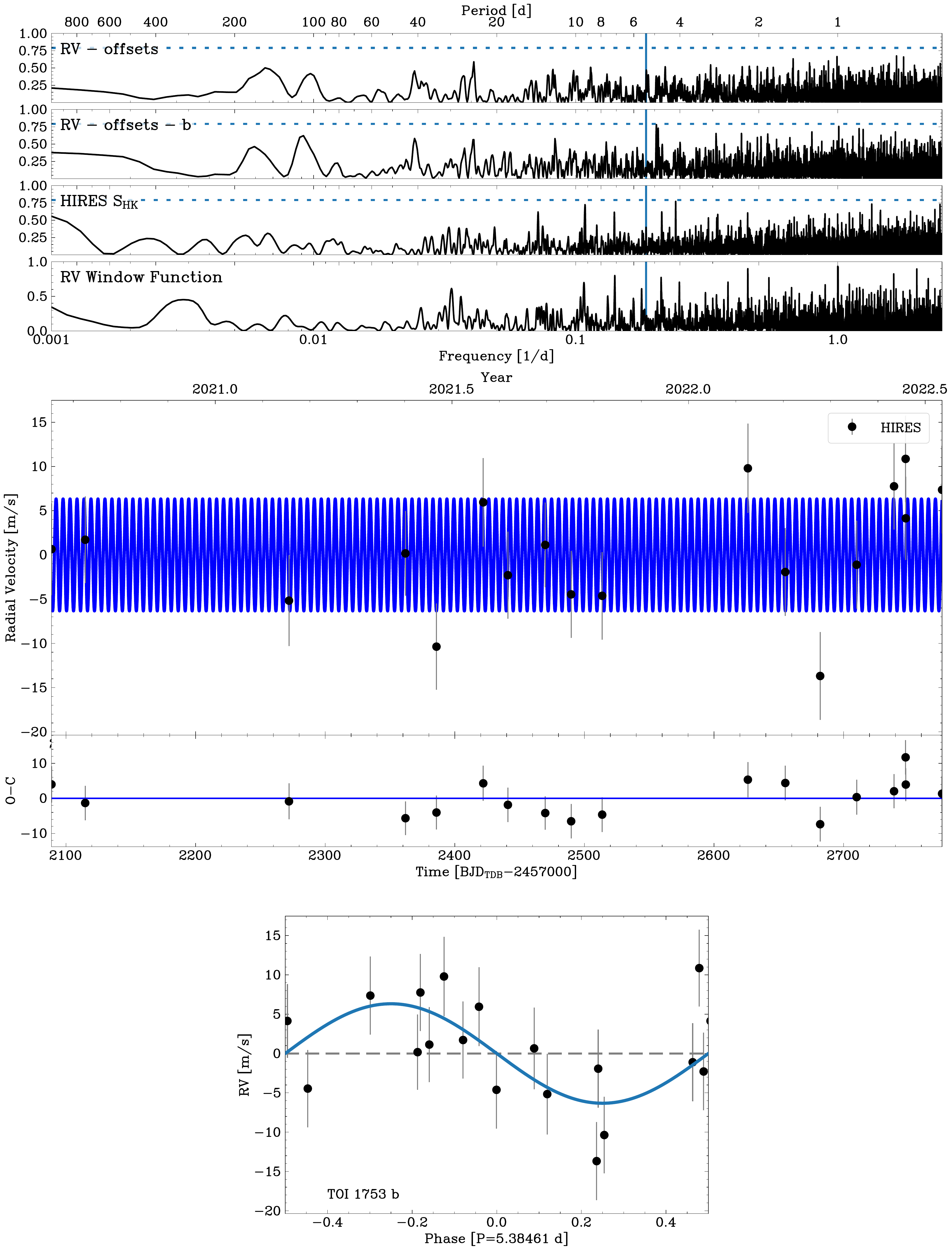}
     \caption{\textbf{Top:} Periodogram analysis of all RVs for TOI--1753 b in our dataset first subtracting any offsets, then removing the best fit planet signal (first two panels). We also show a periodogram of the S-indices and the RV window function (last two panels). The planet period is given is shown with the blue vertical line while the 1\% false alarm probability (FAP) is the horizontal blue dashed line. \textbf{Middle:} The best fit Keplerian orbital model (blue line). We add in quadrature the jitter terms in Table \ref{tab:fitted} with the measurement errors for each RV. Residuals to the 1-planet fit are show in the bottom panel. \textbf{Bottom:} RVs phase-folded to the orbital ephemeris of planet b.}
     \label{fig:T001753_rv_panel}
\end{figure*}

\begin{figure*}
     \centering
     \includegraphics[width=0.9\textwidth]{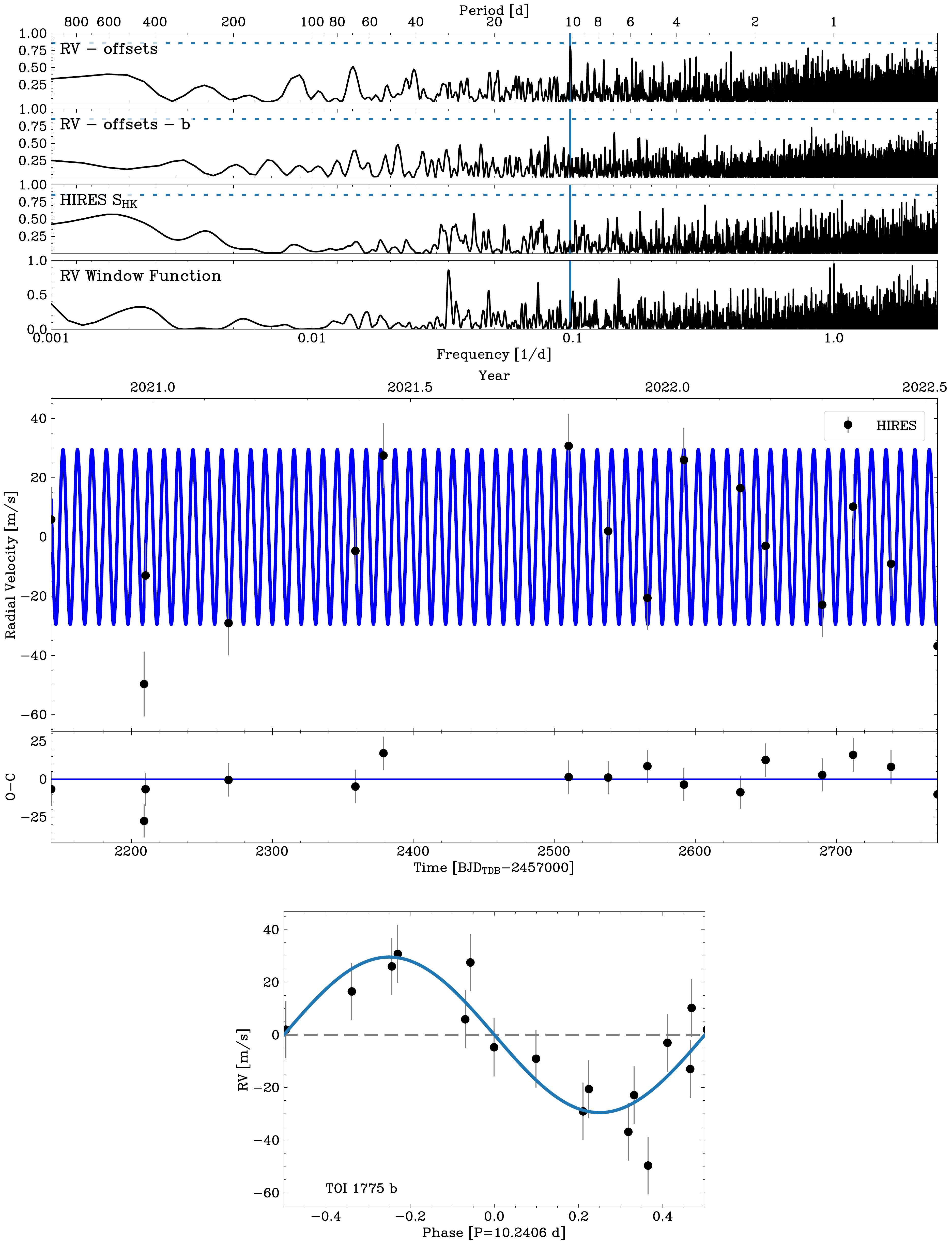}
     \caption{\textbf{Top:} Periodogram analysis of all RVs for TOI--1775 b in our dataset first subtracting any offsets, then removing the best fit planet signal (first two panels). We also show a periodogram of the S-indices and the RV window function (last two panels). The planet period is given is shown with the blue vertical line while the 1\% false alarm probability (FAP) is the horizontal blue dashed line. \textbf{Middle:} The best fit Keplerian orbital model (blue line). We add in quadrature the jitter terms in Table \ref{tab:fitted} with the measurement errors for each RV. Residuals to the 1-planet fit are show in the bottom panel. \textbf{Bottom:} RVs phase-folded to the orbital ephemeris of planet b.}
     \label{fig:T001775_rv_panel}
\end{figure*}

\begin{figure*}
     \centering
     \includegraphics[width=0.9\textwidth]{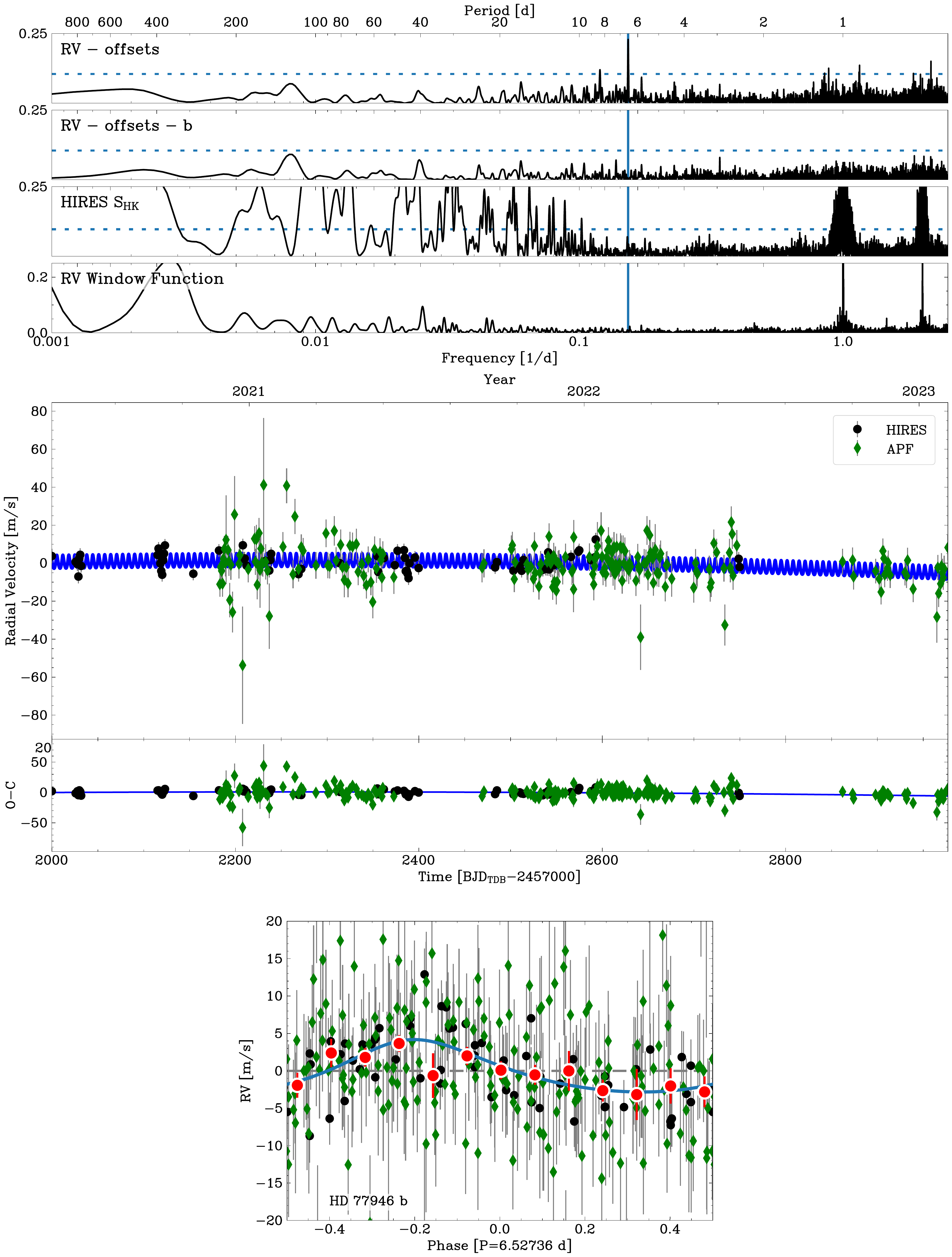}
     \caption{\textbf{Top:} Periodogram analysis of all RVs for HD 77946 b in our dataset first subtracting any offsets, then removing the best fit planet signal (first two panels). We also show a periodogram of the S-indices and the RV window function (last two panels). The planet period is given is shown with the blue vertical line while the 1\% false alarm probability (FAP) is the horizontal blue dashed line. \textbf{Middle:} The best fit Keplerian orbital model (blue line). We add in quadrature the jitter terms in Table \ref{tab:fitted} with the measurement errors for each RV. Residuals to the 1-planet fit are show in the bottom panel. \textbf{Bottom:} RVs phase-folded to the orbital ephemeris of planet b.}
     \label{fig:77946_rv_panel}
\end{figure*}

\begin{figure*}
     \centering
     \includegraphics[width=0.9\textwidth]{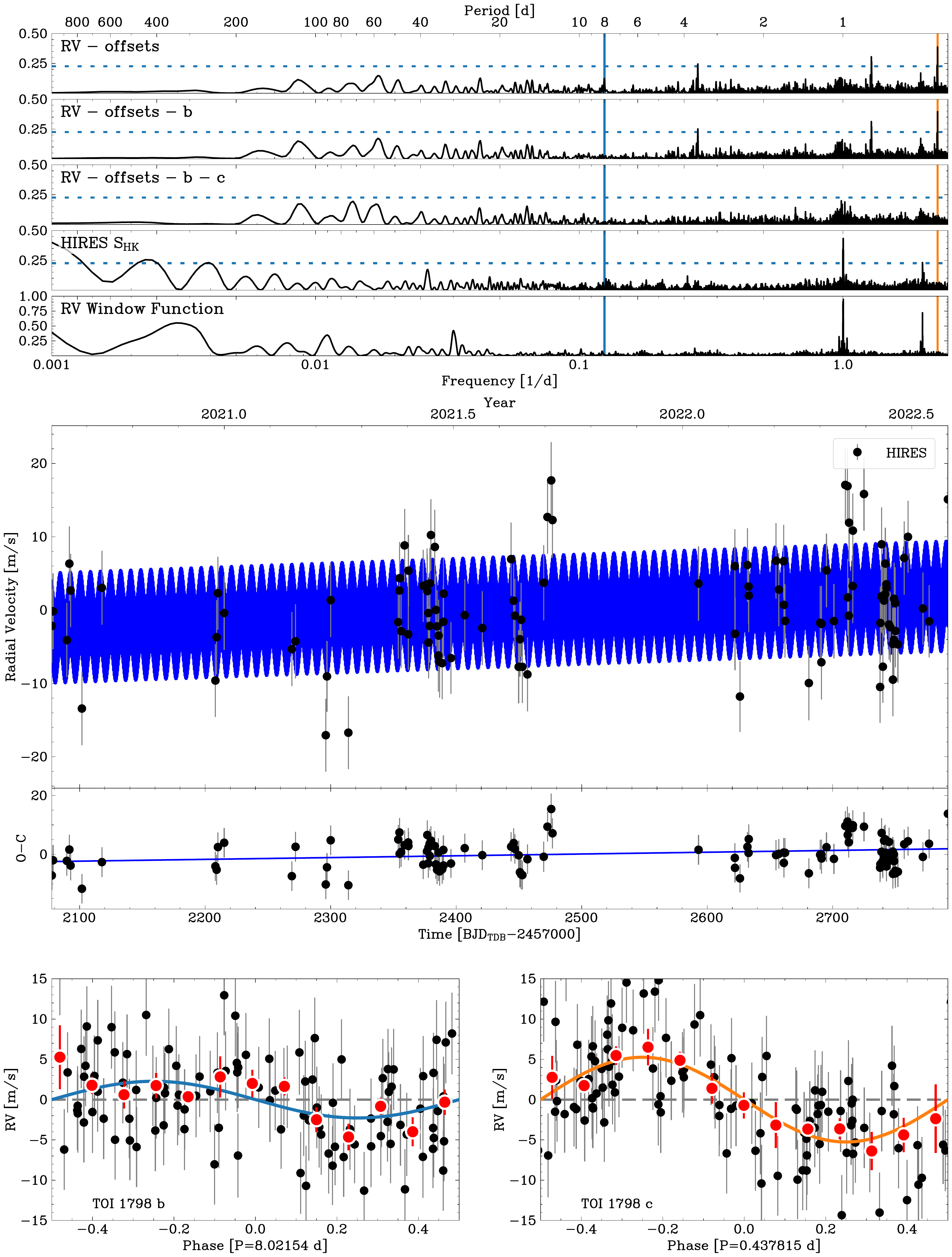}
     \caption{\textbf{Top:} Periodogram analysis of all RVs for TOI--1798 b and c in our dataset first subtracting any offsets, then removing the best fit planet signals (first three panels). We also show a periodogram of the S-indices and the RV window function (last two panels). The periods of planet b and c are shown with the blue and orange vertical lines respectively. The 1\% false alarm probability (FAP) is the horizontal blue dashed line. \textbf{Middle:} The best fit Keplerian orbital model (blue line). We add in quadrature the jitter terms in Table \ref{tab:fitted} with the measurement errors for each RV. Residuals to the 2-planet fit are shown in the bottom panel. \textbf{Bottom:} RVs phase-folded to the orbital ephemeris of planet b and planet c.}
     \label{fig:T001798_rv_panel}
\end{figure*}

\begin{figure*}
     \centering
     \includegraphics[width=0.9\textwidth]{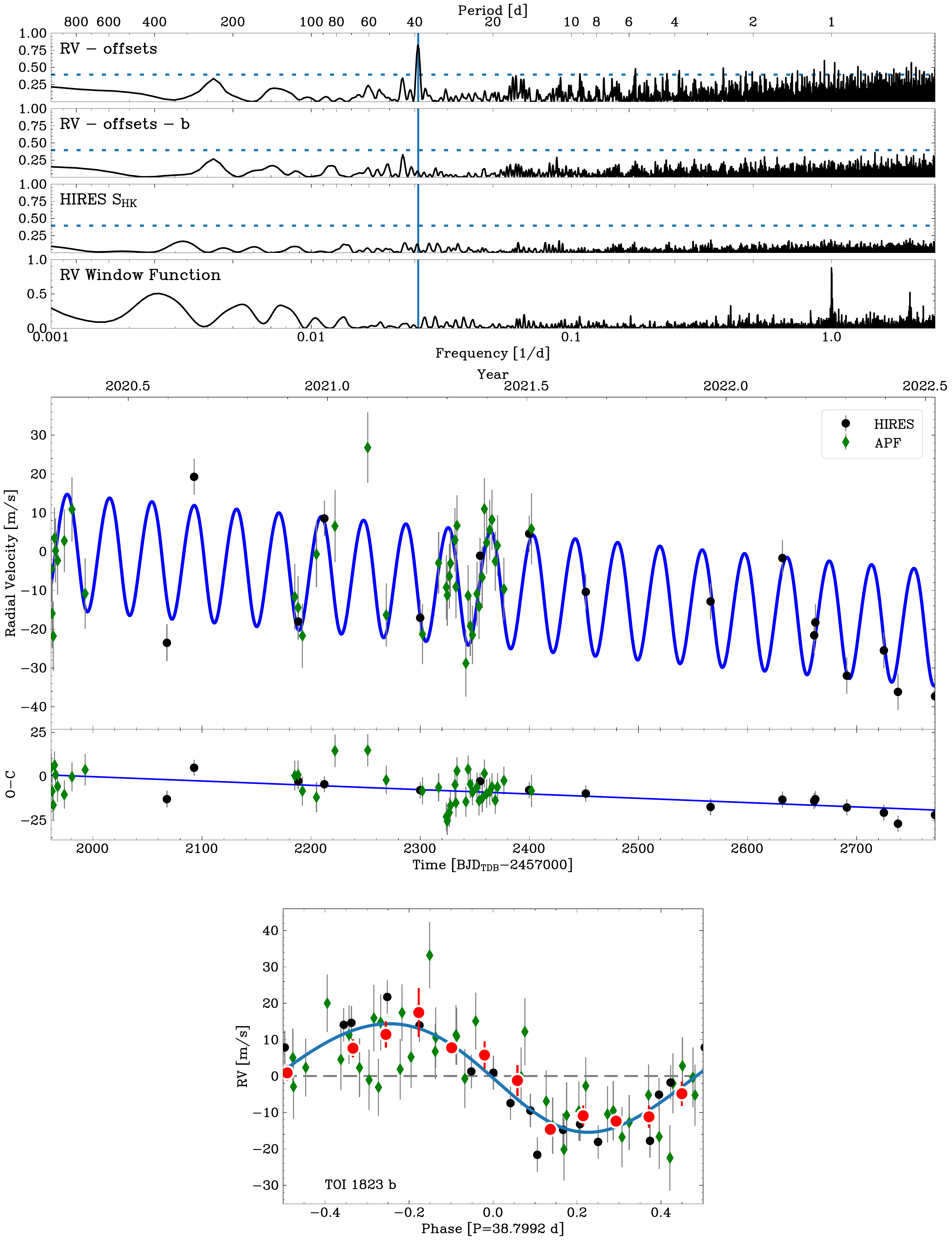}
     \caption{\textbf{Top:} Periodogram analysis of all RVs for TOI--1823 b in our dataset first subtracting any offsets, then removing the best fit planet signal (first two panels). We also show a periodogram of the S-indices and the RV window function (last two panels). The planet period is given is shown with the blue vertical line while the 1\% false alarm probability (FAP) is the horizontal blue dashed line. \textbf{Middle:} The best fit Keplerian orbital model (blue line). We add in quadrature the jitter terms in Table \ref{tab:fitted} with the measurement errors for each RV. Residuals to the 1-planet fit are show in the bottom panel. \textbf{Bottom:} RVs phase-folded to the orbital ephemeris of planet b.}
     \label{fig:TIC142381532_rv_panel}
\end{figure*}

\shorttitle{}
\shortauthors{}

\renewcommand{\arraystretch}{1.0}
\startlongtable
\begin{longrotatetable}
\begin{deluxetable*}{c | c c c c c | c c c c c c}
\movetabledown=10mm
\tabletypesize{\tiny}
\tablecaption{Fitted Parameters \label{tab:fitted}}
\tablehead{
  \colhead{TOI (TKS ID)} &
  \colhead{$P$} &
  \colhead{$T_{\text{0}}$} &
  \colhead{$K$} &
  \colhead{$e$} &
  \colhead{$\omega$} &
  \colhead{$\dot{\gamma}$} &
  \colhead{$\ddot{\gamma}$} &
  \colhead{$\gamma_{\text{HIRES}}$} &
  \colhead{$\gamma_{\text{APF}}$} &
  \colhead{$\sigma_{\text{HIRES}}$} &
  \colhead{$\sigma_{\text{APF}}$} \\
  \colhead{} &
  \colhead{(days)} &
  \colhead{(BJD)} &
  \colhead{(m s$^{-1}$)} &
  \colhead{} &
  \colhead{(radians)} &
  \colhead{(m s$^{-1}$ d$^{-1}$)} &
  \colhead{(m s$^{-1}$ d$^{-2}$)} &
  \colhead{(m s$^{-1}$)} &
  \colhead{(m s$^{-1}$)} &
  \colhead{(m s$^{-1}$)} &
  \colhead{(m s$^{-1}$)}
}
\startdata
TOI 260 (HIP1532) & - & - & - & - & - & $\equiv 0.0$ & $\equiv 0.0$ & -1.17(0.91) & - & 1.84(0.32) & -  \\
b & 13.475815(4.7e-05) & 1392.2944(0.0021) & 1.32(0.43) & $\equiv 0.0$ & $\equiv \pi/2$ & - & - & - & - & - & -  \\
\hline
TOI 266 (HIP8152) & - & - & - & - & - & $\equiv 0.0$ & $\equiv 0.0$ & -1.47(0.36) & - & 3.12(0.29) & -  \\
b & 10.751014(5.5e-05) & 1393.0855(0.0031) & 2.71(0.51) & $\equiv 0.0$ & $\equiv \pi/2$ & - & - & - & - & - & -  \\
c & 19.60562(0.00025) & 1398.2869(0.0064) & 2.76(0.55) & 0.2(0.1) & 2.0(2.8) & - & - & - & - & - & -  \\
\hline
TOI 329 (T000329) & - & - & - & - & - & $\equiv 0.0$ & $\equiv 0.0$ & 0.03(0.86) & - & 4.93(0.69) & -  \\
b & 5.70439(0.00013) & 2090.7955(0.0038) & 15.1(1.3) & 0.39(0.033) & 3.0(3.0) & - & - & - & - & - & -  \\
\hline
TOI 465 (WASP156) & - & - & - & - & - & $\equiv 0.0$ & -1.3e-05(1.1e-05) & 1.23(0.98) & - & 3.04(0.74) & -  \\
b & 3.8361623(1.5e-06) & 1414.13606(0.00031) & 18.7(1.1) & $\equiv 0.0$ & $\equiv \pi/2$ & - & - & - & - & - & -  \\
\hline
TOI 469 (42813) & - & - & - & - & - & 0.0144(0.0046) & 2.08e-05(6.1e-06) & 0.37(0.73) & - & 3.01(0.29) & -  \\
b & 13.630828(2.4e-05) & 1474.569(0.001) & 1.63(0.56) & $\equiv 0.0$ & $\equiv \pi/2$ & - & - & - & - & - & -  \\
\hline
TOI 480 (39688) & - & - & - & - & - & $\equiv 0.0$ & $\equiv 0.0$ & -2.7(0.85) & - & 6.4(0.63) & -  \\
b & 6.866196(1.4e-05) & 1469.5659(0.0015) & 5.9(1.2) & $\equiv 0.0$ & $\equiv \pi/2$ & - & - & - & - & - & -  \\
\hline
TOI 509 (63935) & - & - & - & - & - & 0.0055(0.0031) & -1.17e-05(4.5e-06) & -1.23(0.43) & 1.53(0.82) & 3.55(0.29) & 9.61(0.77)  \\
b & 9.058807(1.5e-05) & 1494.44674(0.00094) & 3.37(0.48) & $\equiv 0.0$ & $\equiv \pi/2$ & - & - & - & - & - & -  \\
c & 21.4027(0.0018) & 1504.13(0.064) & 2.66(0.45) & $\equiv 0.0$ & $\equiv \pi/2$ & - & - & - & - & - & -  \\
\hline
TOI 554 (25463) & - & - & - & - & - & $\equiv 0.0$ & $\equiv 0.0$ & -1.47(0.51) & - & 7.7(0.4) & -  \\
b & 7.0491423(9.5e-06) & 1442.61823(0.00084) & 2.49(0.76) & $\equiv 0.0$ & $\equiv \pi/2$ & - & - & - & - & - & -  \\
c & 3.04405(1e-05) & 1438.473(0.002) & 1.6(0.73) & $\equiv 0.0$ & $\equiv \pi/2$ & - & - & - & - & - & -  \\
\hline
TOI 561 (T000561) & - & - & - & - & - & 0.00197(0.00086) & $\equiv 0.0$ & -2.2(0.5) & - & 2.78(0.25) & -  \\
b & 10.778849(3.2e-05) & 1527.0606(0.0023) & 2.41(0.42) & $\equiv 0.0$ & $\equiv \pi/2$ & - & - & - & - & - & -  \\
c & 0.44656895(7.8e-07) & 1517.94528(0.00072) & 2.99(0.43) & $\equiv 0.0$ & $\equiv \pi/2$ & - & - & - & - & - & -  \\
d & 25.71255(0.00014) & 1521.8815(0.0041) & 2.86(0.45) & $\equiv 0.0$ & $\equiv \pi/2$ & - & - & - & - & - & -  \\
e & 77.9(1.5) & 1536.2(7.3) & 1.48(0.54) & $\equiv 0.0$ & $\equiv \pi/2$ & - & - & - & - & - & -  \\
\hline
TOI 669 (T000669) & - & - & - & - & - & $\equiv 0.0$ & $\equiv 0.0$ & -2.26(0.43) & - & 2.95(0.35) & -  \\
b & 3.945152(1.8e-05) & 1546.1419(0.0025) & 4.26(0.56) & $\equiv 0.0$ & $\equiv \pi/2$ & - & - & - & - & - & -  \\
\hline
TOI 849 (T000849) & - & - & - & - & - & $\equiv 0.0$ & $\equiv 0.0$ & -2.02498874322401(0.0) & - & 0.0(3.5) & -  \\
b & 0.765548(5.3e-05) & 1871.6812(0.0083) & 29.7(1.5) & $\equiv 0.0$ & $\equiv \pi/2$ & - & - & - & - & - & -  \\
\hline
TOI 1136 (T001136) & - & - & - & - & - & $\equiv 0.0$ & $\equiv 0.0$ & 8.6(6.5) & 0.4(0.8) & 12.5(2.4) & 21.8(1.8)  \\
b & 4.17278(0.00022) & 1684.7(0.1) & 3.2(2.7) & $\equiv 0.0$ & $\equiv \pi/2$ & - & - & - & - & - & -  \\
c & 6.2569(0.0002) & 1688.71(0.11) & 2.6(1.7) & $\equiv 0.0$ & $\equiv \pi/2$ & - & - & - & - & - & -  \\
d & 12.51861(1.1e-05) & 1686.0628(0.00043) & 3.3(1.9) & $\equiv 0.0$ & $\equiv \pi/2$ & - & - & - & - & - & -  \\
e & 18.80693(0.00011) & 1697.7299(0.0032) & 1.4(1.5) & $\equiv 0.0$ & $\equiv \pi/2$ & - & - & - & - & - & -  \\
f & 26.317822(5.3e-05) & 1699.3777(0.0011) & 1.5(1.7) & $\equiv 0.0$ & $\equiv \pi/2$ & - & - & - & - & - & -  \\
g & 39.5387(0.0035) & 2423.6705(0.0033) & 2.2(1.7) & $\equiv 0.0$ & $\equiv \pi/2$ & - & - & - & - & - & -  \\
\hline
TOI 1173 (T001173) & - & - & - & - & - & $\equiv 0.0$ & $\equiv 0.0$ & -1.49(0.99) & - & 3.59(0.88) & -  \\
b & 7.06456(2.5e-06) & 1688.71523(0.00021) & 9.8(1.3) & $\equiv 0.0$ & $\equiv \pi/2$ & - & - & - & - & - & -  \\
\hline
TOI 1174 (T001174) & - & - & - & - & - & -0.068(0.011) & 0.0001(4e-05) & -12.7(4.9) & - & 0.0(8.3) & -  \\
b & 8.953458(2.1e-05) & 1690.057(0.001) & 0.3(3.5) & $\equiv 0.0$ & $\equiv \pi/2$ & - & - & - & - & - & -  \\
\hline
TOI 1180 (T001180) & - & - & - & - & - & $\equiv 0.0$ & $\equiv 0.0$ & -0.5(1.5) & - & 6.0(1.2) & -  \\
b & 9.686753(1.2e-05) & 1691.0488(0.00085) & 3.6(1.9) & $\equiv 0.0$ & $\equiv \pi/2$ & - & - & - & - & - & -  \\
\hline
TOI 1181 (T001181) & - & - & - & - & - & $\equiv 0.0$ & $\equiv 0.0$ & -2.9(2.7) & - & 19.0(2.0) & -  \\
b & 2.10319365(4.1e-07) & 1957.82136(0.00011) & 145.8(3.9) & $\equiv 0.0$ & $\equiv \pi/2$ & - & - & - & - & - & -  \\
\hline
TOI 1184 (T001184) & - & - & - & - & - & $\equiv 0.0$ & $\equiv 0.0$ & -0.07(0.77) & - & 5.72(0.58) & -  \\
b & 5.748432(3.9e-06) & 1684.35945(0.00047) & 3.0(1.0) & $\equiv 0.0$ & $\equiv \pi/2$ & - & - & - & - & - & -  \\
\hline
TOI 1194 (T001194) & - & - & - & - & - & $\equiv 0.0$ & $\equiv 0.0$ & 9.3(1.1) & - & 5.07(0.79) & -  \\
b & 2.3106446(5.4e-07) & 1684.92352(0.00014) & 58.4(1.6) & $\equiv 0.0$ & $\equiv \pi/2$ & - & - & - & - & - & -  \\
\hline
TOI 1244 (T001244) & - & - & - & - & - & $\equiv 0.0$ & $\equiv 0.0$ & 0.85(0.89) & - & 3.2(3.1) & -  \\
b & 6.4003(1.3e-05) & 1684.9485(0.0016) & 2.7(1.6) & $\equiv 0.0$ & $\equiv \pi/2$ & - & - & - & - & - & -  \\
\hline
TOI 1246 (T001246) & - & - & - & - & - & 0.0009(0.0012) & 2.2e-05(5.4e-06) & -1.5(0.5) & - & 2.81(0.29) & -  \\
b & 4.307438(4.9e-05) & 1686.56608(0.00074) & 3.42(0.48) & $\equiv 0.0$ & $\equiv \pi/2$ & - & - & - & - & - & -  \\
c & 5.904137(1.4e-05) & 1683.4663(0.0018) & 3.48(0.47) & $\equiv 0.0$ & $\equiv \pi/2$ & - & - & - & - & - & -  \\
d & 18.654874(5.5e-05) & 1688.975(0.002) & 1.41(0.49) & $\equiv 0.0$ & $\equiv \pi/2$ & - & - & - & - & - & -  \\
e & 37.92548(0.00016) & 1700.6958(0.0035) & 3.0(0.46) & $\equiv 0.0$ & $\equiv \pi/2$ & - & - & - & - & - & -  \\
f & 94.0(0.2) & 1702.2(3.8) & 5.0(0.72) & 0.263(0.098) & -0.74(0.37) & - & - & - & - & - & -  \\
\hline
TOI 1247 (135694) & - & - & - & - & - & 0.0032(0.0012) & $\equiv 0.0$ & -0.6(0.5) & -0.61(0.66) & 3.82(0.36) & 7.42(0.52)  \\
b & 15.92346(3.8e-05) & 1687.6496(0.0011) & 1.66(0.49) & $\equiv 0.0$ & $\equiv \pi/2$ & - & - & - & - & - & -  \\
\hline
TOI 1248 (T001248) & - & - & - & - & - & $\equiv 0.0$ & $\equiv 0.0$ & -2.2(0.9) & - & 3.44(0.85) & -  \\
b & 4.3601561(1.1e-06) & 1687.12116(0.00016) & 11.5(1.4) & $\equiv 0.0$ & $\equiv \pi/2$ & - & - & - & - & - & -  \\
\hline
TOI 1249 (T001249) & - & - & - & - & - & 0.016(0.0063) & 2.6e-05(1.3e-05) & -1.4(1.3) & - & 4.25(0.89) & -  \\
b & 13.079151(5.4e-05) & 1694.3804(0.0021) & 3.2(1.4) & $\equiv 0.0$ & $\equiv \pi/2$ & - & - & - & - & - & -  \\
\hline
TOI 1255 (HIP97166) & - & - & - & - & - & 0.0079(0.0011) & $\equiv 0.0$ & -5.6(0.68) & -54.03(0.83) & 2.7(0.3) & 5.92(0.41)  \\
b & 10.288914(6.4e-06) & 1691.65438(0.00037) & 6.18(0.45) & 0.29(0.072) & 1.71(0.18) & - & - & - & - & - & -  \\
c & 16.432(0.041) & 1694.0(1.3) & 2.12(0.41) & $\equiv 0.0$ & $\equiv \pi/2$ & - & - & - & - & - & -  \\
\hline
TOI 1269 (T001269) & - & - & - & - & - & $\equiv 0.0$ & $\equiv 0.0$ & 0.87(0.96) & - & 2.7(3.7) & -  \\
b & 4.2529913(6.8e-06) & 1686.60586(0.00092) & 2.7(1.2) & $\equiv 0.0$ & $\equiv \pi/2$ & - & - & - & - & - & -  \\
c & 9.237885(2.6e-05) & 1685.977(0.0022) & 2.2(1.4) & $\equiv 0.0$ & $\equiv \pi/2$ & - & - & - & - & - & -  \\
\hline
TOI 1272 (T001272) & - & - & - & - & - & $\equiv 0.0$ & $\equiv 0.0$ & 1.6(1.3) & - & 6.0(2.0) & -  \\
b & 3.31599(1.8e-05) & 1713.02552(0.00036) & 13.5(1.3) & 0.35(0.055) & 2.46(0.27) & - & - & - & - & - & -  \\
c & 8.6811(0.0069) & 1885.39(0.29) & 7.4(1.2) & 0.12(0.0) & -1.57(0.0) & - & - & - & - & - & -  \\
\hline
TOI 1279 (T001279) & - & - & - & - & - & -0.0079(0.0021) & $\equiv 0.0$ & 0.44(0.64) & - & 3.1(0.53) & -  \\
b & 9.61419(4e-05) & 1717.4814(0.0022) & 3.5(0.82) & $\equiv 0.0$ & $\equiv \pi/2$ & - & - & - & - & - & -  \\
\hline
TOI 1288 (T001288) & - & - & - & - & - & $\equiv 0.0$ & $\equiv 0.0$ & 3.22(0.69) & - & 3.43(0.54) & -  \\
b & 2.6998279(4.9e-06) & 1712.35877(0.00025) & 20.89(0.93) & $\equiv 0.0$ & $\equiv \pi/2$ & - & - & - & - & - & -  \\
c & 416.0(12.0) & 1894.4(17.0) & 7.54(0.99) & $\equiv 0.0$ & $\equiv \pi/2$ & - & - & - & - & - & -  \\
\hline
TOI 1294 (T001294) & - & - & - & - & - & -0.002(0.016) & -0.000113(3.1e-05) & 15.4(1.9) & - & 4.77(0.81) & -  \\
b & 3.915292(1.5e-05) & 2393.00711(0.00066) & 23.4(1.3) & $\equiv 0.0$ & $\equiv \pi/2$ & - & - & - & - & - & -  \\
c & 159.9(2.5) & 2194.4(3.1) & 15.7(1.3) & $\equiv 0.0$ & $\equiv \pi/2$ & - & - & - & - & - & -  \\
\hline
TOI 1296 (T001296) & - & - & - & - & - & $\equiv 0.0$ & $\equiv 0.0$ & 4.71(0.74) & - & 3.18(0.64) & -  \\
b & 3.9443736(1.6e-06) & 1930.75531(0.00022) & 35.0(1.1) & 0.025(0.016) & -0.01(0.91) & - & - & - & - & - & -  \\
\hline
TOI 1298 (T001298) & - & - & - & - & - & $\equiv 0.0$ & $\equiv 0.0$ & -7.9(1.2) & - & 4.64(0.96) & -  \\
b & 4.537143(3e-06) & 1934.12245(0.00032) & 33.6(1.6) & 0.027(0.027) & -1.0(2.0) & - & - & - & - & - & -  \\
\hline
TOI 1339 (191939) & - & - & - & - & - & 0.1306(0.0032) & -6.53e-05(3.3e-06) & -36.8(0.6) & -39.28(0.73) & 2.27(0.22) & 4.62(0.35)  \\
b & 8.8803232(4.5e-06) & 1715.35572(0.00028) & 3.4(0.3) & $\equiv 0.0$ & $\equiv \pi/2$ & - & - & - & - & - & -  \\
c & 28.579889(3.4e-05) & 1726.05528(0.00066) & 1.46(0.31) & $\equiv 0.0$ & $\equiv \pi/2$ & - & - & - & - & - & -  \\
d & 38.352562(8.4e-05) & 1743.553(0.0011) & 0.47(0.29) & $\equiv 0.0$ & $\equiv \pi/2$ & - & - & - & - & - & -  \\
e & 101.723(0.078) & 2044.15(0.31) & 18.06(0.31) & $\equiv 0.0$ & $\equiv \pi/2$ & - & - & - & - & - & -  \\
\hline
TOI 1347 (T001347) & - & - & - & - & - & $\equiv 0.0$ & $\equiv 0.0$ & -0.5(3.5) & - & 5.55(0.64) & -  \\
b & 0.84742431(5.8e-07) & 1683.55882(0.00036) & 8.6(1.1) & $\equiv 0.0$ & $\equiv \pi/2$ & - & - & - & - & - & -  \\
c & 4.841968(2.5e-05) & 1683.3485(0.0027) & 0.2(1.1) & $\equiv 0.0$ & $\equiv \pi/2$ & - & - & - & - & - & -  \\
\hline
TOI 1386 (T001386) & - & - & - & - & - & 0.0053(0.0066) & -5e-05(2.1e-05) & 0.9(1.3) & - & 5.16(0.57) & -  \\
b & 25.8401(0.0031) & 1752.321(0.003) & 9.7(1.2) & $\equiv 0.0$ & $\equiv \pi/2$ & - & - & - & - & - & -  \\
c & 232.0(9.3) & 1750.9(34.0) & 8.4(1.7) & 0.38(0.17) & -1.24(0.53) & - & - & - & - & - & -  \\
\hline
TOI 1410 (T001410) & - & - & - & - & - & -0.0034(0.0038) & -1.1e-06(6.8e-06) & -2.06(0.73) & - & 2.84(0.36) & -  \\
b & 1.21687(3e-05) & 1739.72931(0.00071) & 8.67(0.63) & $\equiv 0.0$ & $\equiv \pi/2$ & - & - & - & - & - & -  \\
c & 47.56(0.15) & 1980.4(1.2) & 5.57(0.63) & $\equiv 0.0$ & $\equiv \pi/2$ & - & - & - & - & - & -  \\
\hline
TOI 1411 (GJ9522A) & - & - & - & - & - & 0.0065(0.0056) & 2.7e-05(1.3e-05) & -4.0(1.2) & - & 4.2(0.52) & -  \\
b & 1.4520527(1.7e-06) & 1739.47404(0.00046) & 1.55(0.93) & $\equiv 0.0$ & $\equiv \pi/2$ & - & - & - & - & - & -  \\
\hline
TOI 1422 (T001422) & - & - & - & - & - & 0.0131(0.0029) & $\equiv 0.0$ & 2.1(1.3) & - & 4.77(0.73) & -  \\
b & 12.9967(0.0017) & 1745.9221(0.0035) & 3.3(1.2) & $\equiv 0.0$ & $\equiv \pi/2$ & - & - & - & - & - & -  \\
\hline
TOI 1430 (235088) & - & - & - & - & - & 0.0148(0.0057) & -7e-05(2e-05) & 4.0(1.0) & 4.1(2.7) & 3.9(2.5) & 7.2(5.3)  \\
b & 7.434151(1.8e-05) & 1713.08113(0.00087) & 1.53(0.87) & $\equiv 0.0$ & $\equiv \pi/2$ & - & - & - & - & - & -  \\
\hline
TOI 1436 (T001436) & - & - & - & - & - & $\equiv 0.0$ & $\equiv 0.0$ & -0.12(0.74) & - & 6.2(0.59) & -  \\
b & 0.867617(3e-06) & 1711.8963(0.0013) & 0.0(2.0) & $\equiv 0.0$ & $\equiv \pi/2$ & - & - & - & - & - & -  \\
\hline
TOI 1437 (154840) & - & - & - & - & - & $\equiv 0.0$ & $\equiv 0.0$ & -0.2(0.5) & -0.22(0.83) & 4.33(0.39) & 6.4(0.84)  \\
b & 18.84078(0.00012) & 1700.7353(0.0026) & 2.5(0.6) & $\equiv 0.0$ & $\equiv \pi/2$ & - & - & - & - & - & -  \\
\hline
TOI 1438 (T001438) & - & - & - & - & - & 0.0824(0.0057) & $\equiv 0.0$ & -1.5(1.4) & - & 5.0(1.1) & -  \\
b & 5.1396625(7.1e-06) & 1683.6265(0.00071) & 4.4(2.4) & $\equiv 0.0$ & $\equiv \pi/2$ & - & - & - & - & - & -  \\
c & 9.42807(1e-05) & 1689.91613(0.00097) & 1.8(1.7) & $\equiv 0.0$ & $\equiv \pi/2$ & - & - & - & - & - & -  \\
\hline
TOI 1439 (T001439) & - & - & - & - & - & $\equiv 0.0$ & $\equiv 0.0$ & -0.3(0.71) & - & 4.62(0.56) & -  \\
b & 27.644(0.00012) & 1703.4747(0.0029) & 7.0(1.0) & 0.157(0.083) & 2.0(2.7) & - & - & - & - & - & -  \\
\hline
TOI 1443 (T001443) & - & - & - & - & - & 0.029(0.0044) & $\equiv 0.0$ & -3.5(1.4) & - & 5.0(1.0) & -  \\
b & 23.540678(5.8e-05) & 1693.25(0.05) & 1.4(2.2) & $\equiv 0.0$ & $\equiv \pi/2$ & - & - & - & - & - & -  \\
\hline
TOI 1444 (T001444) & - & - & - & - & - & $\equiv 0.0$ & $\equiv 0.0$ & -0.3(0.5) & - & 3.76(0.38) & -  \\
b & 0.4702743(1.1e-06) & 1711.3675(0.0011) & 3.02(0.62) & $\equiv 0.0$ & $\equiv \pi/2$ & - & - & - & - & - & -  \\
c & 16.067(0.017) & 713.2(1.5) & 2.67(0.72) & $\equiv 0.0$ & $\equiv \pi/2$ & - & - & - & - & - & -  \\
\hline
TOI 1451 (T001451) & - & - & - & - & - & $\equiv 0.0$ & $\equiv 0.0$ & -0.09(0.61) & -0.22(0.76) & 4.58(0.52) & 6.39(0.73)  \\
b & 16.537944(4.5e-05) & 1694.3115(0.0015) & 3.82(0.68) & $\equiv 0.0$ & $\equiv \pi/2$ & - & - & - & - & - & -  \\
\hline
TOI 1456 (332231) & - & - & - & - & - & $\equiv 0.0$ & $\equiv 0.0$ & 1.13(0.98) & 1.0(1.0) & 3.85(0.89) & 10.4(0.81)  \\
b & 18.712024(2.3e-05) & 1692.2582(0.00071) & 16.0(1.0) & 0.069(0.034) & 0.04(0.87) & - & - & - & - & - & -  \\
\hline
TOI 1467 (T001467) & - & - & - & - & - & $\equiv 0.0$ & $\equiv 0.0$ & -0.75(0.64) & - & 4.6(0.5) & -  \\
b & 5.97199(0.00083) & 1766.985(0.003) & 0.0(1.8) & $\equiv 0.0$ & $\equiv \pi/2$ & - & - & - & - & - & -  \\
\hline
TOI 1471 (12572) & - & - & - & - & - & -0.058(0.0011) & $\equiv 0.0$ & -13.31(0.59) & -3.86(0.38) & 3.61(0.45) & 5.02(0.31)  \\
b & 20.772858(4.7e-05) & 1767.4228(0.0013) & 2.01(0.41) & $\equiv 0.0$ & $\equiv \pi/2$ & - & - & - & - & - & -  \\
c & 52.56363(0.00016) & 1779.1904(0.0014) & -0.1(0.5) & $\equiv 0.0$ & $\equiv \pi/2$ & - & - & - & - & - & -  \\
\hline
TOI 1472 (T001472) & - & - & - & - & - & $\equiv 0.0$ & $\equiv 0.0$ & -0.3(1.3) & - & 5.9(1.1) & -  \\
b & 6.36381(0.00091) & 1765.6074(0.0022) & 6.0(1.9) & $\equiv 0.0$ & $\equiv \pi/2$ & - & - & - & - & - & -  \\
\hline
TOI 1473 (6061) & - & - & - & - & - & -0.018(0.0061) & 2.7e-05(1.2e-05) & -2.6(0.7) & -2.0(1.9) & 5.21(0.54) & 11.5(1.4)  \\
b & 5.2549(0.0011) & 1769.7863(0.0027) & 3.63(0.87) & $\equiv 0.0$ & $\equiv \pi/2$ & - & - & - & - & - & -  \\
\hline
TOI 1601 (T001601) & - & - & - & - & - & $\equiv 0.0$ & $\equiv 0.0$ & 1.2(1.2) & - & 6.95(0.99) & -  \\
b & 5.33175(0.00014) & 1793.2753(0.0016) & 107.8(1.9) & $\equiv 0.0$ & $\equiv \pi/2$ & - & - & - & - & - & -  \\
\hline
TOI 1611 (207897) & - & - & - & - & - & 0.00165(0.00095) & 3.4e-07(1.8e-07) & 0.52(0.39) & 442.69(0.55) & 2.39(0.28) & 4.11(0.56)  \\
b & 16.20166(1.6e-05) & 1796.49559(0.00067) & 4.26(0.38) & $\equiv 0.0$ & $\equiv \pi/2$ & - & - & - & - & - & -  \\
\hline
TOI 1669 (T001669) & - & - & - & - & - & -0.03(0.006) & $\equiv 0.0$ & -3.7(1.4) & - & 4.3(1.1) & -  \\
b & 500.0(13.0) & 2325.0(9.6) & 15.5(1.8) & $\equiv 0.0$ & $\equiv \pi/2$ & - & - & - & - & - & -  \\
c & 2.6800535(6.2e-06) & 1816.94479(0.00083) & 0.1(1.9) & $\equiv 0.0$ & $\equiv \pi/2$ & - & - & - & - & - & -  \\
\hline
TOI 1691 (T001691) & - & - & - & - & - & $\equiv 0.0$ & $\equiv 0.0$ & 0.48(0.88) & - & 3.74(0.74) & -  \\
b & 16.7369(3e-05) & 1818.091(0.001) & 3.8(1.4) & $\equiv 0.0$ & $\equiv \pi/2$ & - & - & - & - & - & -  \\
\hline
TOI 1694 (T001694) & - & - & - & - & - & $\equiv 0.0$ & $\equiv 0.0$ & -2.8(0.9) & - & 1.2(2.9) & -  \\
b & 3.770107(8.5e-05) & 1817.26629(0.00061) & 14.3(1.1) & $\equiv 0.0$ & $\equiv \pi/2$ & - & - & - & - & - & -  \\
c & 393.1(4.7) & 2170.5(4.3) & 28.84(0.98) & $\equiv 0.0$ & $\equiv \pi/2$ & - & - & - & - & - & -  \\
\hline
TOI 1710 (T001710) & - & - & - & - & - & -0.0125(0.0032) & $\equiv 0.0$ & -2.3(1.4) & 0.7(0.69) & 4.9(1.2) & 7.92(0.49)  \\
b & 24.283377(2.6e-05) & 1836.96292(0.00051) & 4.87(0.86) & $\equiv 0.0$ & $\equiv \pi/2$ & - & - & - & - & - & -  \\
\hline
TOI 1716 (237566) & - & - & - & - & - & $\equiv 0.0$ & $\equiv 0.0$ & -0.7(0.5) & -1.2(0.76) & 3.68(0.41) & 6.4(0.7)  \\
b & 8.0824(4.5e-05) & 1843.853(0.0028) & 1.47(0.62) & $\equiv 0.0$ & $\equiv \pi/2$ & - & - & - & - & - & -  \\
\hline
TOI 1723 (T001723) & - & - & - & - & - & $\equiv 0.0$ & $\equiv 0.0$ & -0.6(1.8) & -0.8(1.1) & 7.1(1.4) & 7.11(0.92)  \\
b & 13.72641(0.00046) & 1852.7024(0.0017) & 2.7(1.3) & $\equiv 0.0$ & $\equiv \pi/2$ & - & - & - & - & - & -  \\
\hline
TOI 1726 (63433) & - & - & - & - & - & $\equiv 0.0$ & $\equiv 0.0$ & -0.4(7.9) & -5.2(9.8) & 15.0(29.0) & 5.0(1.1)  \\
b & 7.1079384(6.6e-06) & 1845.37353(0.00055) & 12.5(3.2) & $\equiv 0.0$ & $\equiv \pi/2$ & - & - & - & - & - & -  \\
c & 20.543847(2.1e-05) & 1844.05899(0.00058) & 4.1(2.3) & $\equiv 0.0$ & $\equiv \pi/2$ & - & - & - & - & - & -  \\
\hline
TOI 1736 (T001736) & - & - & - & - & - & -0.183(0.002) & $\equiv 0.0$ & -131.05(0.76) & -89.89(0.43) & 4.33(0.43) & 6.2(0.3)  \\
b & 7.073076(1.7e-05) & 1792.795(0.001) & 4.01(0.44) & $\equiv 0.0$ & $\equiv \pi/2$ & - & - & - & - & - & -  \\
c & 571.25(0.46) & 1701.81(0.79) & 197.55(0.59) & 0.3685(0.0021) & 2.8312(0.0077) & - & - & - & - & - & -  \\
\hline
TOI 1742 (156141) & - & - & - & - & - & 0.0392(0.0015) & -4.05e-05(5.2e-06) & 5.39(0.45) & -0.56(0.51) & 3.1(0.3) & 6.57(0.32)  \\
b & 21.269084(5.1e-05) & 1725.3518(0.0021) & 2.25(0.44) & 0.3(0.1) & -0.64(0.56) & - & - & - & - & - & -  \\
\hline
TOI 1751 (146757) & - & - & - & - & - & $\equiv 0.0$ & $\equiv 0.0$ & 0.33(0.63) & -0.4(0.7) & 4.7(0.5) & 5.42(0.74)  \\
b & 37.46852(0.00013) & 1733.635(0.002) & 4.15(0.73) & 0.327(0.083) & 2.1(2.9) & - & - & - & - & - & -  \\
\hline
TOI 1753 (T001753) & - & - & - & - & - & $\equiv 0.0$ & $\equiv 0.0$ & -0.2(1.3) & - & 5.1(1.1) & -  \\
b & 5.3846104(9.9e-06) & 1684.503(0.002) & 6.0(2.0) & $\equiv 0.0$ & $\equiv \pi/2$ & - & - & - & - & - & -  \\
\hline
TOI 1758 (T001758) & - & - & - & - & - & $\equiv 0.0$ & $\equiv 0.0$ & 0.1(1.2) & - & 5.0(1.0) & -  \\
b & 20.705127(5.6e-05) & 1806.6974(0.0013) & 1.8(1.7) & $\equiv 0.0$ & $\equiv \pi/2$ & - & - & - & - & - & -  \\
\hline
TOI 1759 (T001759) & - & - & - & - & - & $\equiv 0.0$ & $\equiv 0.0$ & -0.5(2.2) & - & 2.9(0.44) & -  \\
b & 18.85018(0.00031) & 1745.4658(0.0011) & 0.0(1.8) & $\equiv 0.0$ & $\equiv \pi/2$ & - & - & - & - & - & -  \\
\hline
TOI 1775 (T001775) & - & - & - & - & - & $\equiv 0.0$ & $\equiv 0.0$ & 0.1(3.3) & - & 12.3(2.5) & -  \\
b & 10.2405549(9.6e-06) & 1877.5645(0.00047) & 29.8(4.6) & $\equiv 0.0$ & $\equiv \pi/2$ & - & - & - & - & - & -  \\
\hline
TOI 1776 (95072) & - & - & - & - & - & 0.0019(0.0018) & -1.61e-05(5.4e-06) & 0.18(0.68) & -1.78(0.55) & 4.01(0.44) & 3.56(0.51)  \\
b & 2.799868(2.3e-05) & 1871.4976(0.0038) & 0.68(0.51) & $\equiv 0.0$ & $\equiv \pi/2$ & - & - & - & - & - & -  \\
\hline
TOI 1778 (77946) & - & - & - & - & - & -0.0072(0.0019) & -1.34e-05(5.5e-06) & -2.18(0.63) & 178.1(0.6) & -3.38(0.39) & 5.55(0.55)  \\
b & 6.527363(2.7e-05) & 1875.9989(0.0021) & 3.4(0.53) & 0.211(0.072) & -0.16(0.67) & - & - & - & - & - & -  \\
\hline
TOI 1794 (T001794) & - & - & - & - & - & $\equiv 0.0$ & $\equiv 0.0$ & -0.3(0.8) & - & 3.2(0.73) & -  \\
b & 8.765528(3.5e-05) & 1715.312(0.002) & 2.8(1.3) & $\equiv 0.0$ & $\equiv \pi/2$ & - & - & - & - & - & -  \\
\hline
TOI 1797 (93963) & - & - & - & - & - & $\equiv 0.0$ & $\equiv 0.0$ & -3.8(1.6) & 0.4(1.7) & 10.02(0.98) & 10.99(0.97)  \\
b & 1.0376109(8.5e-06) & 2635.926(0.003) & 1.6(1.4) & $\equiv 0.0$ & $\equiv \pi/2$ & - & - & - & - & - & -  \\
c & 3.6451389(4.4e-06) & 1902.874(0.00068) & 7.2(1.2) & $\equiv 0.0$ & $\equiv \pi/2$ & - & - & - & - & - & -  \\
\hline
TOI 1798 (T001798) & - & - & - & - & - & 0.0061(0.0025) & $\equiv 0.0$ & -0.7(0.5) & - & 4.73(0.39) & -  \\
b & 8.02154(2.8e-05) & 1741.5942(0.0022) & 2.3(0.7) & $\equiv 0.0$ & $\equiv \pi/2$ & - & - & - & - & - & -  \\
c & 0.4378146(1.4e-06) & 1738.6343(0.00097) & 5.24(0.66) & $\equiv 0.0$ & $\equiv \pi/2$ & - & - & - & - & - & -  \\
\hline
TOI 1799 (96735) & - & - & - & - & - & $\equiv 0.0$ & $\equiv 0.0$ & -1.65(0.43) & 0.0(3.3) & 3.16(0.35) & 8.0(11.0)  \\
b & 7.085738(8.3e-05) & 1904.8318(0.0072) & 1.4(0.6) & $\equiv 0.0$ & $\equiv \pi/2$ & - & - & - & - & - & -  \\
\hline
TOI 1801 (HIP57099) & - & - & - & - & - & -0.0688(0.0081) & 9.3e-05(3.4e-05) & -10.1(3.8) & - & 9.0(2.0) & -  \\
b & 10.643985(2.6e-05) & 1903.548(0.002) & -2.2(3.2) & $\equiv 0.0$ & $\equiv \pi/2$ & - & - & - & - & - & -  \\
\hline
TOI 1807 (HIP65469) & - & - & - & - & - & $\equiv 0.0$ & $\equiv 0.0$ & 2.1(7.5) & - & 1.5(2.3) & -  \\
b & 0.54937097(7.2e-07) & 1899.345(0.00038) & 2.18(0.51) & $\equiv 0.0$ & $\equiv \pi/2$ & - & - & - & - & - & -  \\
\hline
TOI 1823 (TIC142381532) & - & - & - & - & - & -0.025(0.0052) & $\equiv 0.0$ & 14.9(2.9) & 5.1(1.9) & 5.3(1.5) & 6.9(1.4)  \\
b & 38.8(0.01) & 2645.70502(0.00071) & 14.4(1.7) & 0.062(0.057) & 1.3(2.2) & - & - & - & - & - & -  \\
\hline
TOI 1824 (T001824) & - & - & - & - & - & -0.007(0.0031) & $\equiv 0.0$ & -0.1(0.98) & 0.68(0.94) & 1e-06(0.28) & 5.85(0.46)  \\
b & 22.808533(5.5e-05) & 1879.5467(0.0011) & 3.93(0.62) & $\equiv 0.0$ & $\equiv \pi/2$ & - & - & - & - & - & -  \\
\hline
TOI 1836 (148193) & - & - & - & - & - & -0.017(0.019) & -7e-05(3.6e-05) & 2.3(2.4) & - & 6.04(0.76) & -  \\
b & 20.380845(2.8e-05) & 1933.16603(0.00078) & 5.0(1.2) & $\equiv 0.0$ & $\equiv \pi/2$ & - & - & - & - & - & -  \\
c & 1.7727471(7.9e-06) & 2739.67(0.003) & -0.3(1.3) & $\equiv 0.0$ & $\equiv \pi/2$ & - & - & - & - & - & -  \\
\hline
TOI 1842 (T001842) & - & - & - & - & - & -0.0269(0.0073) & $\equiv 0.0$ & -0.13(0.97) & - & 5.13(0.77) & -  \\
b & 9.5739181(9.9e-06) & 1933.33614(0.00081) & 21.7(1.9) & 0.276(0.064) & 1.18(0.17) & - & - & - & - & - & -  \\
\hline
TOI 1898 (83342) & - & - & - & - & - & $\equiv 0.0$ & $\equiv 0.0$ & -6.9(0.7) & -15.0(3.0) & 3.76(0.59) & 4.9(9.9)  \\
b & 45.52234(6.1e-05) & 1894.2511(0.0008) & 22.9(1.1) & 0.485(0.039) & 1.239(0.088) & - & - & - & - & - & -  \\
\hline
TOI 2019 (T002019) & - & - & - & - & - & $\equiv 0.0$ & $\equiv 0.0$ & 0.57(0.66) & - & 3.93(0.55) & -  \\
b & 15.3444(0.0057) & 1942.942(0.009) & 7.97(0.95) & 0.099(0.069) & 1.0(1.0) & - & - & - & - & - & -  \\
\hline
TOI 2045 (T002045) & - & - & - & - & - & -0.099(0.014) & $\equiv 0.0$ & 0.0(0.0) & - & -2.4(8.5) & -  \\
b & 9.077534(8.5e-05) & 1765.5949(0.0014) & 54.7(4.6) & $\equiv 0.0$ & $\equiv \pi/2$ & - & - & - & - & - & -  \\
\hline
TOI 2076 (T002076) & - & - & - & - & - & $\equiv 0.0$ & $\equiv 0.0$ & -0.5(7.4) & -13.0(19.0) & 17.3(4.3) & 14.0(4.0)  \\
b & 10.355096(1.8e-05) & 1743.744(0.001) & 5.2(2.7) & $\equiv 0.0$ & $\equiv \pi/2$ & - & - & - & - & - & -  \\
c & 21.015327(2.3e-05) & 1748.68944(0.00063) & 5.2(2.9) & $\equiv 0.0$ & $\equiv \pi/2$ & - & - & - & - & - & -  \\
d & 35.125686(8.6e-05) & 1938.2931(0.0014) & 0.3(2.2) & $\equiv 0.0$ & $\equiv \pi/2$ & - & - & - & - & - & -  \\
\hline
TOI 2088 (T002088) & - & - & - & - & - & $\equiv 0.0$ & $\equiv 0.0$ & -0.2(1.6) & - & 7.1(1.3) & -  \\
b & 124.72997(0.00065) & 1769.6066(0.0027) & 0.26(0.92) & $\equiv 0.0$ & $\equiv \pi/2$ & - & - & - & - & - & -  \\
\hline
TOI 2114 (T002114) & - & - & - & - & - & $\equiv 0.0$ & $\equiv 0.0$ & -4.0(10.0) & - & 21.0(28.0) & -  \\
b & 6.20973(0.00019) & 2719.0476(0.0011) & 10.0(14.0) & $\equiv 0.0$ & $\equiv \pi/2$ & - & - & - & - & - & -  \\
\hline
TOI 2128 (155060) & - & - & - & - & - & 0.0016(0.0025) & 1.84e-05(7.7e-06) & -0.38(0.61) & 3.09(0.65) & 3.81(0.37) & 6.01(0.51)  \\
b & 16.34136(0.00013) & 1987.2663(0.0017) & 1.1(0.5) & $\equiv 0.0$ & $\equiv \pi/2$ & - & - & - & - & - & -  \\
\hline
TOI 2145 (HIP86040) & - & - & - & - & - & $\equiv 0.0$ & $\equiv 0.0$ & 29.0(4.3) & - & -25.6(3.3) & -  \\
b & 10.261127(1.1e-05) & 2013.2801(0.0006) & 370.7(6.4) & 0.21(0.018) & 1.682(0.049) & - & - & - & - & - & -  \\
\hline
\enddata
\tablecomments{The first line for each target displays fitted parameters for the system as a whole such as RV offset for each instrument, linear and quadratic terms, etc. Values in parenthesis represent the 1-$\sigma$ errors for each value. BJD is given relative to the TESS standard $T_{\text{0}}$-2457000.}
\end{deluxetable*}
\end{longrotatetable}
\renewcommand{\arraystretch}{1.0} 

\renewcommand{\arraystretch}{1.5}
\startlongtable
\begin{deluxetable*}{c|ccccc}
\tabletypesize{\normalsize}
\tablecaption{GP Fitted Parameters \label{tab:GPfitted}}
\tablehead{
  \colhead{TOI (TKS ID)} & 
  \colhead{$\eta_1$ (HIRES)} & 
  \colhead{$\eta_1$ (APF)} &
  \colhead{$\eta_2$} &
  \colhead{$\eta_3$} &
  \colhead{$\eta_4$}}
\startdata
1759 (T001759) & $5.1^{+1.6}_{-1.1}$ & -- & $178^{+140}_{-171}$ & $36.63^{+0.9}_{-0.53}$ & $0.487^{+0.049}_{-0.049}$ \\
260 (HIP 1532) & $3.12^{+0.72}_{-0.6}$ & -- & $274^{+170}_{-170}$ & $36.6^{+0.8}_{-0.44}$ & $0.474^{+0.051}_{-0.048}$ \\
1174 (T001174) & $6.4^{+5.2}_{-6.3}$ & -- & $145^{+100}_{-99}$ & $11.49^{+0.35}_{-0.35}$ & $0.501^{+0.051}_{-0.05}$ \\
1272 (T001272) & $0.032^{+0.094}_{-0.018}$ & -- & $81^{+210}_{-42}$ & $28.31^{+0.63}_{-0.62}$ & $0.5^{+0.052}_{-0.051}$ \\
2076 (T002076) & $36.7^{+8.7}_{-7.4}$ & $55^{+14}_{-11}$ & $161^{+77}_{-37}$ & $7.367^{+0.02}_{-0.022}$ & $0.381^{+0.057}_{-0.049}$ \\
1473 (6061) & $3.7^{+2.5}_{-3.1}$ & $9.6^{+2.2}_{-2.2}$ & $27^{+30}_{-11}$ & $13.87^{+0.81}_{-0.87}$ & $0.488^{+0.052}_{-0.05}$ \\
1430 (235088) & $2.1^{+3.0}_{-1.9}$ & $0.95^{+4.8}_{-0.76}$ & $15.5^{+300}_{-7.7}$ & $7.6^{+1.6}_{-2.3}$ & $0.498^{+0.051}_{-0.05}$ \\
1726 (63433) & $15^{+16}_{-15}$ & $44.7^{+7.4}_{-5.9}$ & $91^{33}_{-40}$ & $6.382^{+0.021}_{-0.016}$ & $0.367^{+0.045}_{-0.05}$ \\
1807 (HIP 65469) & $33^{+5}_{-4}$ & -- & $64.6^{+7.9}_{-7.6}$ & $8.728^{+0.021}_{-0.022}$ & $0.333^{+0.028}_{-0.027}$ \\
1347 (T001347) & $12.4^{+2.9}_{-1.9}$ & -- & $111^{+76}_{-40}$ & $16.47^{+0.12}_{-0.17}$ & $0.460^{+0.052}_{-0.05}$ \\
1824 (T001824) & $5.16^{+0.68}_{-0.58}$ & $4.68^{+0.73}_{-0.68}$ & $10.3^{2.9}_{-2.3}$ & $8.36^{+0.47}_{-0.51}$ & $0.518^{+0.048}_{-0.046}$ \\
\enddata
\tablecomments{Not all systems have APF data. \\ $\eta_1$ - GP amplitude for given instrument.\\ $\eta_2$ - GP exponential scaling.\\ $\eta_3$ - GP period. \\$\eta_4$ - GP periodic scaling }
\end{deluxetable*}
\renewcommand{\arraystretch}{1.0}

\renewcommand{\arraystretch}{1.2}
\startlongtable
\begin{deluxetable*}{l | c}
\tablecaption{Summary of HRI Observations \label{tab:hri_obs}}
\tablehead{
  \colhead{TOI (TKS ID)} & 
  \colhead{Telescope/Instrument (Filter)}  
}
\startdata
260 (HIP 1532)& Gemini/'Alopeke (562 nm, 832 nm) WIYN/NESSI (562 nm, 832 nm) \\&Palomar/PHARO (Br$\gamma$, H-continuous) Gemini/Zorro (832 nm, 562 nm) \\&Keck2/NIRC2 (Br$\gamma$)  \\
\hline
1173 (T001173) & Gemini/'Alopeke (562 nm, 832 nm) Keck2/NIRC2 (Br$\gamma$) \\&SAI-2.5m/Speckle (I$_{c}$)  \\
\hline
1174 (T001174) & Gemini/'Alopeke (562 nm, 832 nm) 2.2m@CAHA/AstraLux ($z$) \\&Shane/ShARCS (J, K-short)  \\
\hline
1180 (T001180) & 2.2m@CAHA/AstraLux ($z$) WIYN/NESSI (832 nm) SAI-2.5m/Speckle (I$_{c}$)  \\
\hline
1184 (T001184) & WIYN/NESSI (562 nm, 832 nm) Shane/ShARCS (J, K-short) \\&2.2m@CAHA/AstraLux ($z$) Gemini/NIRI (Br$\gamma$) \\
\hline
1194 (T001194) & Shane/ShARCS (K-short, J) Gemini/'Alopeke (562 nm, 716 nm) \\
\hline
1244 (T001244) & WIYN/NESSI (562 nm, 832 nm) Shane/ShARCS (K-short, J) \\&2.2m@CAHA/AstraLux ($z$) SAI-2.5m/Speckle (I$_{c}$) \\&Keck2/NIRC2 (K)  \\
\hline
1248 (T001248) & Gemini/'Alopeke (562 nm, 832 nm) Shane/ShARCS (J, K-short) \\&Palomar/PHARO (Br$\gamma$)  \\
\hline
1249 (T001249) & Gemini/'Alopeke (832 nm, 562 nm) Shane/ShARCS (J, K-short) \\&2.2m@CAHA/AstraLux ($z$)  \\
\hline
1269 (T001269) & Shane/ShARCS (K-short, J) WIYN/NESSI (562 nm, 832 nm) \\&Palomar/PHARO (Br$\gamma$) 2.2m@CAHA/AstraLux ($z$) \\
\hline
1279 (T001279) & Gemini/NIRI (K, Br$\gamma$) Gemini/'Alopeke (832 nm, 562 nm) \\&Shane/ShARCS (J, K-short) Palomar/PHARO (Br$\gamma$) \\&2.2m@CAHA/AstraLux ($z$)  \\
\hline
1436 (T001436) & Gemini/'Alopeke (832 nm, 562 nm) 2.2m@CAHA/AstraLux ($z$) \\&Keck2/NIRC2 (Br$\gamma$)  \\
\hline
1438 (T001438) & Gemini/'Alopeke (562 nm, 832 nm) WIYN/NESSI (832 nm, 562 nm) \\&SAI-2.5m/Speckle (I$_{c}$)  \\
\hline
1443 (T001443) & Gemini/NIRI (Br$\gamma$, K) SAI-2.5m/Speckle (I$_{c}$)  \\
\hline
1451 (T001451) & Gemini/'Alopeke (562 nm, 832 nm) Gemini/NIRI (Br$\gamma$, K) \\&Palomar/PHARO (Br$\gamma$) 2.2m@CAHA/AstraLux ($z$) \\
\hline
1467 (T001249) & Gemini/'Alopeke (562 nm, 832 nm) Keck2/NIRC2 (K) \\
\hline
1472 (T001467) & SAI-2.5m/Speckle (I$_{c}$) Gemini/'Alopeke (562 nm, 832 nm) \\&Shane/ShARCS (K-short) Keck2/NIRC2 (Br$\gamma$) \\
\hline
1669 (T001669) & Gemini/'Alopeke (832 nm, 562 nm) WIYN/NESSI (832 nm, 562 nm) \\&2.2m@CAHA/AstraLux ($z$) SAI-2.5m/Speckle (880 nm) \\
\hline
1691 (T001691) & Gemini/'Alopeke (832 nm, 562 nm) 2.2m@CAHA/AstraLux ($z$)  \\
\hline
1716 (237566) & Gemini/'Alopeke (562 nm, 832 nm) Shane/ShARCS (K-short) \\&SAI-2.5m/Speckle (880 nm) Palomar/PHARO (Br$\gamma$) \\
\hline
1723 (T001723) & WIYN/NESSI (832 nm, 562 nm) 2.2m@CAHA/AstraLux ($z$) \\&SAI-2.5m/Speckle (880 nm) Shane/ShARCS (K-short) \\
\hline
1742 (156141) & Gemini/'Alopeke (562 nm, 832 nm) Keck2/NIRC2 (Br$\gamma$)  \\
\hline
1753 (T001753) & Gemini/'Alopeke (562 nm, 832 nm) Shane/ShARCS (J, K-short) \\&Palomar/PHARO (Br$\gamma$)  \\
\hline
1758 (T001758) & SAI-2.5m/Speckle (625 nm) Gemini/'Alopeke (832 nm, 562 nm) \\&2.2m@CAHA/AstraLux ($z$)  \\
\hline
1775 (T001775) & WIYN/NESSI (832 nm, 562 nm) SAI-2.5m/Speckle (I$_{c}$) \\&Shane/ShARCS (K-short) Palomar/PHARO (Br$\gamma$) \\
\hline
1776 (95072) & Palomar/PHARO (Br$\gamma$) Shane/ShARCS (Br$\gamma$, J) \\&Gemini/'Alopeke (562 nm, 832 nm)  \\
\hline
1778 (77496) & Gemini/'Alopeke (562 nm, 832 nm) Shane/ShARCS (K-short) \\&Palomar/PHARO (Br$\gamma$) SAI-2.5m/Speckle (880 nm)  \\
\hline
1794 (T001794) & Gemini/'Alopeke (832 nm, 562 nm) Palomar/PHARO (Br$\gamma$, H-continuous) \\&Shane/ShARCS (K-short) 2.2m@CAHA/AstraLux ($z$) \\
\hline
1798 (T001798) & Gemini/'Alopeke (832 nm, 562 nm) Keck2/NIRC2 (Br$\gamma$)  \\
\hline
1799 (96735) & Palomar/PHARO (Br$\gamma$) Shane/ShARCS (K-short, J) \\&Gemini/'Alopeke (562 nm, 832 nm)  \\
\hline
1823 (TIC142381532) & Shane/ShARCS (J, K-short) Gemini/'Alopeke (562 nm, 832 nm) \\&Palomar/PHARO (Br$\gamma$)  \\
\hline
2045 (T002045) & Shane/ShARCS (K-short) SAI-2.5m/Speckle (I$_{c}$) \\
\hline
2088 (T002088) & 2.2m@CAHA/AstraLux ($z$) SAI-2.5m/Speckle (I$_{c}$) \\&WIYN/NESSI (832 nm)  \\
\hline
2114 (T002114) & Shane/ShARCS (Ks, J) Palomar/PHARO (Br$\gamma$)  \\
\hline
2128 (155060) & SAI-2.5m/Speckle (625 nm) Shane/ShARCS (Br$\gamma$, J) \\&2.2m@CAHA/AstraLux ($z$) Palomar/PHARO (Br$\gamma$, H-continuous) \\
\hline
\enddata
\end{deluxetable*}
\renewcommand{\arraystretch}{1.0}

\bibliography{bib.bib}

\end{document}